\documentclass[twocolumn]{aastex63}
\usepackage{multirow}

\newcommand{\kms}{km\,s$^{-1}$}
\newcommand{\SN}{S$/$N}

\newcommand{\hii}{\ion{H}{2}}

\newcommand{\hei}{\ion{He}{1}}

\newcommand{\Ni}{[\ion{N}{1}]}
\newcommand{\nii}{[\ion{N}{2}]}
\newcommand{\oi}{[\ion{O}{1}]}
\newcommand{\OI}{\ion{O}{1}}
\newcommand{\oii}{[\ion{O}{2}]}
\newcommand{\oiii}{[\ion{O}{3}]}
\newcommand{\sii}{[\ion{S}{2}]}
\newcommand{\siii}{[\ion{S}{3}]}
\newcommand{\ha}{H$\alpha$} 
\newcommand{\hb}{H$\beta$} 
\newcommand{\feii}{[\ion{Fe}{2}]}
\newcommand{\cliii}{[\ion{Cl}{3}]}
\newcommand{\ariii}{[\ion{Ar}{3}]}
\newcommand{\SiII}{\ion{Si}{2}}

\newcommand{\Ha}{$\rm{H}\alpha$}
\newcommand{\Hb}{$\rm{H}\beta$}

\newcommand{\EWha}{W$_{{\rm H}\alpha}$}
\usepackage{makecell}

\received{June 1, 2019}
\revised{January 10, 2019}
\accepted{\today}
\submitjournal{AJ}

\shorttitle{The AMUSING++ sample}
\shortauthors{Lopez-Coba et al.}
\graphicspath{{./}{figures/}}

\begin{document}

\title{The AMUSING++ Nearby Galaxy Compilation: \\ I. Full Sample Characterization and Galactic--Scale Outflows Selection.}

\correspondingauthor{C. L\'opez-Cob\'a}
\email{clopez@astro.unam.mx}

\author[0000-0003-1045-0702]{Carlos ~L\'opez-Cob\'a}
\affiliation{Instituto de Astronom\'ia, Universidad Nacional Aut\'onoma de  M\'exico \\
Circuito Exterior, Ciudad Universitaria, Ciudad de M\'exico 04510,  Mexico}

\author[0000-0001-6444-9307
]{Sebasti\'an~F.~S\'anchez}
\affiliation{Instituto de Astronom\'ia, Universidad Nacional Aut\'onoma de  M\'exico \\
Circuito Exterior, Ciudad Universitaria, Ciudad de M\'exico 04510,  Mexico}

\author[0000-0003-0227-3451]{Joseph~P.~Anderson}
\affiliation{European Southern Observatory, Alonso de C\'ordova 3107, Vitacura, Casilla 190001, Santiago, Chile}

\author[0000-0002-2653-1120
]{Irene~Cruz-Gonz\'alez}
\affiliation{Instituto de Astronom\'ia, Universidad Nacional Aut\'onoma de  M\'exico \\
Circuito Exterior, Ciudad Universitaria, Ciudad de M\'exico 04510,  Mexico}

\author[0000-0002-1296-6887]{Llu\'is~Galbany}
\affiliation{Departamento de F\'isica Te\'orica y del Cosmos, Universidad de Granada, E-18071 Granada, Spain}

\author[0000-0001-6984-4795]{Tom\'as~Ruiz-Lara}
\affiliation{Instituto de Astrof\'isica de Canarias, E-38200 La Laguna, Tenerife, Spain}
\affiliation{Departamento de Astrof\'isica, Universidad de La Laguna, E-38205 La Laguna, Tenerife, Spain}

\author{Jorge~K.~Barrera-Ballesteros}
\affiliation{Instituto de Astronom\'ia, Universidad Nacional Aut\'onoma de  M\'exico \\
Circuito Exterior, Ciudad Universitaria, Ciudad de M\'exico 04510,  Mexico}

\author[0000-0003-1072-2712]{Jos\'e~L.~Prieto}
\affiliation{N\'ucleo de Astronom\'ia de la Facultad de Ingenier\'ia y Ciencias, Universidad Diego Portales, Av. Ej\'ercito 441 Santiago, Chile}
\affiliation{Millennium Institute of Astrophysics, Santiago, Chile}

\author{Hanindyo~Kuncarayakti}
\affiliation{Tuorla Observatory, Department of Physics and Astronomy, FI-20014 University of Turku, Finland}
\affiliation{Finnish Centre for Astronomy with ESO (FINCA), FI-20014 University of Turku, Finland}




\begin{abstract}
We present here AMUSING\textrm{++}; the largest compilation of nearby galaxies observed with the MUSE integral field spectrograph so far. This collection consists of 635 galaxies from different MUSE projects covering the redshift interval $0.0002<z<0.1$. The sample and its main properties are characterized and described in here. It includes galaxies of almost all morphological types, with a good coverage in the color-magnitude diagram, within the stellar mass range between 10$^8$ to 10$^{12}$M$_\odot$, and with properties resembling those of a diameter-selected sample. The AMUSING++ sample is therefore suitable to study, with unprecendent detail, the properties of nearby galaxies at global and local scales, providing us with more than 50 million individual spectra. We use this compilation to investigate the presence of galactic outflows. We exploit the use of combined emission-line images to explore the shape of the different ionized components and the distribution along classical diagnostic diagrams to disentangle the different ionizing sources across the optical extension of each galaxy. We use the cross correlation function to estimate the level of symmetry of the emission lines as an indication of the presence of  shocks and/or active galactic nuclei. We uncovered a total of 54 outflows, comprising $\sim$8\% of the sample. A large number of the discovered outflows  correspond to those driven by active galactic nuclei ($\sim$60\%), suggesting some bias in the selection of our sample. No clear evidence was found that outflow host galaxies are highly star-forming, and outflows appear to be found within all galaxies around the star formation sequence.
\end{abstract}



\section{Introduction} \label{sec:intro}

{ 
Galactic outflows are phenomena that are predicted by theoretical models of galaxy evolution and observed in a wide variety of galaxies at many different redshifts. 
They can be driven either by supernova (SN) explosions or by an active galactic nucleus (AGN), through mechanisms that inject energy into both  the interstellar and intergalactic medium (hereafter the ISM and IGM, respectively). They are, indeed, the most common explanation to the observed metal enrichment of the intergalactic medium  \citep[e.g.,][]{Pettini1998,Ferrara2000}. How the energy released in these processes is dissipated through the disks and how much gas mass is expelled, are key questions to explain whether outflows are capable of quenching the SF in galaxies, and therefore explain the transition to the observed retired population of galaxies \citep[e.g.][]{Bower2006,hopkins09}. On the other hand, some studies have suggested that outflows can inject positive feedback and trigger galaxy SF, instead of halting it \citep[e.g.,][]{silk13,Zubovas2013,Maiolino2017,Gallagher2019}.

Outflows driven by SF have been clearly identified  in nearby galaxies, particularly in Luminous and Ultra Luminous Infrared Galaxies (LIRGs and ULIRGs) and starbursts  \cite[e.g.,][]{Heckman2000,Aguirre2001,rupke2005a,Rupke2005b,Rupke2005c}, although they are neither ubiquitous nor exclusive of galaxies with high rates of star formation \citep[SF; see][]{Ho2014,clc2018}. 
 It is believed that the precense of this type of outflows is closely related to the amount of SFR in a galaxy. Depending on how intense and efficient the SF is in producing massive stars, via the initial mass function (IMF), those stars will eventually produce supernovae explosions in few mega years, injecting energy to its surroundings. Its eventual expansion into the ISM produces typically ionized cones result of the stratified density between the disk and the gaseous halo.

Despite these outflows are usually found in low mass galaxies \citep[e.g.,][]{Veilleuxetal2005}, where they apparently prevent the formation and growth of dwarf galaxies \citep[e.g.,][]{Silk1998}, recent studies have shown they are also present in galaxies as massive as $\sim10^{11}\,\mathrm{M}_\sun$ or more. 

On the other hand, super massive black holes in the center of galaxies are responsible of launching powerful radio jets, sweeping the surrounding ISM to form outflows. The energy source of this is the accretion of material  onto the central black hole of the galaxies. Most massive galaxies tend to host massive black holes. Therefore the produced energy when active is some orders of magnitude (assuming a high efficiency  mass--energy conversion,  typically 0.1 ), larger than that produced by SN explosions, surpassing in some cases the binding energy of a hole galaxy \citep{Veilleuxetal2005,Harrison2018}. These outflows are usually found in the most luminous AGN.

Regardless of its origin, the warm phase of outflows (T\,$\approx 10^{4}$ K) is the most accessible part to explore them given their strong emission in optical emission lines. In particular, the high-excitation \oiii$\lambda 5007$\, line (hereafter \oiii) traces AGN winds in general, while \ha\,+\,\nii$\lambda 6584$\, (hereafter \nii) traces SF outflows produced by supernova explosions \citep{Veilleuxetal2005,Sharp2010}. Therefore, in outflows emerging from the disk an increase is expected of these emission lines with respect to hydrogen recombination lines along the semi-minor axis. 
These lines reveal typically ionized gas with conical structures as the result of the expansion of the gas and its interaction with the ISM \citep[e.g.,][]{carlos16}, which filamentary shapes. The extension of an outflow ranges from a few parsecs to a few kiloparsecs \citep{Heckman1990}, depending on how intense and efficient in the SF in producing massive stars or in how luminous is the  AGN.
}

Before any detailed study of the physical conditions of outflows, it is necessary to have large samples of bonafide outflows across a broad range of galaxy properties. The study of outflows has been addressed using different observational techniques and at different wavelength ranges, from X-ray to radio wavelengths \citep[e.g.,][]{Husemann2019}.

The methodology applied to detect and study galactic outflows has been improved by the implementation of modern observational techniques, going from an incomplete vision provided by long slit spectroscopy to the fully spatially resolved picture provided by Integral Field Spectroscopy (IFS). This technique provides a spatial and spectral description of galaxies, limited only by the specifications of the spectrographs (and telescopes). However, to date there is a lack of robust methods to detect ionized gas outflows in large samples of galaxies.

Even though the outflowing ionized gas is more or less constrained by models \citep[e.g.,][]{mappings}, further information, such as velocity dispersion, multiple kinematic components, distance to the mid plane, and morphology of the ionized gas, are required to identify shocks beside the use of pure ionization diagnostic diagrams \citep{clc2018,Agostino2019}.

Recent large IFS galaxy surveys (IFS-GS), like MaNGA \citep{mangaOLD}, CALIFA \citep{sanchez12a} and SAMI \citep{samiOLD} have enabled investigations into the presence of outflows at kiloparcec scales in the nearby Universe 
\citep[e.g.,][]{Ho2014,clc2018,Pino2019}. Our understanding of this phenomena and its impact on the overall evolution of galaxies would improve with the detection of larger and less biased samples of host galaxies \citep{clc2018}. However, all those explorations have been limited by the spatial resolution of the above surveys ($\sim$2.5$\arcsec$).

New instruments, like the Multi Unit Spectroscopic Explorer \citep[MUSE;][]{MUSE} with its unprecedented combination of high spatial and spectral resolutions, provides a new way to study galaxies at scales of hundreds of parsecs. While there does not exist a MUSE galaxy survey with a large sample size that matches the numbers from previous IFS-GSs, there are now available multiple distinct projects (with public data) from which it is possible to create a synthetic compilation sample.

In the present work we compile the `AMUSING++' nearby galaxy sample and use this to identify and study galactic outflows at sub-kiloparsec scales over a large number of different galaxies. The paper is structured as follows: the presentation of the AMUSING++ sample is presented in  \S2; the data analysis in \S3; the  methodology used to select outflows is in \S4, and the presentation of the outflows sample is in \S5; some scaling relations of the sample are presented in \S6.

Throughout the paper we adopt the standard $\Lambda$CDM cosmology with H$_0$\,=\,70 \kms\,Mpc$^{-1}$, $\Omega_\mathrm{m}$\,=\,0.3, and  $\Omega_\mathrm{\Lambda}$\,=\,0.7.

\section{The AMUSING++ sample}\label{sec:amusing_presentation}

The MUSE instrument provides  a wide field of view (FoV) of $1\arcmin\times1\arcmin$, with a spatial sampling of $0.\arcsec2\times0.\arcsec2$ per spaxel, thus, the spatial resolution is seeing limited. MUSE covers the whole optical range from 4750 \AA\ to 9300 \AA, with a spectral sampling of 1.25 \AA, and a full width at half maximum (FWHM) that depends slightly on the wavelength \citep{Bacon2017}, being $\sim 2.4$ \AA\  at the red part of the spectrum (at 7500 \AA). Although the instrument was designed to study intermediate/high redshift objects, MUSE is an excelent instrument to study with unprecedented detail, the structural components of nearby galaxies \citep[e.g.][]{laura18}.

As indicated before, the study of galactic outflows has been limited by the coarse spatial resolution of the previous IFS-GS (FWHM $\sim2.5\arcsec$). 
The seeing limited resolution of MUSE (FWHM $\sim1\arcsec$), allows selecting galaxies where the spatial resolution is of the order of sub--kiloparsec scales at similar redshifts. Therefore, we selected galaxies observed with this instrument
from the European Southern Observatory (ESO) archive, acquired until August 2018 with redshifts below $z< 0.1$. This is basically the highest redshift covered by the previously quoted IFS-GS. We perform a visual inspection to select galaxies that fit into the FoV of MUSE. A more detailed diameter selection cannot be applied given the lack of information of the R$_{25}$  parameter for a large fraction of these Southern galaxies (based on a scan through the Hyperleda database). Nevertheless, as we will argue later, our final galaxy collection broadly resemble a diameter selection sample (see Fig.~\ref{fig:redshift}).

Nearby galaxies ($z\sim0$), where the optical extension is not covered entirely by the MUSE FoV were treated separately.
In those cases we selected galaxies where at least the optical nuclei is covered, since galactic--outflows are nuclear processes. 
Partial pointings, where just a small fraction of a galaxy is observed, are excluded from the sample (i.e., spiral arms, bars, tails), except in cases where it was possible to do mosaics in order to cover larger areas of a galaxy. 
All together the current compilation comprises a total of 635 galaxies observed with  MUSE at the VLT, covering the redshift interval  $0.0002\,<\,z\,<\,0.1$, with a mean value of $\sim0.019$. 
The full list of ESO MUSE programs used in the present galaxy compilation is presented in the acknowledgement section.

Galaxies from different MUSE projects were collected by using the previous selection criterion. The projects with larger data contributions to the final AMUSING++ collection are briefly described below (more details should be found in the presentation articles of each project):

(i) The All-weather MUse Supernova Integral-field of Nearby Galaxies \citep[AMUSING;][]{AMUSING} survey. AMUSING is an ongoing project at the ESO, that aims to study the environments of supernovae and the relation with their host galaxies. For details about the observation strategy and data reduction we refer to \citet{AMUSING}. So far it comprises $\sim$328 galaxies, being the core of the current compilation. This sample has been used to explore different science topics: (i) the radial profiles of the oxygen abundances in galaxies \citep{laura18}, and its azimuthal variation \citep{censusHII,laura16b}; (ii) extended ionized gas fillaments associated to galaxy interactions \citep{prieto16}; (iii) the discovery of new strong lenses \citep{galbany18}, and the optical counterpart of a radio-jet \citep{clc17}; (iv) the derivation of main galaxy kinematic parameters such as velocity and velocity dispersion by different approaches \citep{Bellocchi2019}; (v)  individual type II supernova \citep{Meza2019}; (vi) ionized gas tails \citep{Boselli2018}; besides (vi) the local environment of supernovae, i.e., the major goal of the survey \citep{AMUSING,Kruhler2017}.

 (ii) CALIFA galaxies observed with MUSE. In order to compare with previous analysis \citep[e.g.,][]{clc2018}, we selected all those galaxies observed within the CALIFA survey \citep{sanchez12a} covered with MUSE. This sample comprises 6 galaxies so far. In addition we searched through the ESO archive looking for any galaxy within the footprint of the CALIFA selection (redshift, magnitude, diameter), relaxing the declination limits to include all Southern galaxies, this results in 41 galaxies.
 
 (iii) The  GAs Stripping Phenomena in galaxies with MUSE \citep[GASP;][]{GASP}. This project has observed 114 stripping candidates galaxies at redshifts $0.04<z<0.07$. GASP aims to study the gas removal process in galaxies due to this physical process, i.e., when galaxies fall into clusters. They also observe a comparison sample of field galaxies. In this study 26 galaxies are included from this sample. 
 
 (iv) The MUSE Atlas of Disks \citep[MAD;][]{MAD}. This is an ongoing project that studies star forming galaxies at very low redshift. 
 So far, MAD has observed 38 galaxies. MAD is focused in  the study of the properties of the ionized gas, such as oxygen abundances, star formation rates, in local disks at scales of hundreds of pc. In the present study  22  of the 38 galaxies from this survey are included.
 
 (v) The Close AGN Reference Survey \citep[CARS,][]{CARS}. CARS aims to explore the AGN-host galaxy connection over a sample of  40 nearby unobscured AGN ($0.01<z<0.06$), and thus establish a connection towards  high-redshift AGNs. Our compilation includes 12 CARS galaxies.
 
 (vi) The Time Inference with MUSE in Extragalactic Rings \citep[TIMER;][]{TIMER}. TIMER is a project that observed 24 nearby  barred galaxies (z $<0.0095$), with rings or inner disks. The goal of TIMER is to understand when the disk galaxies settle dynamically. The target galaxies present isophotal  sizes slightly larger than the FoV of the instrument ($D_{25} > 1\arcmin$). Seven galaxies from this project are included in our study.
 
As indicated before nearly two thirds of the galaxy compilation were extracted from the AMUSING survey, and for this reason we named it AMUSING++. 

\begin{figure}
    \centering
    \includegraphics[width=\columnwidth]{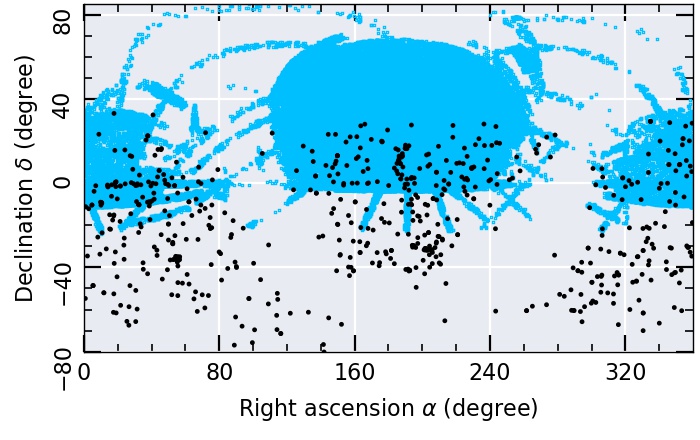}
    \caption{Distribution in the sky of AMUSING++ galaxies. Black dots represent the 634 galaxies analysed in this paper. For comparison we have added the distribution of galaxies in the NASA-Sloan Atlas catalogue (NSA), shown with cyan dots.}
    \label{fig:alpha_delta}
\end{figure}

Figure~\ref{fig:alpha_delta} shows the distribution on the sky for all the galaxies analysed in this study on top of the NASA-Sloan Atlas (NSA) catalogue \citep{nsa}, for comparison. As the VLT is located in the Southern hemisphere, just a few galaxies coincide with the Sloan Digital Sky Survey \citep[SDSS,][]{york00}. This limits our ability to extract useful information from this exquisite survey, like photometry, comparison of spectroscopy at the same aperture, or even perform an estimation of the volume correction just assuming a random sub-selection of the targets \citep[like the one presented by][]{Sanchez2018}. The declination limits of our sample reflect the sky visibility of the VLT, $-80^{\circ}\,<\,\delta\,<\,40^{\circ}$. On the other hand the sample is distributed randomly around any right accession (R.A.), once considering the region coincident with the Milky Way disk. Therefore, it is well suitable for any further survey-mode exploration along the year with a telescope or antennae in the Southern Hemisphere.

\begin{figure*}
    \centering
        \includegraphics[]{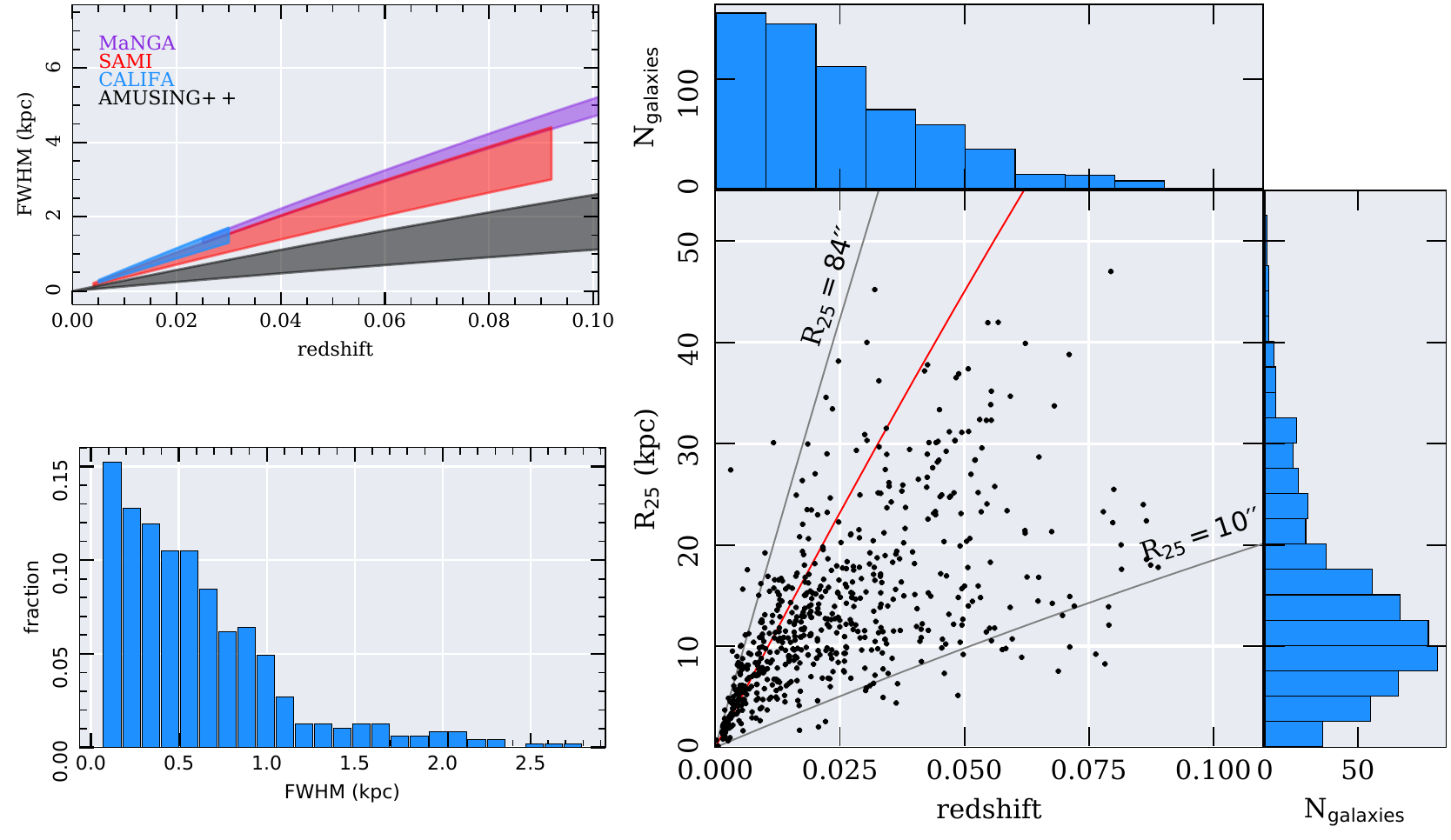}
    \caption{{\it Top-left panel:} PSF spatial resolution (FWHM) vs. redshift for different IFU galaxy surveys together with the AMUSING++ compilation. The shaded color regions accounts for 1$\sigma$ error in the reported FWHMs. The purple region shows the distribution of the MaNGA survey ($2.54\arcsec<\mathrm{FWHM}<2.8\arcsec$, $0.025<\mathrm{z}<0.15$) \citep{Yan2016b}, the red corresponds to SAMI ($\mathrm{FWHM} = 2.16\arcsec \pm 0.41\arcsec, 0.004<\mathrm{z}<0.092$) \citep{Green2018}, in blue is CALIFA ($\mathrm{FWHM} = 2.5\arcsec \pm 0.34\arcsec, 0.005<\mathrm{z}<0.03$) \citep{CALIFA3}. Finally in black the AMUSING++ compilation is shown with $0.6\arcsec<\mathrm{FWHM_{DIMM}}<1.4\arcsec$ covering a redshift interval  $0.0002 <z< 0.1$. { {\it Bottom-left panel:} Distribution of the physical spatial resolution of the data normalized to the total number of galaxies.}  {\it Right panel:} distribution of the $\mathrm{R_{25}}$ parameter along the redshift for the AMUSING++ objects. Diagonal lines confine galaxies between $10\arcsec<\mathrm{R_{25}}<84\arcsec$. The red line indicates $\mathrm{R_{25}} = 42\arcsec$, that is, the maximum radius that fits into FoV of MUSE. Over each axis are plotted histograms of the corresponding $z$ and $\mathrm{R_{25}}$ distributions.    }
    \label{fig:redshift}
\end{figure*}

The redshift coverage of AMUSING++ spans over the range covered by other large IFS-GS (such as MaNGA, CALIFA and SAMI, see  Fig.~\ref{fig:redshift}, top panel). The physical spatial resolution was derived
for each object by extracting the DIMM seeing along the observations from the header of each datacube and shifting it to the corresponding cosmological distance. The average seeing of the sample is $1.0\arcsec$ with a standard deviation of $0.4 \arcsec$. This corresponds to typical physical resolution of $\sim$400\,pc for the average redshift of the sample, although it ranges from 10 pc (for the lowest redshift galaxy) to $\sim$3 kpc (for the highest redshift ones). Fig.~\ref{fig:redshift} demonstrates that at any redshift interval, the AMUSING++ sample offers a better spatial physical resolution with respect to the IFS-GS mentioned above. 
Thus, spatial resolution is clearly one of the major advantages of the considered dataset. However, we stress that the current dataset does not comprise a homogeneously selected and well defined sample, being a collection of different galaxies observed with MUSE. 

Given the redshift range of the sample, the optical diameter (D$_{25}$) is not covered completely  by the FoV of MUSE for all galaxies of the sample. In a few cases (3 galaxies) there are multiple pointings available for the same galaxy, from which we performed mosaics joining together the datacubes to cover the maximum extension for those galaxies. In order to estimate which fraction of the optical extension of each galaxy is covered by our IFS data, we perform an isophotal analysis on the V-band images extracted from the datacubes, deriving the position angle, ellipticity and R$_{25}$ isophotal radii
(the semi-major axis length at which a surface brightness of 25 mag arcsec$^{-2}$ is reached). For this purpose, we adopt the publicly available isophote fitting tool {\sc Photutils} \citep{photoutils} as part of the python based package {\sc Astropy} \citep{Astropy}. This routine mimics the standard procedures implemented in the {\sc SExtractor} package \citep[e.g.,][]{sextractor}. An interactive tool was developed for this propose, including a visual masking of foreground stars and/or close companion galaxies, a selection of the centroid of the galaxy, and certain tuning of the background level, in order to derive those parameters for all galaxies. In Appendix~\ref{appendix:isophote_params} the  procedure is described in more details, and the derived parameters are presented (see Table~\ref{tab:all_amusing_table}).      

The bottom panel of Fig.~\ref{fig:redshift} shows the distribution of R$_{25}$ as a function of redshift for our compilation. Unexpectedly, the distribution of galaxies seems to be grouped in a narrow region in this diagram. Indeed, it resembles a diameter selected sample, limited by $10\arcsec<\mathrm{R_{25}}<84\arcsec$ for 92\% of the objects. This property could be used, in principle, to provide  a volume correction, something that will be explored in a forthcoming article. The red line in this figure, traces the maximum isophotal radius that fits into the FoV of MUSE. For $\sim$~80\% of the sample we have a complete coverage of the optical extent of the AMUSING++ galaxies, that is $\mathrm{R}_{25}<42\arcsec$ (extension from the center to the corner of the MUSE FoV).

\begin{figure}[t]
    \centering
    \includegraphics[width=\columnwidth]{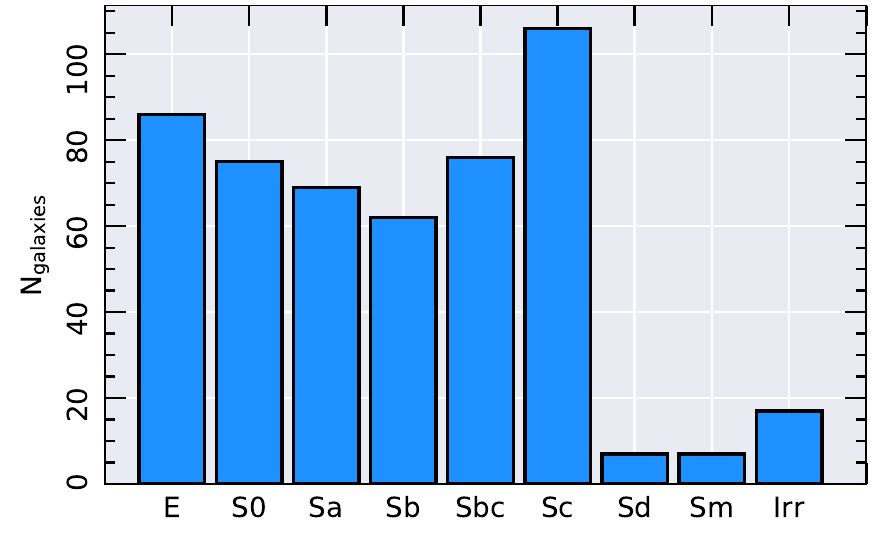}
        \includegraphics[width=\columnwidth]{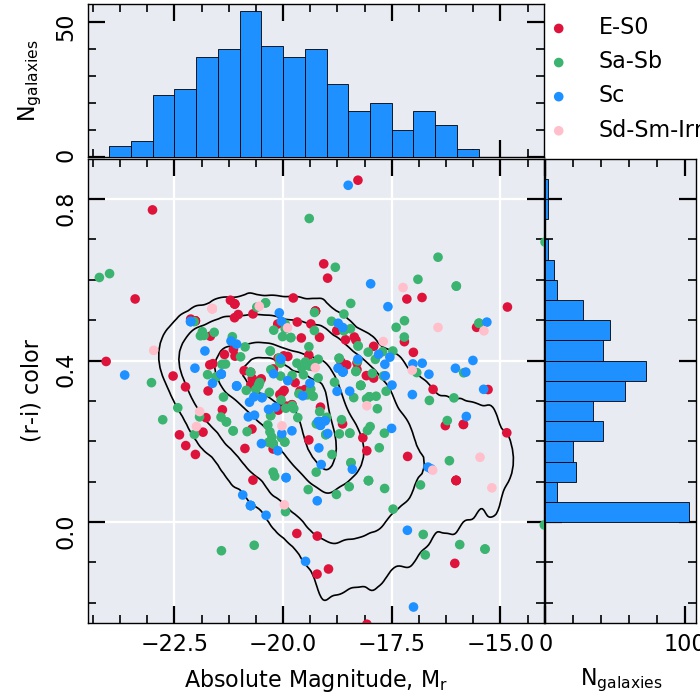}
    \caption{{\it Top:} Morphology distribution of the AMUSING++ sample. This parameter was extracted directly from the Hyperleda database \citep[][\url{http://leda.univ-lyon1.fr/}]{hyperleda}. For $\sim100$ galaxies there is no information about their morphology, therefore they were not included in this plot. {\it Bottom:} $r\mathrm{-}i$ color vs. M$_r$ absolute magnitude diagram for the sample: red circles comprise E-S0s; green circles include Sa-Sb-Sbc; blue circles Sc; and pink circles Sd-Sm-Irr. The contours represent the same distribution for the NSA catalogue, at different density levels (99\% the outermost, 95\%, 80\%, 65\%). Over each axis it is plotted a histogram of the distribution of the $r\mathrm{-}i$ color and M$_r$ absolute magnitude.  }
    \label{fig:morphology}
\end{figure}

The top panel of Fig.~\ref{fig:morphology} shows the morphological distribution of the sample. All types are covered by the compilation, with early types (E+S0) being one third of the total number, and the remaining ones being mostly late types (Sc mainly), with a low fraction of Sd-Sm-Irregulars. The bottom panel of Fig.~\ref{fig:morphology} shows the $g-r$  color vs. $r$-band absolute magnitude distribution. Like in the case of the morphological types, the AMUSING++ compilation covers a substantial fraction of this diagram.
A visual comparison with similar distributions presented by other IFS-GS, in particular CALIFA \citep{walcher14} or MaNGA \citep{Sanchez2018}, does not show any clear/strong difference. Thus, the current compilation does not seem to be biased towards a particular morphological type, color or magnitude.

\begin{figure*}
    \centering
    \includegraphics[width=\textwidth]{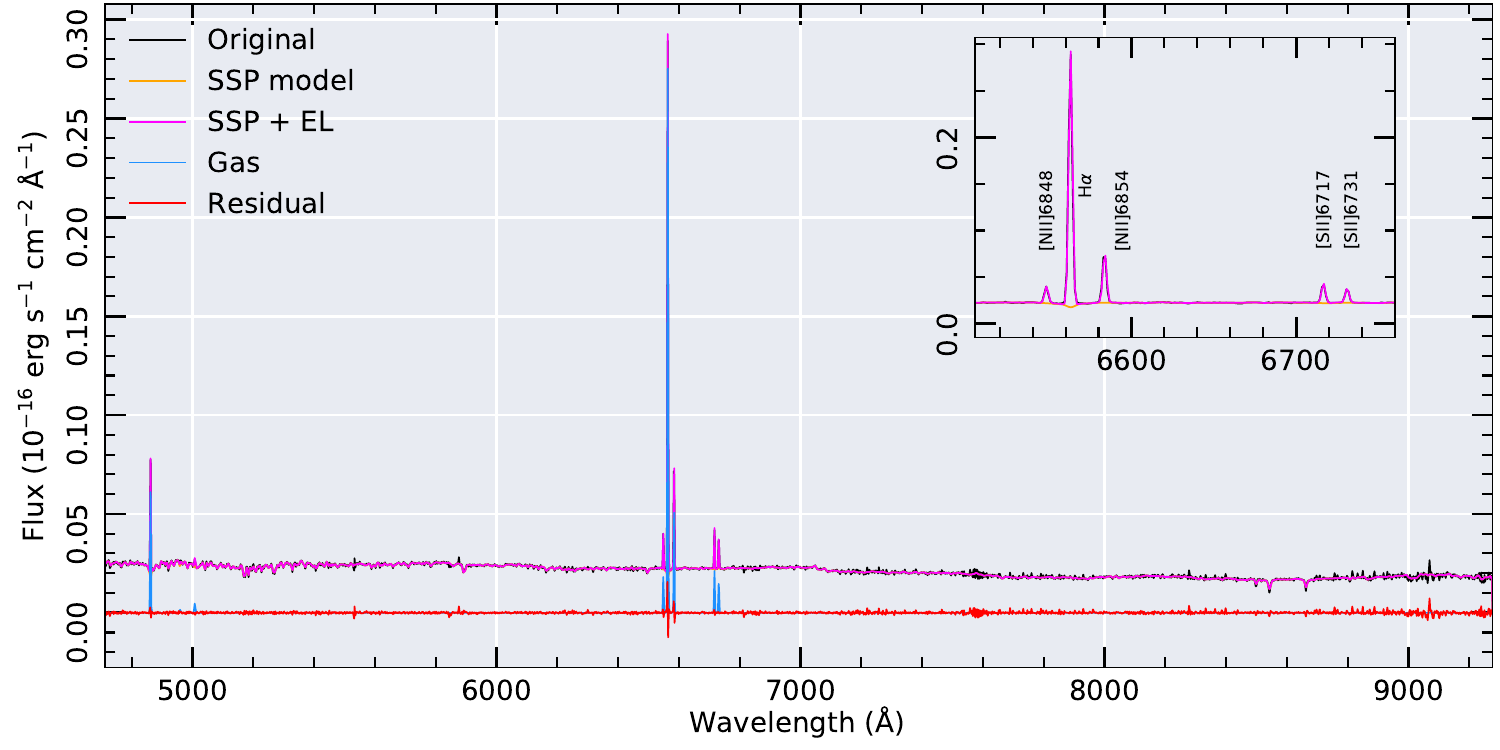}
    \caption{An example of the results of the fitting procedure to recover the best model of the stellar population and emission lines applied to a spectrum extracted from a galaxy within our compilation, NGC\,1762, shifted to the rest frame. In black is shown the spectrum corresponding to a spaxel from the nuclear region. In yellow is shown the best stellar population model. In magenta the best joint model of the multi-SSP fitting and emission lines (EL). In blue the pure gas spectrum after subtraction of the best fit SSP model to the original spectrum. Finally, in red it is shown the residuals of the fitting procedure. {\it Right upper inset:} Zoom of the same figure covering the \ha\, to \sii$\lambda\lambda$6717,6731 spectral window.}
    \label{fig:fitting}
\end{figure*}

\section{Data Analysis}
\label{sec:analysis}

The reduction of the AMUSING raw datacubes was performed with {\sc reflex}
\citep{REFLEX} using version 0.18.5 of the MUSE pipeline
\citep{MUSEpipeline} with default parameters. Also we use the processed datacubes downloaded directly from the ESO archive\footnote{\url{http://archive.eso.org/scienceportal/home}}. At this stage we perform a visual inspection of the reduced datacubes to exclude objects observed under very poor weather conditions (mostly bad seeing), with clear problems in the sky subtraction (plagued of residuals in all the spectral range) or showing problems in the combination of different cubes (vertical/horizontal patterns). In some cases the problems were not evident after performing a preliminary analysis of the data, as the one described below. Altogether poor datacubes correspond to a few percent of the compiled data-set, and they are all excluded from further considerations.

The analysis of the emission lines and the stellar population content of the datacubes was performed using the {\sc Pipe3D} pipeline \citep{Pipe3D_I}, a fitting routine adapted to analyse IFS  data using the package {\sc FIT3d} \citep{Pipe3D_II}. {\sc Pipe3d} has being extensively used in the analysis of datacubes from the main large IFS surveys: CALIFA \citep[e.g.][]{mariana16,laura18}, MaNGA \citep[e.g.][]{ibarra16,bb17,Sanchez2018,thorp19} and SAMI \citep{sanchez19}. This package provides the user with dataproducts that contain information of the emission lines and the stellar continuum.

The  fitting procedure is described in detail by \citet{Pipe3D_I}, here we provide just a brief description. 
The procedure starts by performing a spatial binning on the continuum (V-band) in order to increase the signal-to-noise (S$/$N) in each spectrum of the datacube preserving as much as possible the original shape of the light distribution. After that, all the spectra within each spatial bin are co-added and treated as a single spectrum. First, it is derived the stellar kinematics and stellar dust attenuation, using a limited set of SSPs comprising twelve populations. We adopted a stellar population library extracted from the MIUSCAT templates \citep[e.g.,][]{Vazdekis2012}, that cover the full optical range included in the MUSE spectra. This first step is performed to limit the effects of the degenerancy between metallicity, velocity dispersion and dust attenuation. Once these parameters are recovered, the final stellar population model is derived by performing a similar fitting procedure using an extensive SSP library. The actual {\sc Pipe3D} implementation adopts the GSD156 stellar library, that comprises 39 ages (from 1 Myr to 14 Gyr) and four metallicities (from 0.2 to 1.6 $Z_\odot$), extensively described in \citet{cid-fernandes13}, and used in previous studies \citep[e.g.][]{ibarra16,ellison2018,thorp19}. Then, a model of the stellar continuum in each spaxel is recovered by re-scaling the model within each spatial bin to the continuum flux intensity in the corresponding spaxel. The best model for the continuum is then subtracted to create a {\it pure gas} data-cube (plus noise).

A set of 30 emission lines within the MUSE wavelength range { (HeI$\lambda4922$, \oiii$\lambda5007$, \oiii$\lambda4959$, \Hb, \feii$\lambda4889$, \feii$\lambda4905$, \feii$\lambda5111$, \feii$\lambda5159$, \Ni$\lambda5199$, \feii$\lambda5262$, \cliii$\lambda5518$,  \cliii$\lambda5537$, \OI$\lambda5555$,
\oi$\lambda5577$, \nii$\lambda5754$, HeI$\lambda5876$, \oi$\lambda6300$, \siii$\lambda6312$, \SiII$\lambda6347$, \oi$\lambda6364$, \ha, \nii$\lambda6548$, \nii$\lambda6584$, \hei$\lambda6678$, \sii$\lambda6717$, \sii$\lambda6731$, \ariii$\lambda7136$, \oii$\lambda7325$, \ariii$\lambda7751$\ and \siii$\lambda9069$)}, are fitted spaxel by spaxel for the pure gas cube, by performing a non parametric method based on a moment analysis. We re-cover the main properties of the emission lines, including the integrated flux intensity, line velocity and velocity dispersion. { For this analysis, we assume that all emission lines within a spaxel share the same velocity and velocity dispersion, as an initial guess. For doing so, we select as an initial guess the values derived from the fitting of the usually strongest emission line across the entire FoV, i.e., H$\alpha$, using a simple Gaussian function. Then, we perform a moment analysis weighted by this Gaussian function, as extensively described in \citet{Pipe3D_II}. This way we suppress the possible contribution of adjacent emission lines and derive the properties of considered line without considering a particular shape.} The data products of this procedure are a set of bi-dimensional maps of the considered parameters, with their corresponding errors, for each analysed emission line. Figure~\ref{fig:fitting} shows an example of the results of the fitting procedure for a spectrum extracted from a MUSE cube. 

\begin{figure*}
    \centering
    \includegraphics[]{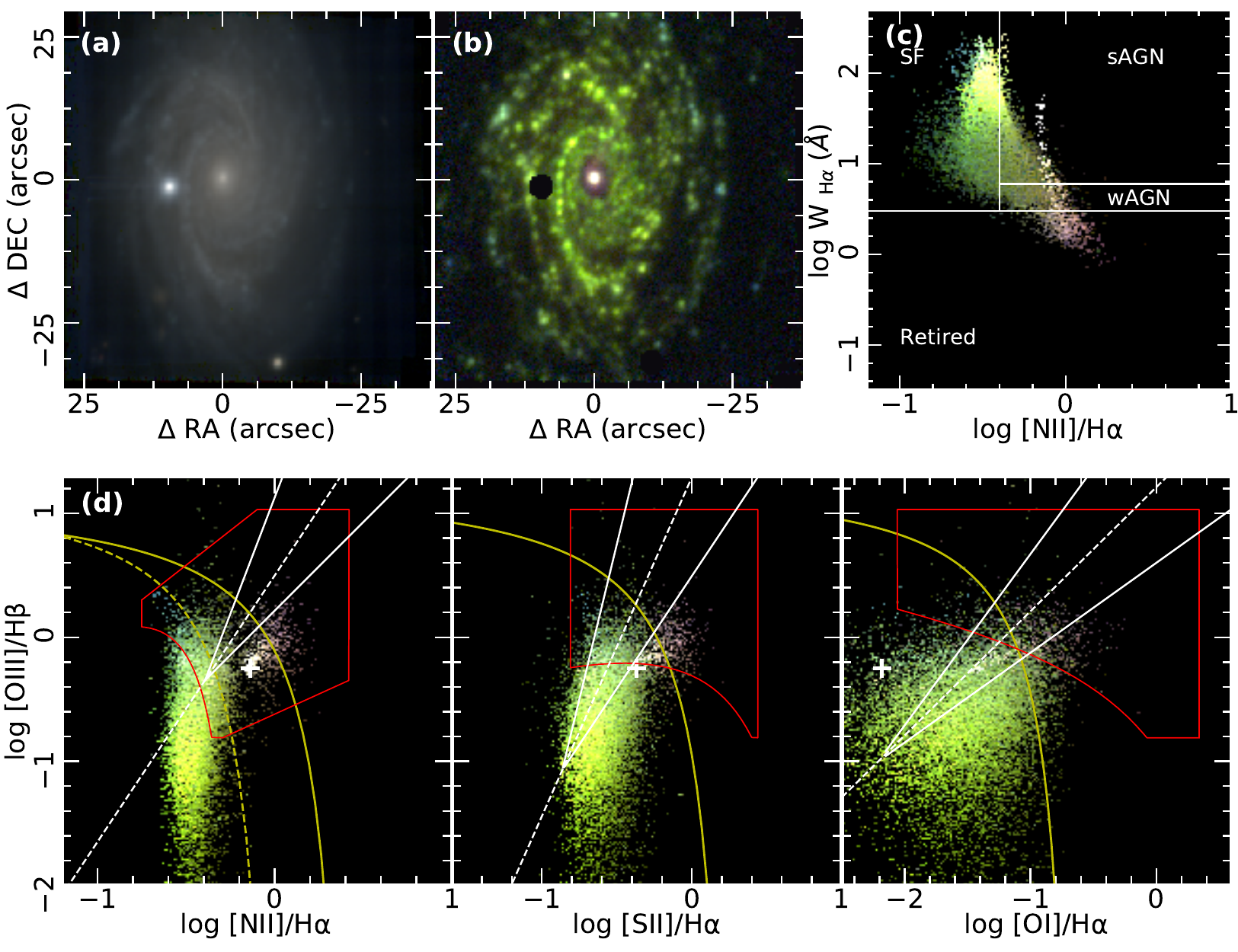}
    \caption{Different maps and diagnostic diagrams derived for NGC\,1762, a star-forming galaxy in the AMUSING++ sample. {\it (a)} {\it gri} color image reconstructed from the MUSE data cube. {\it (b)} Emission-line image constructed with the fluxes of \nii$\lambda 6584$ in red, \ha\ in green and \oiii$\lambda 5007$ in blue. No cut in the signal-to-noise was applied to construct this image. A square root scale has been applied to each filter to enhance the emission of the ionized gas. The three filters are at the same scale flux. Black colors represent regions of very low intensity or no ionized gas. {\it (c)} Spatially resolved WHAN diagram \citep{Cid2011}, \nii/\ha\ vs. \EWha, which separated SF galaxies from strong and weak AGN (sAGN and wAGN respectivelly) and retired sources. Each point corresponds to a single spaxel, showing the same color of that spaxel in the emission-line image presented in figure {\it (b)}.
    {\it (d)} Spatially resolved diagnostic diagrams associated to the ionized gas distribution from {\it (b)} (\nii\ = \nii$\lambda 6584$, \sii\ = \sii $\lambda\lambda6717,6731$ and \oi\ = \oi$\lambda6300$). The color code is the same used in the previous panel. The yellow dashed and discontinuous curves represent the demarcation lines from \citet{kauffmann03} and \citet{kewley01} respectively. { The red lines shows the boundaries of fast and slow shock models grids from {\sc mappings iii} \citep[e.g.,][]{mappings} with different  shock velocities, metallicities and preshock densities computed by \cite{Alatalo2016}.} The white cross represents the central ionization (3\arcsec\,$\times$\,3\arcsec centered in the optical nucleus). The continuous diagonal white lines represent the locus proposed by \citet{Sharp2010} of shock ionization (rightmost) and AGN ionization (leftmost). The white dashed line between them represents the bisector line between shocks and AGN ionization. All demarcation lines have been included for reference.}
    \label{fig:SF_galaxy}
\end{figure*}

\section{Methodology to select outflows}

In order to uncover the presence of outflows in the AMUSING++ compilation, we first need to describe the 
spatially resolved ionization conditions and kinematics for galaxies 
without accompanying outflows (i.e., the majority of the compilation sample). Then we identify which galaxies present extended ionized gas { structures visible morphologically, but presumably} not associated with other sources (e.g., star-formation, post-AGBs, AGNs). We describe the adopted procedure by exploring the properties of { two} galaxies, a methodology later applied to all galaxies in our sample.

\subsection{NGC\,1762: a normal star forming galaxy}

{ The interplay between the ionized gas and stellar continuum emission is closely related to the local conditions of the ISM. In the absence of non-stellar ionization there is a spatial coupling between gas ionization and stars.}
That is, the ionized gas is distributed throughout the stellar disk, with the main source of ionizing photons produced by massive OB stars. The \ha\ emission traces the spiral arms, while \nii\ emission is increased towards the center of spiral galaxies either for the higher abundance in the nuclear regions \citep{VilaCostas:1992p322,sanchez14}, by the presence of non thermal photoionization like shocks \citep[e.g.,][]{Ferland1983}, the existence of an AGN \citep[e.g.,][]{osterbrock89,davies16}, or ionization due to old stars \citep[e.g.][]{binn94,sign13}. On the other hand, elliptical galaxies present weak, or undetected emission lines, with poor or null star-formation activity.  Post-AGB  and evolved stars represent the major contribution to the ionization in retired galaxies \citep{gomes16a}, with the possible presence of AGN ionization in a fraction of them \citep[e.g.][]{Sanchez2018}. 

To illustrate the different contributions to the ionization in a single galaxy we use the spiral galaxy NGC\,1762 (part of the AMUSING++ sample) as an example case. Fig.~\ref{fig:SF_galaxy}{\it a} shows the {\it gri} color image\footnote{The {\it g-}band is only partially covered with MUSE. We took the covered part of the band to construct the RGB continuum image.} of this object extracted from the MUSE data. 
This $60\arcsec\times60\arcsec$ size image shows a nearly face-on late-type galaxy with clear spiral arms. 

In order to visualize the ionized gas distribution across galaxies, we construct a RGB emission--line image where each color represents the flux intensity of a single emission line, R: \nii$\lambda6584$, G: \ha\ and B: \oiii$\lambda5007$. We show the constructed color emission-line image for  NGC\,1762 in Fig.~\ref{fig:SF_galaxy}{\it b}. This image reveals how the spatial distribution of the ionized gas follows the same distribution of the continuum emission. Particularly, note that the \ha\ flux dominates the emission over the two other lines in most of the disk, with \nii\ increasing towards the nucleus and \oiii\ being weak compared with the two other lines almost at any location (apart from the nucleus). At this resolution, one is able to identify many green clumpy structures associated with \hii\ regions. Quantifying the number of \hii\ regions in  MUSE galaxies is important to understand the chemical evolution in galaxies \citep[e.g.,][]{censusHII,laura18}. Finally, the central region presents an almost point-like strong ionized region, with high \nii\ and \oiii\ that most probably corresponds to an AGN. The advantages of displaying the ionized gas component in one RGB image is that we can explore immediately the distribution of different ionization sources by just looking at such color and intensity.

Line ratios sensitive to the ionization are commonly used to explore the ionization source in galaxies. The \nii/\ha\ ratio gives a quantitative assessment of the different physical processes that ionize the gas. This ratio has the advantage (over other lines) of being both accessible in optical spectra and being almost insensitive to dust attenuation due to their small wavelength separation. The equivalent width (EW) of \ha\ (\EWha) has also been used to explore the ionization in galaxies. Ionizing sources that produce weak emission lines present also low equivalent widths (\EWha\ $<3 $\AA). Ionization by evolved stars, as post-AGB, frequently present this kind of line ratios and EWs \citep[e.g.,][]{binn94, stas08,Lacerda2018}. The remaining ionization sources that produce emission lines present higher \EWha\ in general ($>3 $\AA).  Fig.~\ref{fig:SF_galaxy}{\it c} shows the WHAN diagram, introduced by \citet{Cid2011}, that combines both the \nii/\ha\ ratio and the \EWha, for each ionized spaxel from the emission-line image presented previously. Note that the \EWha\ is one of the parameters derived as part of the fitting procedure performed by {\sc Pipe3D}.  

Each pixel in the emission-line image is associated with an unique pair of values \EWha\ and \nii/\ha\ in the WHAN diagram.  
The result is the spatially resolved WHAN diagram shown in Fig.~\ref{fig:SF_galaxy}{\it c}. We note that the gas in the spiral arms is mainly distributed in the SF regions in this diagram as revealed by the green color. Meanwhile the inter-arm  gas and the gas surrounding the nucleus are distributed in the AGN region and in regions associated  to ionization by hot low-mass evolved stars (HOLMES), the main ionization source in retired galaxies.

Excluding the demarcation at \EWha = 3\AA, the transition lines in the WHAN diagram are just the best transposition of the demarcation curves from  \cite{kewley06} and \cite{Stasinka2006} in the classical diagnostic diagram, like the BPT one involving the \oiii/\hb\ vs. \nii/\ha\ line ratios \citep[e.g.,][]{Baldwin1981}. 
The vertical line at log \nii/\ha\ = -0.4 maps the division between SF and AGNs regions, while the horizontal line at \EWha = 6\AA, represent the classical separation between quasars and Seyfert galaxies \citep[e.g.,][]{Baldwin1981}. Since this is a projection, the separation between the different ionizating sources is not as clean as in the classical diagnostic diagrams, and its use is recommended only if the \oiii/\hb\ ratio is not available, as indicated by \citet{Cid2011}.

\cite{Veilleux1987} were the first to introduce diagnostic diagrams based on emission line ratios as a  method to classify entire galaxies. They introduced the \nii/\ha, \sii/\ha\ and \oi/\ha\ versus \oiii/\hb\ diagnostics already presented by \citet{Baldwin1981}. Several demarcation curves have been proposed over these diagrams to try to separate the soft ionization sources, like \hii\ regions, from those with a harder ionization, like AGNs. The most common is the one proposed by \citet{kewley01} (K01) based on photoionization grid models. This curve represents the
maximum envelope in the considered line ratios that can be reached by ionization due to multiple bursts of star-formation. Line ratios above this curve cannot be reproduced by ionizing photons produced by young OB stars. Classically the region above this curve is known as the region populated by AGN, although it is not exclusive of this ionizing source (as is broadly assumed).

The ionization produced by old stars (post-AGBs, HOLMES), commonly found in retired galaxies, at large extraplanar distances in disk galaxies, or in the central and inter-arm regions of galaxies, can also reproduce the line-ratios observed in the LINER region of the BPT diagram \citep[e.g.,][]{binn94, stas08, sign13}. The equivalent widths that produce these sources tend to be much lower compared with that from SF or AGN ionization \citep{sign13,Lacerda2018}. It has been shown that the demarcation at \EWha$<3 $\AA\ is a good indicator for the ionization produced by this kind of stars \citep[e.g.][]{gomes16a}.

Photoionization induced by shocks can also reproduce the line-ratios observed in the AGN/LINER region \citep[e.g.,][]{Alatalo2016}. The combination of three free parameters in shocks (magnetic fields, shock velocities and the pre-shock densities), give rise to a wide range of values of line ratios that may cover an ample region in the diagnostic diagrams, from SF regions to AGN/LINER ones \citep[e.g.,][]{mappings}. As a consequence, for shock ionization a demarcation curve does not exist as for the other ionizing sources. Nevertheless, there has been efforts to constrain certain regions of the diagram where shocks are more frequently found, depending on the origin of the galactic wind i.e., SF-driven or AGN-driven, \citep[e.g.,][]{Sharp2010}.

In Figure~\ref{fig:SF_galaxy}{\it d} we present the spatially resolved diagnostic diagrams for the example galaxy NGC\,1762. They, combined with the emission-line image, provide us with unique information about where the different sources of ionization take place inside a galaxy. The ionized gas located at the spiral arms (greenish in Fig.~\ref{fig:SF_galaxy}{\it b}), is found in the SF regions of the diagrams clearly below the K01 demarcation in the three of them. On the other hand, the gas in the nucleus is located at the AGN/LINER region. If we combine the information provided by these
diagnostic diagrams with the distribution along the WHAN diagram, we can conclude that nuclear regions present two kinds of ionizations. The very center presents a hard ionization with high \EWha (i.e., the signatures of an AGN). However, the surrounding regions present also hard ionization, but with low \EWha\ (i.e., the signature of ionization by old stars). This later one is spatially associated with the optical extension of the bulge.

{ Despite of the several spaxels falling in regions constrained by the shock models grids, and in regions compatible with SF- and AGN-driven winds, it is unlikely that they are associated to shock ionization due to their spatial distribution (as they are concentrated in a nuclear almost point-like emission region). Although small galactic fountains can drive outflows and produce shock line ratios (such as giant \hii\ regions), it is unlikely that this is the main ionization source in galaxy disks. Furthermore, in this work we are interested in kiloparsec scale outflows, instead of galactic fountains.}
\begin{figure*}
    \centering
    \includegraphics[]{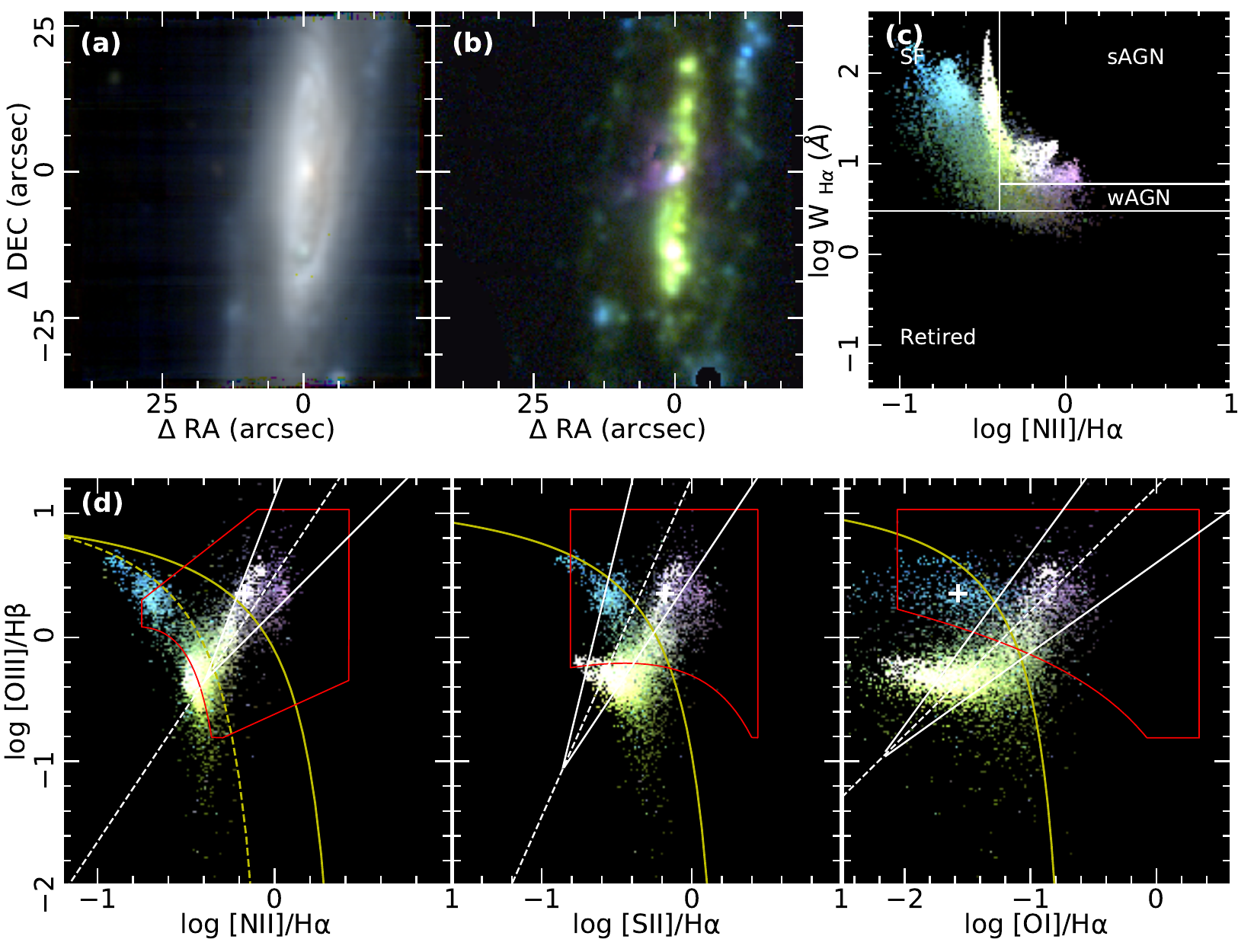}
    \caption{ IC\,1657 an outflow host galaxy in the AMUSING++ sample. All panels are similar to those presented for  NGC\,1762, in Fig.~\ref{fig:SF_galaxy}.}
    \label{fig:eline_SN2012hd}
\end{figure*}

Although the previous analysis was made with a spiral galaxy, the spatial concordance of the ionized gas distribution with the stellar continuum emission also applies for ellipticals, although these might present weaker ionized gas (in the absence of an AGN). For those galaxies most of the ionized regions would be spread from the right-end of the SF-region towards the LINER-like region in the diagnostic diagrams \citep[e.g.][]{Lacerda2018}. They will present little or no evidence of clumpy ionized regions (like the HII regions observed in spirals), with low values of \EWha \citep{gomes16a}, and with an underlying continuum dominated by an old stellar population \citep[e.g., Fig.~3 of ][]{sanchez14}.

\subsection{Outflows and extended emission-line objects in AMUSING++}
\label{sec:sel_out}

Under the presence of a mechanism perturbing the gas, the spatial coincidence with the continuum emission might not necessarily persist. Galactic outflows are one phenomenon that can eject gas out of the galaxies making the emission of ionized gas and the continuum emission become spatially uncoupled. 
The warm phase of outflows (T$\sim 10^4$ K) is directly observable in high spatial resolution images \citep[e.g.,][]{Strickland2004,Mutchler2007}. At this temperature optical emission lines reveals typically hollow conical, biconical or filamentary structures of ionized gas emerging from the nuclear regions \citep[e.g.,][]{Veilleux2002,Strickland2004,carlos16}.  
Therefore, our primary criteria to select galaxies hosting galactic outflows is based on the spatial distribution of the ionized gas:  ionized gas decoupled from a plausible underlying source (young/old stars or an AGN), spatially distributed following bicones, cones or filamentary structures, departing from the inner towards the outer regions of galaxies \citep[e.g.,][]{Heckman1990,Veilleux2002,Strickland2004}.

Based on the emission-line images of the AMUSING++ galaxies  as well as the spatially resolved diagnostic diagrams described before, we select our outflow candidates among those galaxies with extended, filamentary, and conical emission. The location of the extended emission in the diagnostic diagrams must be at least not fully dominated by SF ionization and must be spatially decoupled from the stellar continuum. It is possible that in some cases the extended emission might not be due to the presence of an outflow. We will discuss their nature in section \ref{sec:discussion}. On the other hand, small-scale outflows (below the resolution of our data) could escape to this scrutiny. So far, we will focus on large-scale ones, clearly identified with the current dataset. Finally, we note
galaxies with outflows analysed in this study that have already been reported. However, it will be possible to make comparisons with galaxies not hosting outflows using data of similar quality, a task not yet addresed with MUSE data. Finally, the so-called jellyfish galaxies are excluded from this selection criterion due to the different nature of the extended ionized gas emission.

\begin{figure*}
    \centering
    \includegraphics[]{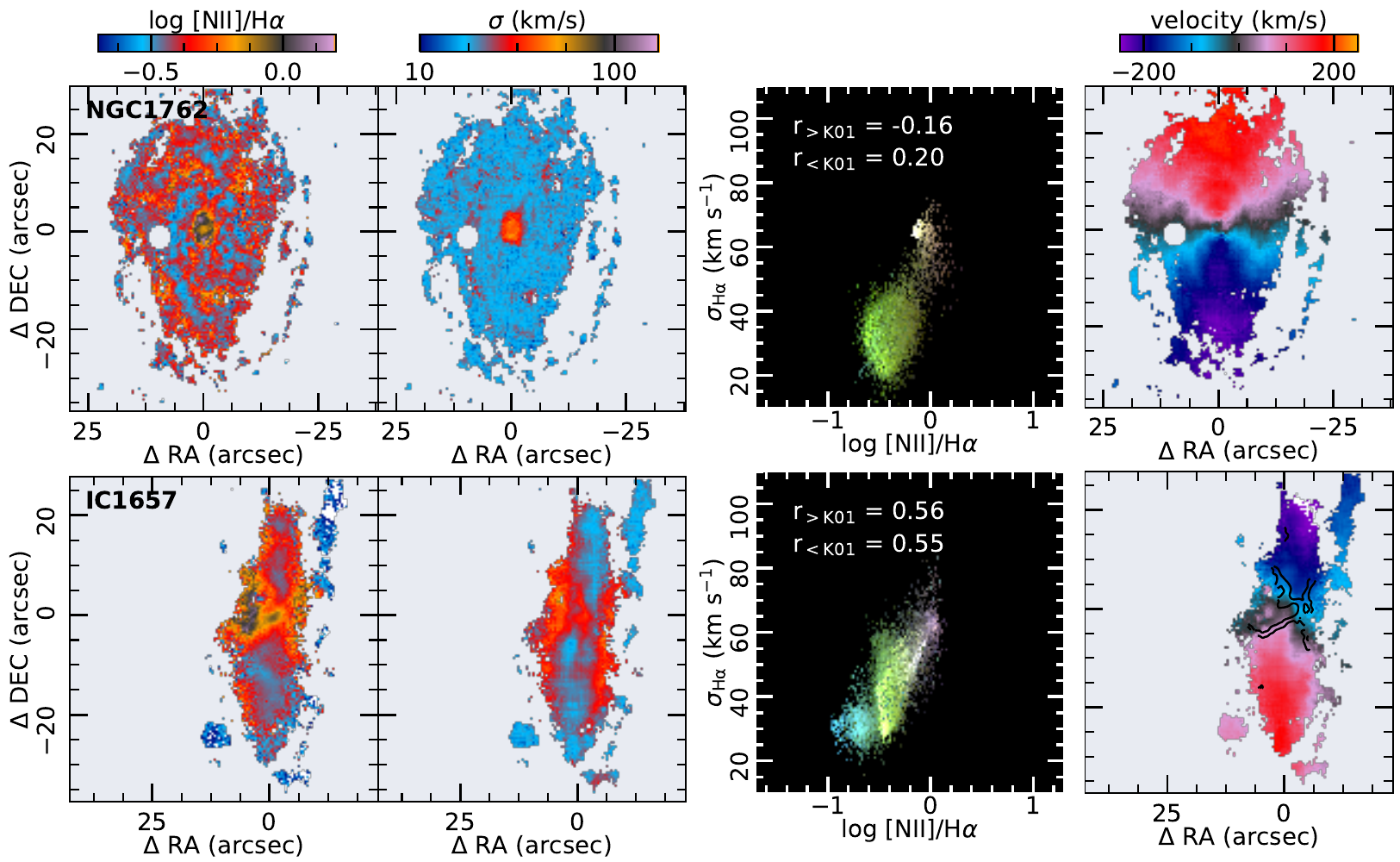}
    \caption{Ionization and kinematic diagrams for the two example galaxies: NGC 1762 (top panels) and IC 1657 (bottom panels). Each panel shows, from left to right: {\it (i)} the spatially resolved \nii/\ha\ line ratio and {\it (ii)} velocity dispersion maps; {\it (iii)} the \nii/\ha\-$-\sigma$ resolved map color coded with the emission-line image from Figs.~\ref{fig:SF_galaxy}{\it b} and \ref{fig:eline_SN2012hd}{\it b} respectively. { Two correlation coefficients $r$, between these variables were computed for spaxels lying above $(r>\mathrm{K01})$ and below $(r<\mathrm{K01})$ the K01 curves}. {\it (iv)} the rightmost panel shows the \ha\ velocity map.  Black contours superimposed in the IC\,1657 velocity map represent the best demarcation of the ionized cone traced with the $\log$~\nii/\ha\ ratio map with levels $-0.3$ and $-0.2$ dex. }
    \label{fig:dispersion}
\end{figure*}

In the next section we show the case of an outflow galaxy detected using the previous technique to illustrate the main features that enabled us the identification of all candidates. 

\subsubsection{The ionized cone in IC\,1657 }

Figure~\ref{fig:eline_SN2012hd}{\it a,b} shows the {\it gri} color image and ionized gas distribution of galaxy IC\,1657. This is a highly inclined ($i=78^\circ$) spiral galaxy. There is no information in the literature about the presence or signature of an outflow in this galaxy, although it is in list of outflows candidates by \cite{Colbert1996} but with no reported analysis of the emission lines. The RGB emission-line image in Fig.~\ref{fig:eline_SN2012hd}{\it b}, reveals what it seems a conical structure of gas (more intense in \nii, i.e., reddish) perpendicular to the disk plane, that looks to be outflowing from the optical nucleus. As indicated before conical structures like this are typical from outflows produced by either SF or AGN. The \ha\ emission reveals the \hii\ regions in the galaxy disk (greenish), while some ionized clumps present a slightly larger \oiii\ emission at the edge of disk (blueish). The \EWha\ and \nii/\ha\ ratio of the gaseous cone component is not compatible with being ionized by SF  (log \nii/\ha\ $< -0.4$ ). It is neither compatible with being ionized by evolved stars (\EWha\ $> 3 $ \AA) as revealed by the WHAN diagram (Fig.~\ref{fig:eline_SN2012hd}{\it c}). 

The spatially resolved  diagnostic diagrams can be interpreted as follows: the clumps with stronger \oiii\,, located at the outskirts  in the emission-line image, are compatible with having low gas metallicities in these diagrams.
Finally, the clumps dominated by \ha\ emission in the center of the disk are located where the high metallicity \hii regions are found. Thus, the emission-line color-image illustrates qualitatively the metallicity gradient observed in galaxies \citep[e.g.,][]{sanchez14}. { On the other hand the gaseous cone, visible morphologically, is well separated in all diagnostic diagrams from the ionization 
most probably due to star-formation (green clumpy structures). It clearly spreads towards regions where a harder ionization source is required to reproduce the observed line ratios. In the WHAN diagram the cone nebulae is identified in regions characteristics of AGN-like ionization. Low values of the \EWha\ ($<3$\AA) are characteristic in extra-planar \citep{Flores2011,Amy2017} and non extra-planar diffuse ionized gas \citep{sign13,Lacerda2018}. Nevertheless, the predominant large values of the \EWha\  exclude the low-mass evolved stars as the main source of ionization in the cone nebulae.  Regarding the line ratio diagnostic diagrams, the spaxels spatially associated to the ionized cone are also located at the classical AGN-ionized region. Indeed, all of them fall within the region occupied by shock ionization according to the predicted line ratios from theoretical models (e.g., {\sc mappings iii}). Moreover, the line ratios at the ionized cone are more compatible with the SF-driven wind scenario according to the empirical demarcations from AGN-driven and starburst-driven winds by \cite{Sharp2010}. Therefore, shock ionization produced by a SF-driven outflow seems to be the most likely explanation to the observed morphology as well as its observed line ratios.

From this example, it is clear that the intrinsic complexity of outflows inhibits its direct identification in diagnostic diagrams. It is just by a discarding process of ionizing sources, considering both line rations and morphologies simultaneously, in which it is possible to obtain hints of shock ionization, indicative of the possible presence of outflows \citep[in aggreement with the recent review by][]{sanchez20}.}

\subsection{Kinematics: velocity dispersion and \ha\, velocity}

Most star-forming galaxies are disk dominated spiral galaxies \citep[e.g.][]{Sanchez2018}, which typically present velocity dispersion ranging from some tens of \kms\ \citep[e.g.][]{bershady10} to $\sim$  100 \kms\ in the case of turbulent or high SF galaxies \citep{Genzel2008,Green2010}. At the wavelength of \ha\ the spectral resolution of MUSE is $\sigma \sim 50$ \kms\,, which allows us to resolve the velocity dispersion of these galaxies in a wide range of galactocentric distances. For early-type galaxies the velocity dispersion is much larger, in general, and therefore is well recovered with this data. Galactic outflows are generally associated with increases in the velocity dispersion, a property used to characterise, detect and confirm them \citep[e.g.,][]{Monreal2010,Rich2011,Rich2015}.

Figure~\ref{fig:dispersion} shows 2D maps of the \nii/\ha\ line ratio and \ha\ velocity dispersion for the two archetype galaxies described throughout this article. In the case of NGC\,1762, there is a clear increase of the line ratios towards the center, as discussed in previous sections. This increase is spatially associated with an increase in the velocity dispersion, that traces clearly the location of the bulge. This reinforces our interpretation that a fraction of the ionization in this region is due to old-stars that dynamically present hot/warm orbits comprising the bulge \citep[e.g.][]{zhu18a,zhu18b}. On the other hand, the velocity dispersion along the disk presents values $\sim$40--70 \kms, i.e., within the expected values for a SF disk galaxy \citep[e.g.][]{Genzel2008,bershady10}.

In the case of the galaxy hosting an outflow,  IC\,1657, the velocity dispersion along the disk is of the same order. However, there is an evident increase of the velocity dispersion associated with an enhancement of the \nii/\ha\ ratio along the semi-minor axis of the galaxy. This enhancement is spatially associated not only with the cone structure observed in the emission-line image (Fig.~\ref{fig:eline_SN2012hd}), but additionally another conical structure in the opposite direction of the main one described above (i.e., behind the disk). A detailed inspection of Fig.~\ref{fig:eline_SN2012hd} shows that indeed this second conical structure is appreciated in there too. The dust attenuation of the disk (c.f.,  Fig.~\ref{fig:eline_SN2012hd}{\it a}) may be causing the partial obscuration of this second cone.

Following a similar procedure as the one adopted to create the diagnostic diagrams (Figs.~\ref{fig:SF_galaxy} and \ref{fig:eline_SN2012hd}), we construct a spatially resolved $\sigma$--\nii/\ha\ diagram. A positive correlation between the ionization strength and velocity dispersion is typically found in the presence of shocks \citep[e.g.,][]{Monreal2010,Rich2011,Ho2014,Rich2015,carlos16}. This is a natural correlation if the emission lines present a broad component, induced by an asymmetry of the line profile, associated with shocks. Velocity dispersions larger than  $90$ \kms\ have been associated to shocks produced by galactic winds \citep[e.g.,][]{Rich2015}.

Figure~\ref{fig:dispersion}, right panel, shows the spatially resolved $\sigma$--\nii/\ha\ diagram, color coded with the emission-line images presented in Figs.~\ref{fig:SF_galaxy}{\it b} and \ref{fig:eline_SN2012hd}{\it b} respectively. In general, low velocity dispersion values ($<50$ \kms) are observed where the SF is the dominant ionization. The nucleus in both cases present high dispersion values ($>50$ \kms). { As a positive correlation between these variables is a signature of shocks \citep{Monreal2010}, we compute the correlation coefficients betweeen both parameters for spaxels dominated by SF ionization (those lying below the K01 curve in the BPT diagram) and for spaxels lying above the K01 curve, presumable mostly dominated by shocks in the presence of outflows. 
In the case of NGC\,1762 the spaxels with higher dispersion (those close the nucleus) present a negative correlation, while those associated to the disk, present a very weak correlation ($r<\mathrm{K01} = 0.20$).
On the other hand, IC\,1657 present a moderate positive correlation for spaxels in the disk and also in the ionized cone ($r<\mathrm{K01} = 0.55$ and $r>\mathrm{K01} = 0.56$ respectively). The positive correlation in the ionized cone may suggests the presence of multiple or broad components, produced most probably by the presence of a shocked layer of gas.}

The right-most panel of Fig.~\ref{fig:dispersion} shows the \Ha\ velocity maps for the two considered galaxies. 
In absence of an external agent perturbing the ISM, a regular  rotation pattern is expected in the gas kinematics. NGC\,1762  shows, indeed, the typical pattern of a rotating disk with symmetrical velocities around the center with a receding (North) and an approaching (South) side. On the other hand the presence of the ionized cone observed in IC\,1657 is producing deviations from the expected velocity pattern around the galaxy semi-minor axis, where the outflow is expanding. { This is also clear in the
distribution of differential velocities between the ionized gas and the stars, i.e., the 
$v_{gas}-v_{\star}$ maps (Fig. \ref{fig:appendix_1} in the appendix section). We observe differences $>60$ \kms\ in the outflow influenced regions between both velocity maps, while in the unperturbed disks, the velocity difference is much smaller (compatible with zero in many cases). In Table \ref{tab:all_amusing_table} we report the W90 value of the absolute difference between both velocities across the FoV of the data, $\Delta v_{gas,\star} = |v_{gas}-v_{\star}|$. In general, spaxels of not outflow host galaxies and with line ratios above the K01 curve tend to present smaller differences in $\Delta v_{gas,\star}$ than in galaxies hosting outflows.}

As part of our candidates selection, beside looking for ionized regions where line ratios cannot be explained by the underlying continuum (stellar or AGN), with filamentary or conical structures, we explore the distribution of the velocity dispersion and its agreement with an enhancement of the \nii/\ha\ (and when feasible of \sii/\ha\ and \oi/\ha\ which are also associated to shocks). In addition, we explore possible perturbations in the velocity maps, again associated with similar enhancements in the considered line ratios and increases in the velocity dispersion.

\begin{figure*}
    \centering
        \includegraphics{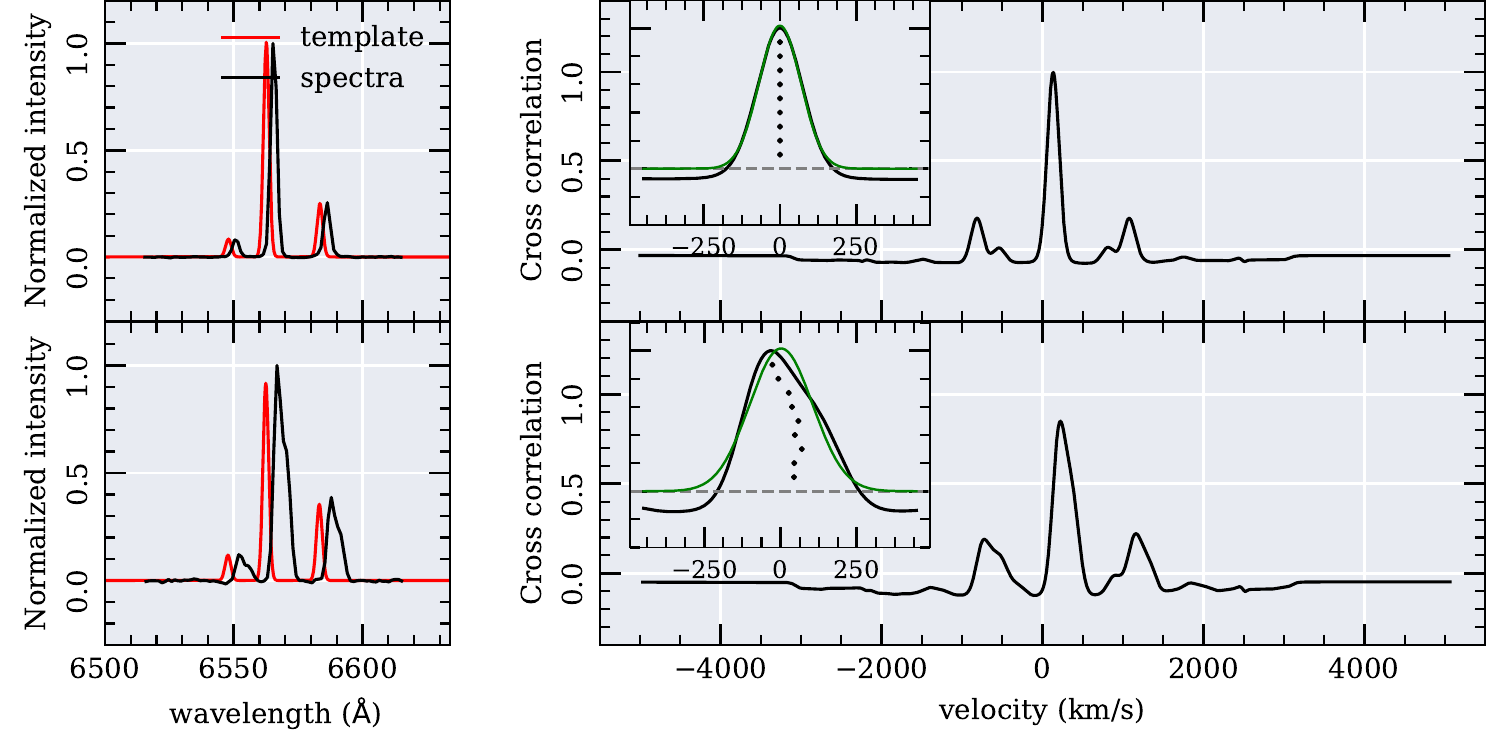}
    \caption{Illustration of the cross-correlation method applied to the emission line spectra to compute the asymmetries of the line profiles. Top (bottom) panels show the cross correlation technique applied to a spectra containing the \ha\ $+$ \nii\ emission lines described in this case by a single (multiple) kinematic component. {\it Top-left:}  The black solid line represents the normalised gas spectrum,  shifted to the rest-frame, for a spaxel located on a spiral arm of NGC\,1762. The red line represents the template used to cross-correlate the observed spectra. The templates are modelled with three Gaussian functions (one for each emission line) with the FWHM fixed to the instrumental resolution. {\it Top-right:} The main panel shows the normalised cross-correlation function (CCF) in velocity space between the model and the spectra for a wide range of velocities (-5500 \kms\ to 5500 \kms). The inset in the right panels shows a zoom for the velocity range (shifted to the zero velocity) where the maximum of the cross-correlation is observed.
    Crosses represent the bisectors of the CCF at different intensity levels relative to the peak, ranging from 10\% to 90\%. The green line represents the best Gaussian fit to the CCF distribution. Bottom panels show the same plots for a spectrum extracted from the conical ionized gas structure detected in IC\,1657 (see Fig.~\ref{fig:eline_SN2012hd}{\it b}), i.e., a clear candidate to galactic outflow.}
    \label{fig:ccf}
\end{figure*}

\begin{figure*}[t]
    \centering
    \includegraphics[width=\textwidth]{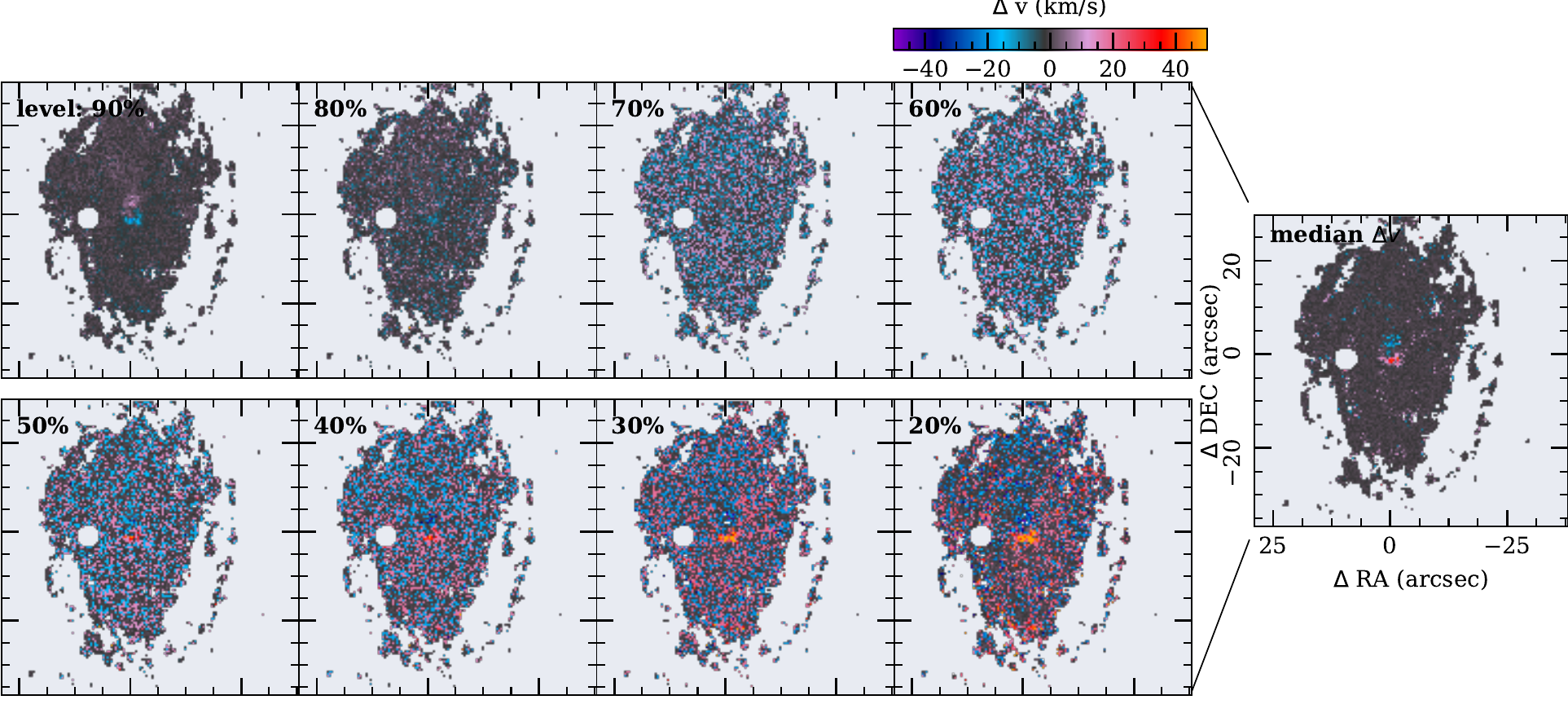}
        \includegraphics[width=\textwidth]{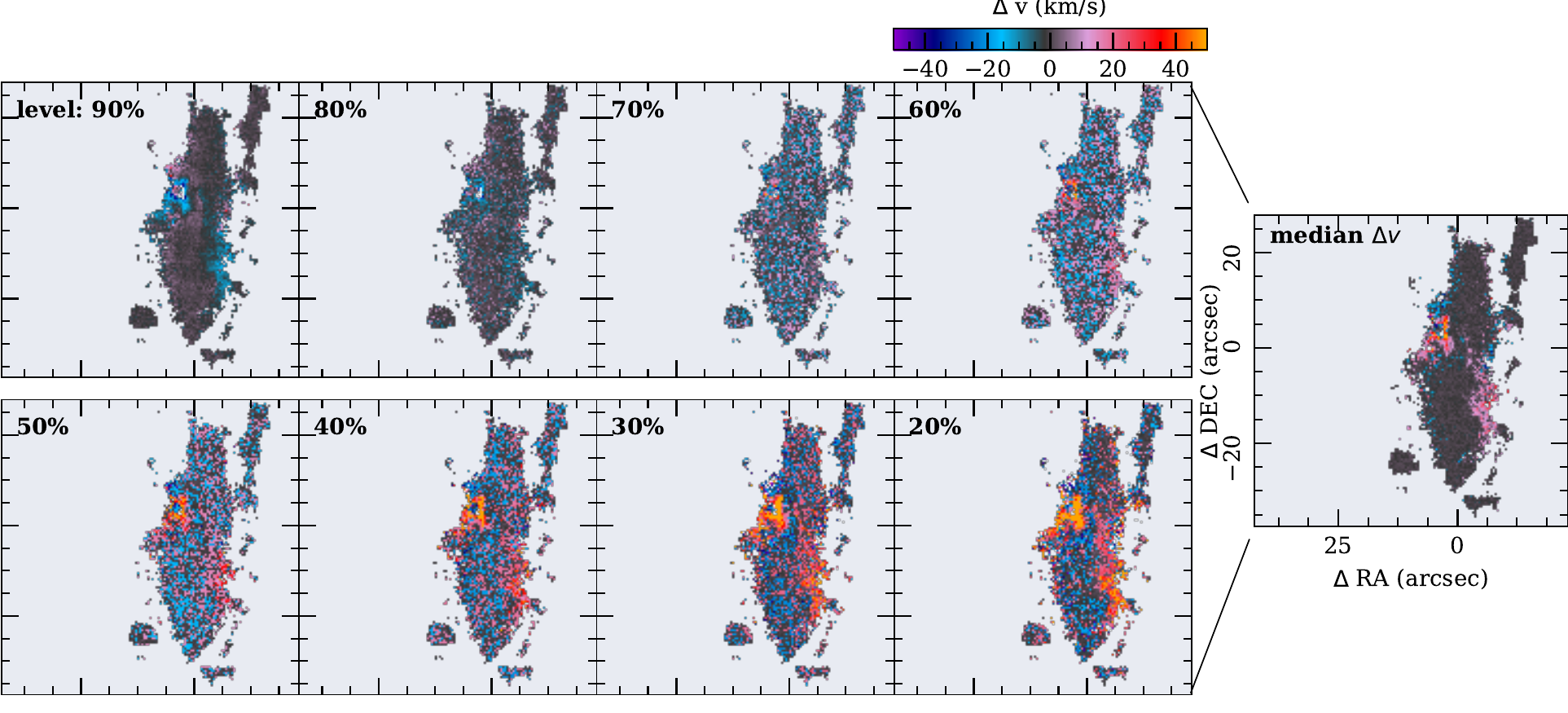}
    \caption{2D maps of the asymmetries at different intensity levels for the two archetype galaxies described in this article: NGC\,1762 and IC\,1657 in the top and bottom panels respectively. Spaxels with \SN $>4$ in \ha\ are shown in these maps. Each panel shows the percentage of the flux with respect to the peak of the CCF at which the asymmetry is estimated (as illustrated in Fig.~\ref{fig:ccf}), ranging from 90\% to 20\% in steps of 10\%. The asymmetry, $\Delta$v is defined as the difference between the bisector velocity at the corresponding intensity level and the velocity at which the peak intensity is found.}
    \label{fig:asymmetries}
\end{figure*}

\begin{figure}
    \centering
    \includegraphics[width=\columnwidth]{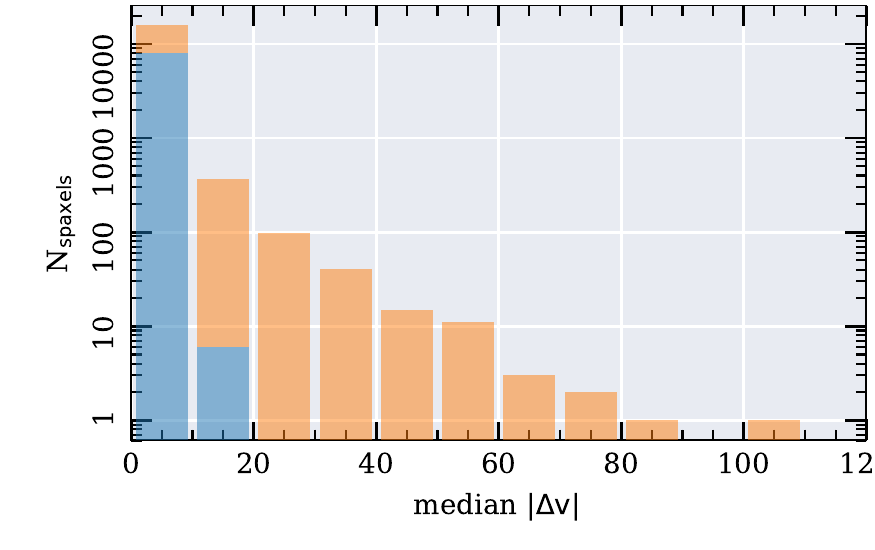}
    \caption{Distribution of the median values of the asymmetries $|\mathrm{\Delta v}|$ in each spaxel. Blue histogram corresponds to  NGC\,1762  and orange to IC\,1657.}
    \label{fig:histogram_asymmetries}
\end{figure}
\begin{figure}
    \centering
    \includegraphics[width=\columnwidth]{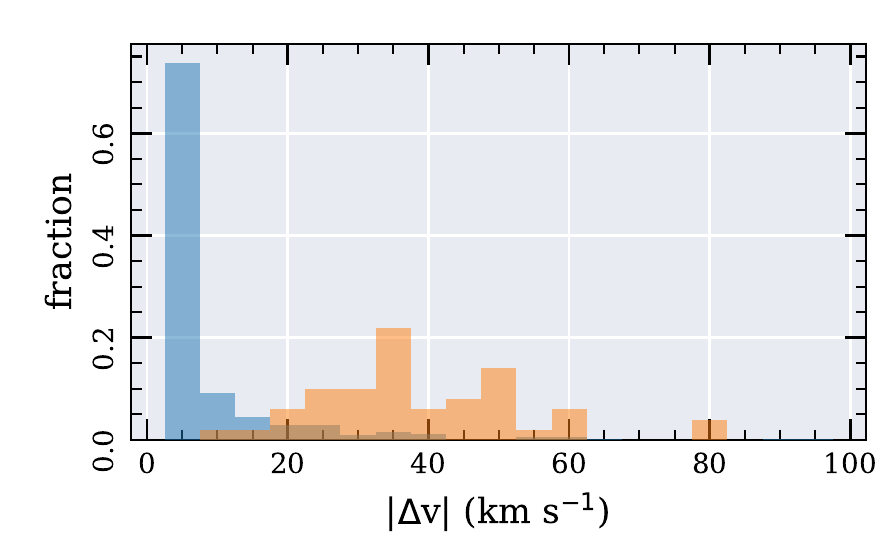}
    \caption{Distribution of asymmetries W90 parameter for all the AMUSING++ sample without outflows (blue colors), and with an outflow (orange colors). W90 was calculated in each galaxy from the median $|\mathrm{\Delta v}|$ map, taking spaxels where the \SN $>4$ in \ha. Histograms are normalised to the total number of galaxies in each sub-sample.}
    \label{fig:fraction_assimetries}
\end{figure}

\subsection{The cross correlation function: emission-line asymmetries}
\label{sec:ccf}

The broad profiles detected at the location of the outflowing regions may indicate the presence of multiple components. Therefore, analysing the shape of the emission lines is important in the identification of these processes. This shape is the result of the sum of all the kinematics components associated with different ionizing processes occurring at each location within a galaxy, integrated along the line-of-sight. Although the typical profile to model emission line at our spectral resolution is a Gaussian function (Voigt functions are used in the case of better resolution), in many cases more complex profiles are required to characterise the observed emission lines. Regardless of the functional form adopted for modelling, and in the absence of any perturbing external mechanism, the emission lines appear to be symmetrical around their intensity peak. 
Line bisectors are the best way to describe the symmetry of a line. The study of asymmetries of line profiles is a technique that was developed for the analysis of stellar spectra to study granulation decades ago \citep[e.g.][]{gray1988lectures}. Although this technique was designed to analyse absorption lines, it is straightforward to adapt it to study emission line profiles. In this case it is useful to derive the cross-correlation using a model profile. This way the contrast is enhanced and it is possible to include several emission lines simultaneously in the analysis.

The cross-correlation technique is an estimation of the similarity of two signals that gives as result a set of correlation coefficients for every lag or offset in the frequency or velocity space (defined as $\tau$). If the two signals are similar but they differ by a certain lag/offset, then the maximum of equivalence between them is reached at $\tau_{r\rm_{max}}$, where $r_{\rm max}$ is the maximum value of the cross correlation, following a symmetrical profile. This technique has already been applied successfully to measure the degree of symmetry of emission lines associated with ionized gas in galaxies \citep[e.g.,][]{Garcia-Lorenzo2013,Begona2015}. The resultant cross correlation function (CCF), i.e., the distribution of correlation coefficients along $\tau$ (in this case the velocity), is a measure of the average profile of the spectrum of reference (in the velocity space).

Following \cite{Garcia-Lorenzo2013}, we compute the CCF in a spectral window that covers multiple emission lines close in wavelength. We use the pure gas spectra (i.e., continuum subtracted, as described in Sec. \ref{sec:analysis}), and a model of all involved emission lines is generated by adopting a set of Gaussian functions with FWHMs equal to the spectral resolution of the data (FWHM $\sim 2.6$ \AA). Preliminary fits to the spectra are performed with the considered model to estimate the intensity of the emission lines involved. The relative intensities of the lines are then passed to the template. Finally the template is shifted to the redshift of the galaxy (previously determined by {\sc Pipe3D}). The cross-correlation is finally performed between this adjusted template and the gas-pure spectra.

Figure~\ref{fig:ccf} shows the cross correlation technique applied to two particular spectra in a spectral window that covers the \ha\ $+$ \nii$\lambda\lambda 6548,6584$ emission lines. This way the effect of the residual-continuum is mitigated. The top panels show the case of a spectrum extracted from an \hii\ region of NGC\,1762. The emission profiles seems to be well described by a single Gaussian component. When it is cross correlated with the appropriate template, the CCF shows multiple peaks at different velocities. However, the  maximum similarity is reached for a peak near to the systemic velocity (i.e., near zero at the scales shown in the figure). We select the CCF at a regime within $\pm$500 \kms\ around this peak, and compute the bisectors at different intensity levels relative to the peak (from 90\% to 20\%, with steps of a 10\%). Then, a fit to the selected range of the CCF is performed to have a better estimation of the peak velocity and velocity dispersion. 
Finally we estimate $\Delta$v$_{level}$, i.e., the velocity difference between the bisector at each intensity level and the corresponding velocity of the peak intensity. The mean of all estimated $\Delta$v$_{level}$ for the different levels is stored as the final estimation of the asymmetry of the lines for the considered spectrum and spectral range (defined as $\Delta$v).

The second example in Fig.~\ref{fig:ccf} corresponds to a spectrum extracted from the outflowing region discovered in IC\,1657. In this case the bisectors show clear deviations from the peak velocity, with an obvious shift to the blue with respect the central velocity.
These kind of asymmetries are typical of outflows \citep[e.g][]{Ho2014,Maiolino2017}. In general, $\Delta$v$_{level}$ represents the velocity with respect to that of the intensity peak, which does not necessarily correspond to the systemic velocity (except in the case that the emission line profiles are described well by a single component).

We apply the described methodology to the whole pure gas cube of each galaxy to obtain a set of asymmetry maps ($\Delta$v$_{level}$, one for each intensity level) and the corresponding mean asymmetry ($\Delta$v), estimated along all the asymmetry levels.  

Figure~\ref{fig:asymmetries} shows the derived asymmetry maps for the different levels and the final mean map for NGC\,1762 and IC\,1657. In the case of NGC\,1762 almost no asymmetry is detected across the entire disk of the galaxy. At the central regions - dominated by the bulge - an asymmetry towards the opposite velocity of the disk is found. This asymmetry may indicate the existence of a central-region counter rotation, or disturbed kinematics that could be associated with the presence of an AGN candidate discussed above. 

On the other hand, the asymmetry maps of IC\,1657 clearly illustrate the complex kinematic structure associated with galactic outflows. The higher values of asymmetry are spatially associated with the velocity perturbations, the increase of velocity dispersion and the enhancement of line ratios found at the bi-conical structure that we describe as a galactic outflow. Following these results we explore the asymmetry maps derived for all the galaxies in the sample and inspect the possible association of high asymmetry values with the other properties describing an outflow. Any galaxy including these properties is selected as a candidate outflow for further inspection.

Figure~\ref{fig:histogram_asymmetries} shows the distribution of the absolute value of asymmetries ($|\mathrm{\Delta v}|$) for all spaxels with a \SN $>4$ in \ha\ for the two archetype galaxies. This figure shows that  NGC\,1762 is dominated in general by low values of asymmetry ($<15$ \kms) while  IC\,1657 presents a tail towards higher values ($>50$ \kms). Although $|\mathrm{\Delta v}|$ does not represent the real velocity of the extra components, it represents a lower limit of the velocity of the shocked gas in the case of outflows. Finally, we derive for each galaxy the $W_{90}$ parameter for $|\mathrm{\Delta v}|$, i.e., the velocity difference between the 5th and 95th percentiles of the distribution of asymmetries for all the spaxels of each datacube. We include in Table \ref{table:properties} this parameter just for spaxels dominated by the outflows in each of the host galaxy candidates. Figure~\ref{fig:fraction_assimetries} shows the distribution of these $W_{90}$ values compared with the same distribution for all galaxies in the sample (and for all spaxels). This figure clearly illustrates how different the asymmetries are in the presence of perturbations like the ones introduced by outflows.

\begin{figure*}
    \centering
    \includegraphics[]{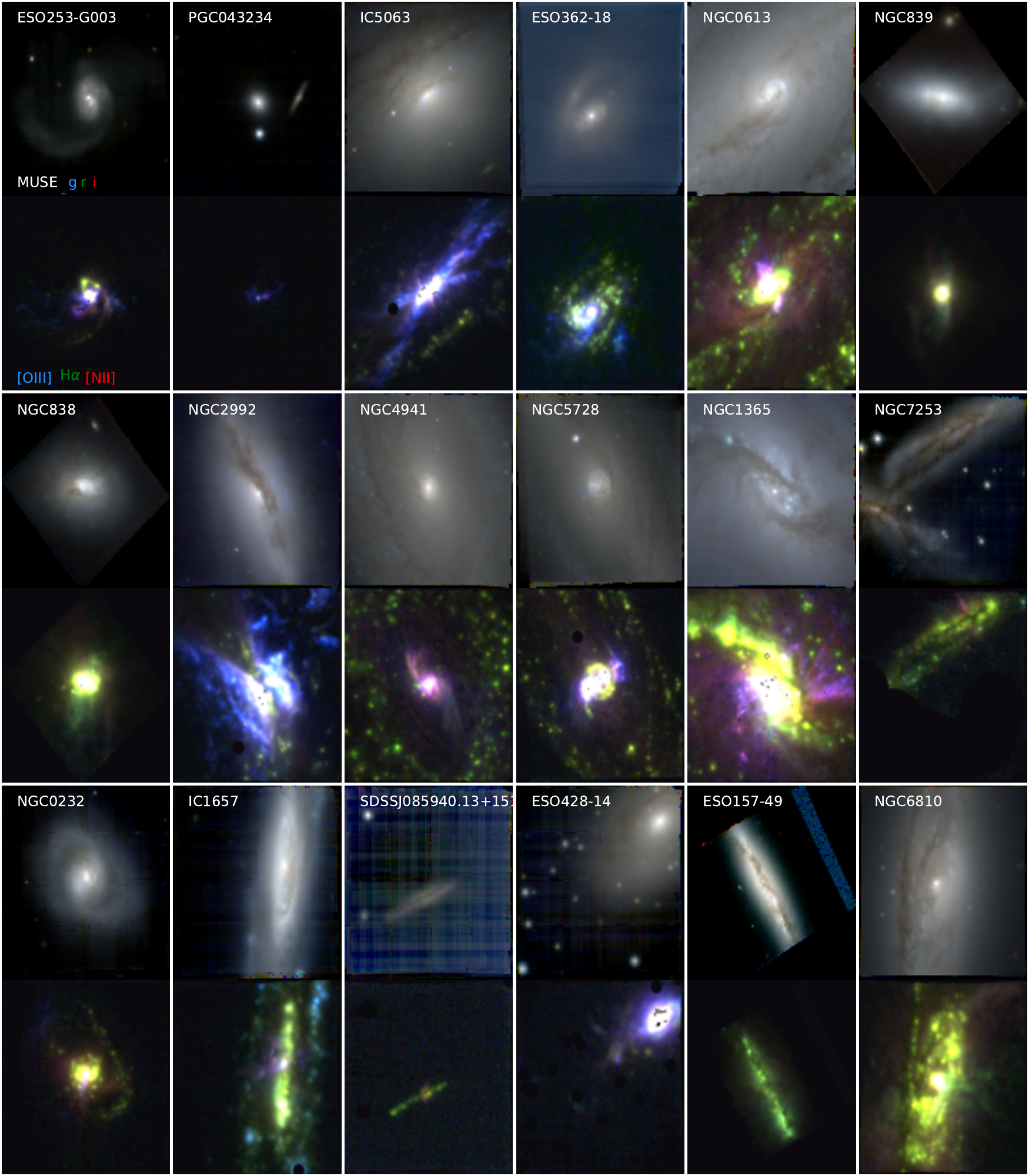}
    \caption{
    Galaxy outflow candidates based on our selection criteria (see text).
    The continuum image and its corresponding emission line-image, as described in Fig.~\ref{fig:SF_galaxy} (panels a and b), are shown for each galaxy. 
    The same FoV of 60$\arcsec \times$ 60$\arcsec$ was selected for each galaxy,  although the scale varies depending the redshift of the object (and the presence of a Mosaic or a single pointing). }
    \label{fig:outflows}

\end{figure*}
\addtocounter{figure}{-1}
\begin{figure*}
    \centering
    \includegraphics[]{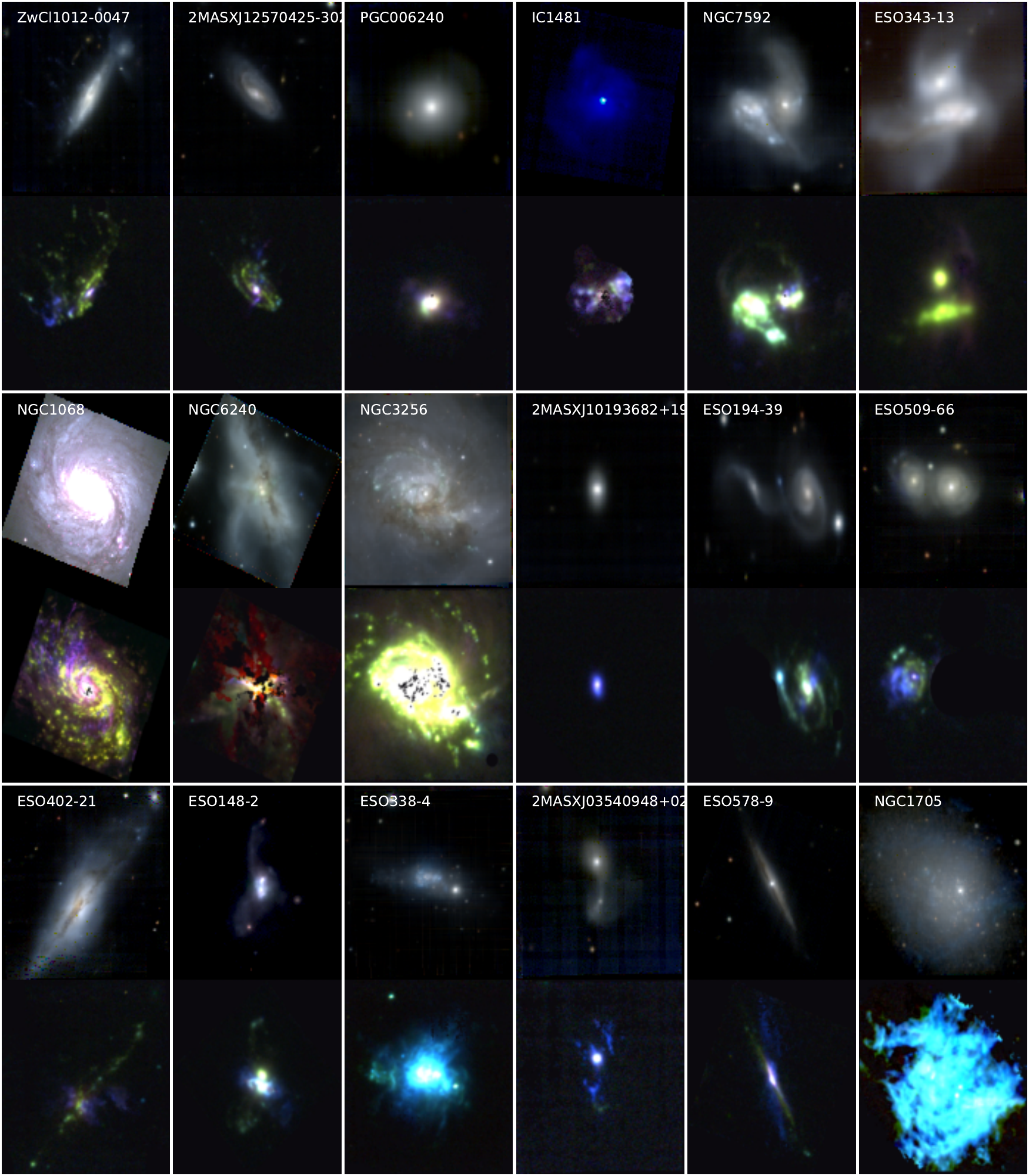}
    \caption{ ({\it continued}) }
    \label{fig:outflows2}
\end{figure*}
\addtocounter{figure}{-1}
\begin{figure*}
    \centering
    \includegraphics[]{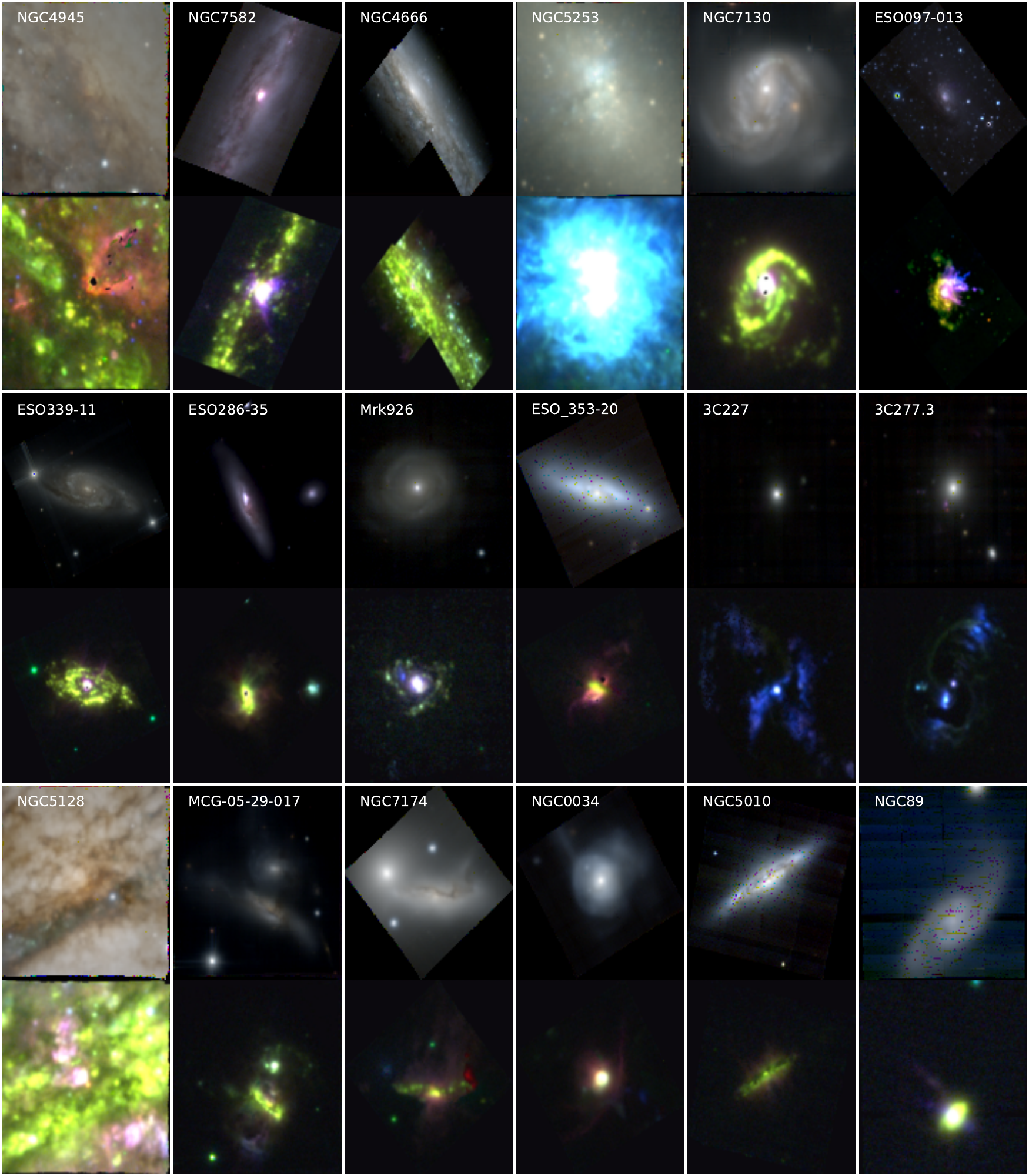}
    \caption{ ({\it continued}) }
    \label{fig:outflows2}
\end{figure*}

\begin{figure*}[t]
    \centering
    \includegraphics[]{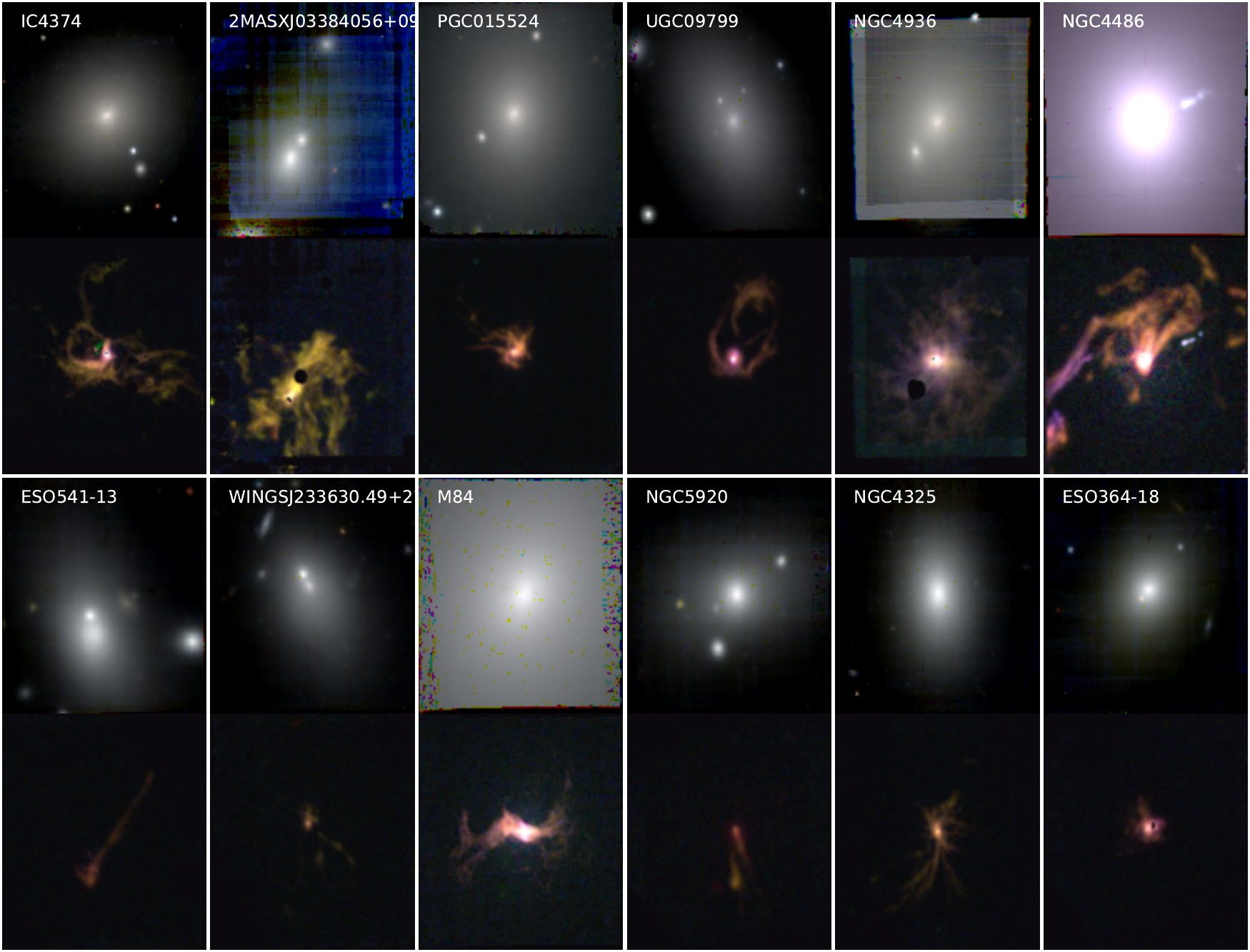}
    \caption{Galaxies with extended emission and filaments, but not fulfilling all the criteria to be selected as {\it bonafide} galactic outflows. Panels show similar images as the one described in Fig.~\ref{fig:outflows}.}
    \label{fig:outflows_E}
\end{figure*}

\section{Results}
\label{sec:results}

Following the examples above, here we present the results of the search for and selection process of galactic outflows in the AMUSING++ compilation.

\subsection{Candidate galactic outflows}
\label{sec:candidates}

Our continuum and emission-line images, the spatially resolved diagnostic diagrams, the kinematic properties of the lines, and their level of asymmetry,
together provide a robust method to select candidate galactic outflows. All galaxies with detected conical/bi-conical emission in AMUSING++ are presented in Fig.~\ref{fig:outflows}, and their main properties are listed in Table~\ref{table:properties}. The reconstructed continuum images as well as those of emission lines, are presented in Fig.~\ref{fig:outflows}. This is our final sample of galaxies hosting a galactic outflow.
Comments on some individual objects are included in Appendix \ref{appendix:individual_objects}. The figures summarizing the whole analysis of the emission lines discussed before (asymmetries maps, kinematics, line-ratios and diagnostic diagrams) for each of these galaxies are included in Appendix  \ref{appendix:outflows_amusing}. The final sample of galactic outflows comprises 54 objects. Similar figures for all the remaining 582 galaxies in the AMUSING++ compilation are included in Appendix \ref{appendix:spatial_resolved_AMUSING} for reference. 

In addition to the objects hosting galactic outflows, in the process of selecting them we found a set of galaxies with extended ionized gas emission, but not fulfilling all the criteria outlined in the selection process. Thus, for these galaxies the ionization appears to be driven by other physical processes. These objects are presented in Fig.~\ref{fig:outflows_E} and their main properties in Table~\ref{table:properties}. We note that all of these galaxies are Elliptical, and in many cases are located at the central regions of galaxy clusters. { These filaments might be associated to the optical counterpart of cooling flows in elliptical galaxies.}

\begin{figure*}
    \centering
    \includegraphics[width=\textwidth]{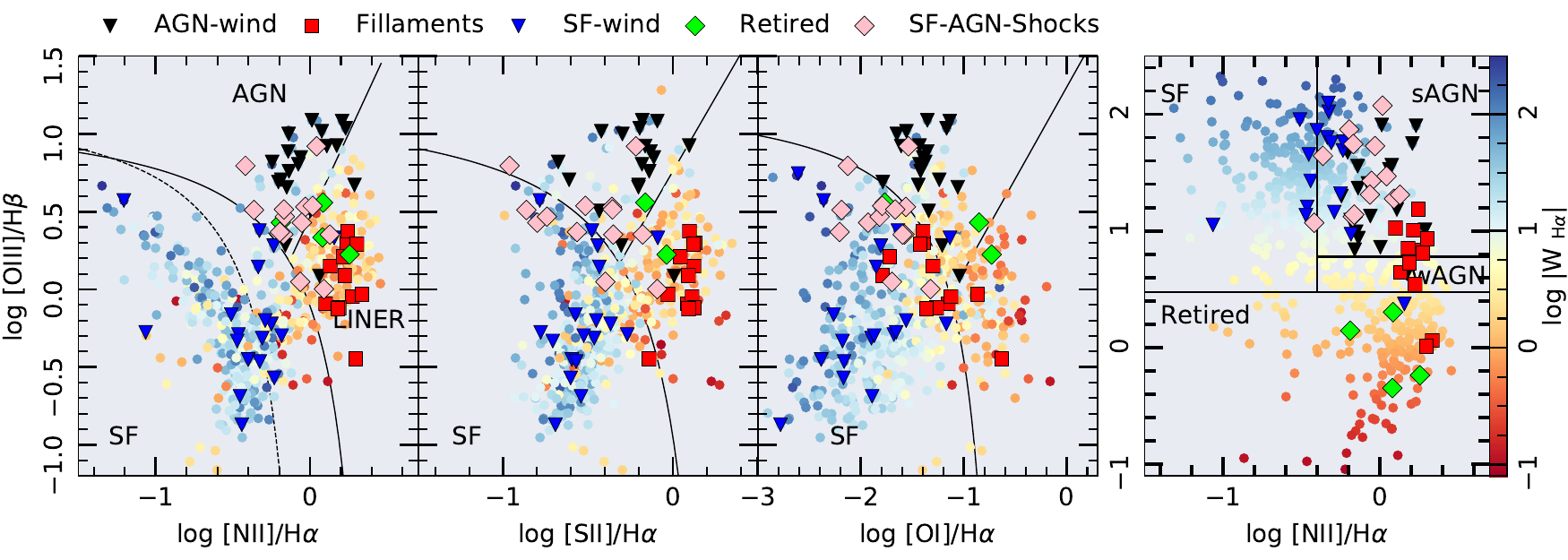}
    \caption{Diagnostic diagrams for the central spectrum of the AMUSING++ galaxies. Line ratios and equivalent widths were extracted from a $3\arcsec\times3\arcsec$ square around the optical nucleus of each galaxy, based on the analysis performed by {\sc Pipe3D}. The \nii/\ha\ vs. \oiii/\hb\ diagnostic diagram is presented in the left panel, with the \sii/\ha\ vs. \oiii/\hb\ one centrally and finally the \oi/\ha\ vs. \oiii/\hb\ in the right. The colored circles represent the AMUSING++ galaxies without outflows. Blue (black) inverted triangles represent galaxies hosting SF--driven (AGN--driven) outflows; pink diamonds are outflow host galaxies whose nuclei are dominated by either SF--AGN or characterized by shock excitation;
    green  diamonds represent  galaxies hosting outflows with central \EWha\,$<3$\,\AA;
    finally, red  squares represent the elliptical  galaxies with extended emission and filaments (see Fig.~\ref{fig:outflows_E}). The black solid curve in the three diagnostic diagrams represents the \citet{kewley01} demarcation curves; the black straight-line represents the \cite{kewley06} Seyfert-LINER demarcation line; the dotted line in the \nii/\ha\ represents the \cite{kauffmann03} demarcation curve. The rightmost panel shows the WHAN diagram \citep[e.g.,][]{Cid2011}, comprising \nii/\ha\ vs. \EWha. The color code represent in all diagrams the $\log$ \EWha\, value.}
    \label{fig:dd_central}
\end{figure*}

\subsection{Global properties of the sample}
\label{sec:global}

In this section we characterise  the main spectroscopic properties of the AMUSING++ sample, in order to (i) understand how the properties of galaxies hosting outflows compare with those of the general population, and (ii) determine which is the most likely physical mechanism driving the observed outflows. 

\subsubsection{Central ionizing source}
\label{sec:bpt}

Figure~\ref{fig:dd_central} shows the distribution of the  \nii/\ha, \sii/\ha, \oi/\ha\, and \oiii/\hb\  line ratios extracted from a $3\arcsec$ aperture centered in the optical nuclei, for those galaxies with detected line emission (601 out of 635), together with the WHAN diagram using the same aperture. 
This figure shows the galaxy distribution over the three classical diagnostic diagrams, which reflects the variety of ionizing sources in the nuclear regions of galaxies. Bluish colors  in these diagrams (\EWha\  $>$ 6~{\AA}),  are associated to a SF nucleus (on the left) or with a strong AGN (on the right), while reddish are associated to retired galaxies or a LINER nucleus.

We define SF nuclei in general, or a SF--driven outflow (for candidates), those galaxy nuclei located simultaneously below the K01 curves and with an \EWha $>$ 6~{\AA}. Moreover, we define as AGN in general or AGN--driven outflows (for candidates) to those galaxy nuclei located simultaneously above the K01 curves and with an \EWha $>$ 3~{\AA}. These include both weak and strong AGN as defined by \citet{Cid2011}. If the central value of \EWha $<$ 3~{\AA}, the galaxy is classified as retired or post-AGB dominated, irrespective of their location in the diagnostic diagrams (for the central ionization). Objects that present an \EWha $>$ 3~{\AA} with some ratio below the K01 curves are either SF--AGN or shock dominated. Thus, they are poorly classified. 

The results of this classification are summarised in Table~\ref{table:classication} and shown in Figure~\ref{fig:dd_central}. There are 19 objects with outflows that lie well below the K01 curves in all diagnostic diagrams (NGC\,839, NGC\,838, NGC\,7253, PS15mb, ESO\,157-49, NGC\,6810, NGC\,7592, ESO\,343-13, NGC\,3256, ESO\,194-39, ESO\,148-IG002, ESO\,338-IG04, NGC\,1705, NGC\,4945, NGC\,5253, ESO\,286-35, MCG-05-29-017, NGC\,7174, NGC\,5010). These outflow galaxies are clearly not driven by an AGN. Other sources lie close to the border between the AGN-SF demarcation, and they could be either classified as AGN- or SF--driven depending on the diagram. In addition, 19 objects (IC\,5063, ESO\,362-18, NGC\,2992, NGC\,4941, NGC5728, ESO\,428-14, JO204, JO135, PGC\,006240, NGC\,1068, NGC\,6240, 2MAS\,XJ10193682+1933131, ESO\,509-66, ESO\,402-21,HE\,0351+0240, ESO\,339-11, Mrk926, 3C277.3, NGC\,5128) are located (in all diagrams) in regions where AGN-dominated ionization is usually found. 

Table \ref{table:properties} lists the results of the galaxy classification based on the properties of the ionized gas in the central region. Assuming that this indicates which is the driving mechanism for the observed outflow we conclude that $\sim 3\,\%$ of them are driven by SF, $\sim 3\,\%$ are driven by an AGN and $\sim 2\,\%$ can be either AGN- or SF--driven. If the AMUSING++ compilation comprises a representative sample of galaxies in the nearby Universe, these numbers would indicate that $\sim8\,\%$ of galaxies host an outflow. This fraction is in agreement with that reported recently using complete and well defined samples extracted from IFS-GS \citep[2--8\,\%][]{Ho2016,clc2018}.
{ Comparing this fraction with the random Poisson noise of the sample, ($\sigma_{Poisson} = \sqrt{N} \sim 25 = 4\%$), the total fraction of outflows found doubles this error. However this is not achieved if it is considered either SF- or AGN-driven outflows separately. Although this is a low value, even in more controlled galaxy samples the reported fraction of outflows is still at the limit of the Poisson noise \citep{{Ho2016,clc2018}}.}

The fraction of AGN host galaxies in the AMUSING++ sample ($52/635\sim 8\,\%$), is nearly double that recently observed in a larger IFS-GS, 4\,\% \citep[e.g.,][]{Sanchez2018}. This may indicate some bias in the selection of the sample towards AGN sources, which is somewhat expected since some of the sub-samples included in this collection comprise only such objects (e.g., CARS).

\begin{deluxetable}{l c c c c }
\tablecaption{Classification of the nuclear ionization of the AMUSING++ galaxies.}\label{table:classication}
\tablewidth{0pt}
\tablehead{
& \colhead{SF} & \colhead{AGN} & \colhead{SF-AGN-Shocks} & \colhead{Retired}
}
\startdata
AMUSING++ & 255 & 52&76 & 217+35$^*$\\
Outflows & 19 & 19 & 13 & 3\\
Extended/Fillaments & 0 & 0 & 10 & 2\\
\enddata
\tablecomments{The BPT and WHAN diagnostic diagrams presented in Fig.~\ref{fig:dd_central} for the central spectrum were used to classify the galaxy nuclear ionization considering both their location with respect to the K01 curves and the central  \EWha\ value. \\
$^*$ Note also that 35 galaxies do not present \ha\ or \hb\ emission and they are catalogued as retired due to the lack of ionized gas.}
\end{deluxetable}

Interestingly, the elliptical galaxies with extended ionized regions not classified as outflows present a nuclear ionization incompatible with that of retired galaxies (i.e., ionized by old-stars). Indeed, they present ionizations that would correspond to weak AGN or simply by ionization due to shocks. We cannot rule out the presence of an AGN in these galaxies. Some of them show clear evidence of AGN activity since they resent radio jets or central radio sources, as is the case of UGC\,09799 \citep[e.g.,][]{Morganti1993} and M87 \citep[e.g.,][]{Owen1989}. However a visual exploration of Fig.~\ref{fig:appendix_eliptics} suggests that the dominant ionization for these objects seems to be more related to shocks: they present filamentary and highly perturbed ionized gas structures. Some authors have already reported that a few of these objects have remnants of a past nuclear activity or recent merging processes or the final end of IGM streams (like cooling-flows), which could produce the observed ionization \citep[e.g.,][]{Balmaverde2018}.

{ At this stage we need to mention that the fractions presented in this section may not be representative of the full population of galaxies at considered redshift range, since they are based on a compilation of data.}

\begin{figure*}
    \centering
    \includegraphics[]{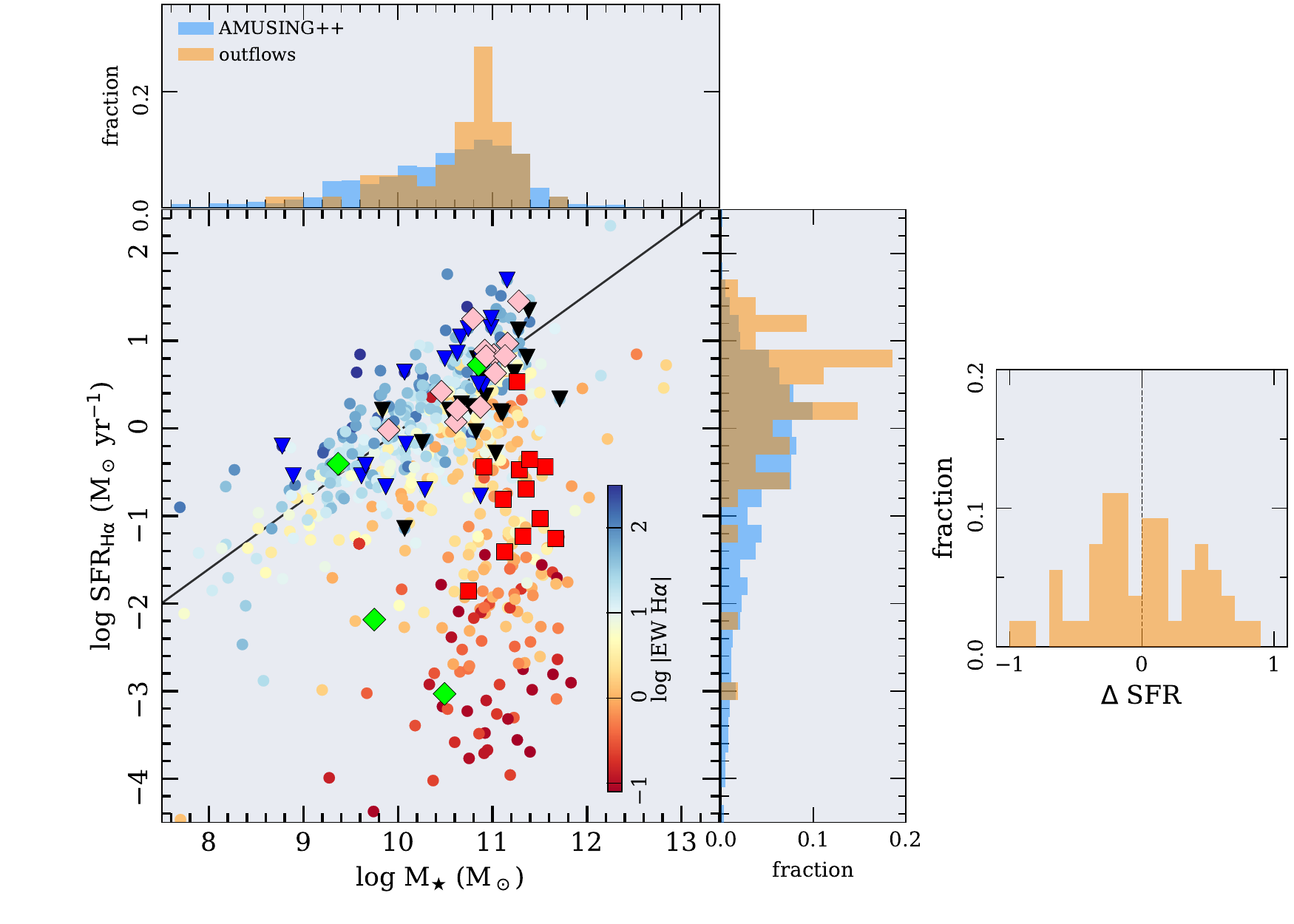}    \caption{Distribution along the SFR-M$_*$ diagram for galaxies in the AMUSING++ compilation and the outflow host galaxies. The SFR was derived from integrated \ha\ flux. The color-code represents the central ($3\arcsec\times3\arcsec$) \EWha\ extracted for each object. The black straight line represents the best fit for the SF objects. Galaxies hosting outflows are labelled with the same symbols included in Fig.~\ref{fig:dd_central}. Histograms of the respective fraction are shown over each axis. For AMUSING++ the histograms are normalized to the total sample (635 galaxies). In the case of the outflows, the distribution is normalized to the 54 outflows detected (i.e., excluding the extended objects). The right-most panel shows the relative difference  ($\Delta$SFR) between the observed SFR of the outflow sources and the expected SFR given its stellar mass with respect to the SFMS.}
    \label{fig:sfms}
\end{figure*}

\subsubsection{Distribution along the SFR-M$_*$ diagram}
\label{sec:SFMS}

It is well known that SFGs follow a tight relation when they are plotted in the SFR-M$_*$ diagram (in logarithm scale). This relation is known as the star formation main sequence (SFMS), and it presents a dispersion of $\sigma_\mathrm{SFMS}\sim$0.25 dex \citep[e.g.,][]{Cano2016}. It has been widely studied at different redshifts although the relation at $z\sim0$ is the most frequently explored  \citep[e.g.][]{Brinchmann2004,Salim2007,Noeske2007,speagle14}. On the other hand, retired galaxies are located well below the SFMS ($>$ 2$\sigma_\mathrm{SFMS}$), conforming a second trend or cloud covering a range of specific star-formation rates (sSFR) broadly corresponding to \EWha$\sim$ 1 \AA\ \citep[based on the relation between both parameters, e.g.][]{sanchez14,belf16b}. Finally, AGN hosts are usually located in the less populated region between these two major groups, known as the Green Valley \citep[e.g.][]{schawinski+2010,Sanchez2018}.

Figure~\ref{fig:sfms} shows the distribution of the integrated SFR along the M$_*$ values and color-coded by the central values of the \EWha\ for the AMUSING++ galaxies.
The M$_*$ was derived using the stellar population decomposition described in Sec. \ref{sec:analysis}, following the prescriptions extensively described in \cite{Pipe3D_II} and \cite{Sanchez2018}. The integrated SFR was derived from the dust corrected \ha\ luminosity (assuming the \citealt{cardelli89} extinction law; $R_\mathrm{v}=3.1$; \ha/\hb\ = 2.86 corresponding to case B of recombination, \citealt[e.g.,][]{osterbrock89}), and applying the \cite{kennicutt98} relation.
In both cases a Salpeter IMF was assumed \citep{Salpeter1955}. Galaxies with high \EWha\ values populate the upper region of this diagram, contrary to galaxies with much lower \EWha\ lying well below the SF objects. The SFMS was obtained with SF galaxies with central \EWha\ $> 6$ \AA\ and line ratios below the K01 curve in the BPT diagram. It follows a log-linear relation between the two parameters, with best fitted parameters from a linear regression to:
$$\rm \log ({\rm SFR}_{\rm H\alpha}) = -7.73{\pm 0.33} \,+\, 0.77{\pm 0.03} \log ({\rm M}_*) $$ 
with $\sigma=0.43$. This relation is very similar those recently reported \citep[e.g.,][]{mariana16}.
For galaxies where \ha\ emission is not associated to SF, the plotted SFR corresponds just to a linear transformation of that luminosity. Thus, the plotted SFR should be considered just as an upper limit to the real SFR (if any). As expected, in the case of retired galaxies (RGs), they are distributed in a cloud well separated from the location traced by the star forming galaxies (SFGs), as mentioned before.

Most of the galaxies hosting a galactic outflow are located within 2$\sigma$ of the loci of the SFMS, independently of the central driving mechanism (SF, AGN, SF--AGN). Most AGN-driven outflow galaxies lie in the Green Valley region of this diagram as has been found in several works \citep[e.g.,][]{Sanchez2018}.
Particularly interesting is PGC\,043234 (green diamond in the retired sequence), although its nuclear  ionization is compatible to host a (weak) AGN, it has a central \EWha~$<$ 3\AA. This object has been presented recently in \cite{prieto16}. These authors favor photoionization by a recent AGN to the observed \oiii\ filaments in this galaxy.
Thus, most galaxies hosting an outflow do actually present active SF, and therefore, the presence of gas, the main ingredient for a galactic wind. It is worth noting that the elliptical galaxies with extended filamentary ionized gas structures lie all below the SFMS, with no evidence of SF activity. This suggests that the nature of ionized gas - in these cases - is most probably external to the object itself, or a remnant of an earlier event.

\begin{figure*}
    \centering
    \includegraphics[]{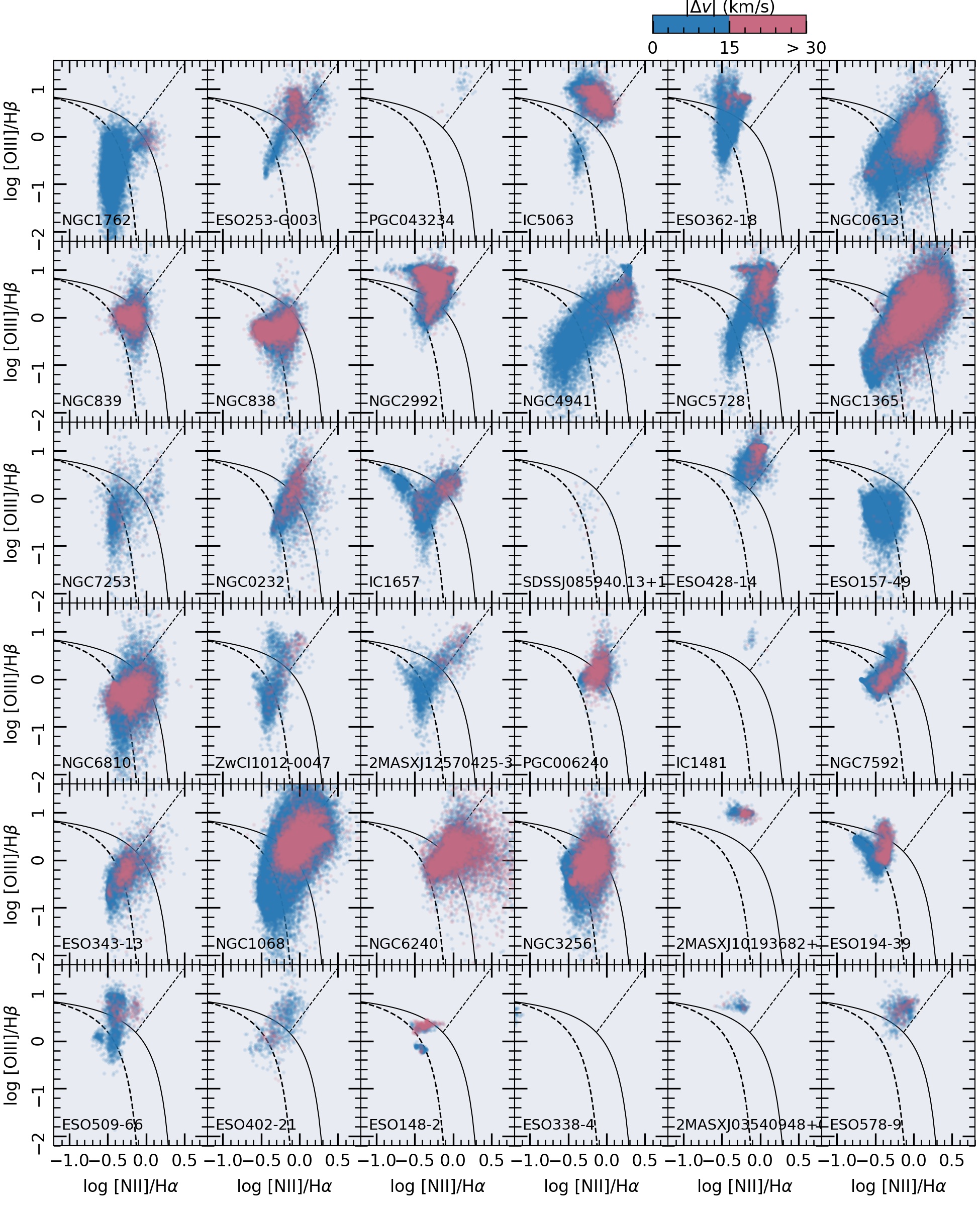}
    \caption{Spatially resolved BPT diagrams for each of the galaxies hosting a galactic outflow and elliptical galaxies with extended emission. We applied a \SN\, $>4$ to all the emission lines involved in the diagrams. The color-code represents the absolute value of the asymmetry ($|\mathrm{\Delta v|}$) derived at each spaxel. Blue points indicate asymmetry values smaller than 15 \kms\ while red points indicate  asymmetry values $>30$ \kms.}
    \label{fig:outflows_bpt_asimmetries1}
\end{figure*}
\addtocounter{figure}{-1}
\begin{figure*}
    \centering
    \includegraphics[trim={0 4cm 0 0},clip]{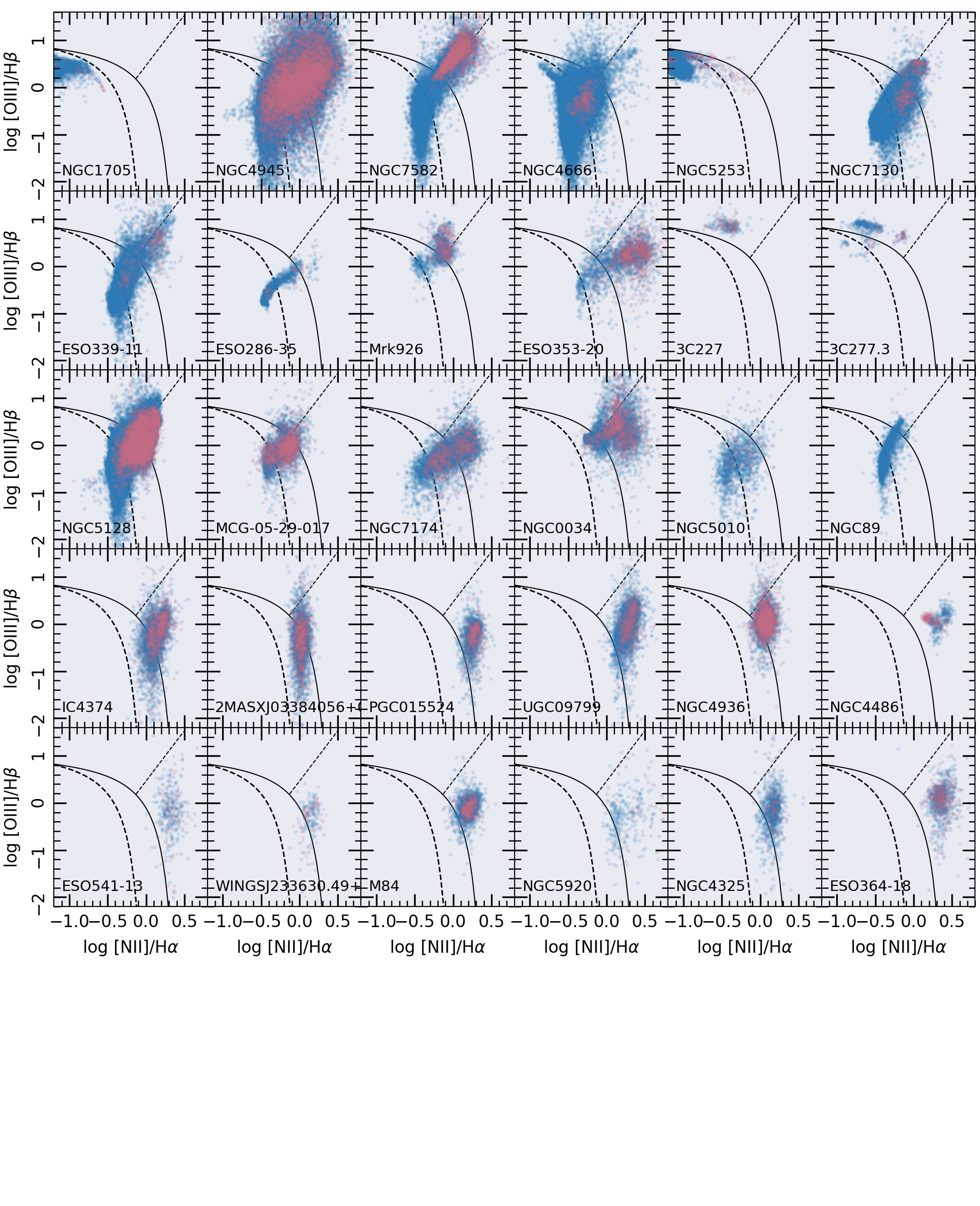}
    \caption{({\it continued})}
    \label{fig:outflows_bpt_asimmetries2}
\end{figure*}


\subsection{Outflows in diagnostic diagrams}
\label{sec:out_diag}

As outlined earlier, a common characteristic of galactic outflows is the presence of multiple components in the emission lines. The number of components needed to fit an emission line depends on the complexity of the wind, and the capacity to resolve them. Typically, when an outflow is observed along the line-of-sight of a galactic disk, the wider components with larger dispersion are associated to the outflow, while the narrow ones are associated to the disk emission. The latter follows the dynamics traced by the gravitational potential of the object (i.e., it traces the systemic velocity). This was already illustrated in Fig.~\ref{fig:dispersion}. In order to study the properties of the outflows it is necessary to decouple the different components in the emission lines. This analysis is beyond the scope of the current paper, and will be addressed in a forthcoming article. 

It is broadly accepted that diagnostic diagrams provide information about the ionizing source in a galaxy or a region within a galaxy. 
However, their in-distinctive use may lead to serious misinterpretations of the physical process occurring across a galaxy. A hint of this was observed in the analysis of Figs \ref{fig:SF_galaxy} and \ref{fig:eline_SN2012hd}, showing that the so-called composite and LINER-like region in the diagrams may be populated by ionization associated with outflows. Moreover, areas well below the usual demarcation lines adopted to select SF regions may be easily populated by shock ionization. Thus, the real origin of the ionization cannot be uncovered by the position in the diagnostic diagrams alone: further information is required.

As described previously, {\sc Pipe3D} performs a moment analysis to estimate the flux of the emission lines. This means that multi-components in the emission lines are treated equally as single-components in the analysis. In the case of outflows, a moment analysis reconstructs the full intensity of the emission lines better than fitting the lines with single Gaussians. In Fig.~\ref{fig:outflows_bpt_asimmetries1}  we show the distribution along the BPT for the individual spaxels with detected emission lines in the galaxies with galactic outflows color-coded by the corresponding values derived for the asymmetry. A fundamental difference with respect to the values shown in the spatially resolved diagnostic diagrams in Appendix \ref{appendix:outflows_amusing}, is that here we select only those spaxels in which all the emission lines in the diagram have at least a \SN $>4$. This selection excludes regions with low signal, most probably removing the diffuse ionized regions that in general have low surface brightness \citep{Zhang2017}. However, it avoids including regions where the asymmetries are influenced by the residuals of the continuum modelling and subtraction.

We include the color-coded BPT digram for NGC\,1762 in the first panel of Fig.~\ref{fig:outflows_bpt_asimmetries1} for comparison purposes. It is clearly seen that the asymmetry is very low ($<15$ \kms) for the entire disk of this galaxy, i.e., for the areas dominated by SF ionization. Only in the nuclear regions, where the ionization is due to post-AGBs and maybe tracing the presence of an AGN, the asymetry rises to values between 15 and 30~\kms. In contrast, outflow host galaxies present clear asymmetric profiles in regions where the outflow in detected, thus are dominated by shock induced ionization. This can be appreciated in more detail in the spatially resolved diagram for each galaxy in Appendix~\ref{appendix:outflows_amusing}. One of the immediate results from this is that the spectroscopically  unresolved (so-far) multi-components are one of the reasons of shifting the line ratios towards the composite-LINER/AGN regions in the BPT diagram. The flux contribution associated to shocks in the emission lines depends on the shock velocity in combination with the pre-shock density \citep[e.g.,][]{mappings}. The combination of shocks+SF emission can cover a wide area in the diagnostic diagrams as shown by \citet{Alatalo2016}. This is obvious from the distribution of regions with high asymmetries in the BPT diagrams. They span through regions usually associated with ionization by young stars, crossing the so-called composite region and reaching areas usually associated with post-AGB/HOLMES or AGN ionization. On the contrary, spaxels  with low asymmetry values lie in general in the SF region, below the K01 curve in the BPT diagram. 

It is clear from this result that without a detailed exploration of the spatially resolved information together with and analysis of the shape of the profiles, a pure exploration of the BPT (or any other diagnostic diagram) can lead to miss-interpretations on the ionizing source.

\section{Discussion and Conclusions}
\label{sec:discussion}

AMUSING++ is the largest compilation of galaxies observed with MUSE so far. The current compilation comprises a total of 635 galaxies, covering the redshift interval $0.0002< z <0.15$, with a mean value of $\sim 0.019$. This compilation, although not complete, does resemble a sample that is selected by diameter given by the limits $10\arcsec<\mathrm{R_{25}}<84\arcsec$. Moreover, for 80\% of the objects we have a complete coverage of the optical extension up to R$_{25}$, i.e., their isophotal radii are smaller than $45\arcsec$. The compilation covers all morphological types and widely populates the color magnitude diagram. Thus, it is well suited to study the spatially resolved properties of galaxies in the nearby universe.

In this paper, we have analysed the ionized gas content across the FoV of the instrument (1 arcmin$^2$) for all the galaxies in the sample. For the vast majority of these objects we could explore the incidence of sub-kiloparsec scale outflows. Indeed, the combination of the high spatial resolution together with the spectral resolution of MUSE provides an opportunity to renew ionization diagnostics, with the addition of spatial information as a new parameter.

We developed a new methodology to detect gas outflows ionized by shocks. It relies on: (i) the visual inspection of the exquisite emission-line images and their lack of association with stellar continuum or AGN; (ii) the search for filamentary and conical/bi-conical structures in the ionized gas and the location of the line ratios in those structures in the BPT and WHAN diagrams; and (iii) the association of those structures with high velocity dispersions, velocity perturbations and strong asymmetries.

The search for outflows and objects with extended emission line regions is straightforward when all this information is combined. 
We found 54 galaxies with evidence of hosting galactic outflows. From this sample 19 objects (3\,\% of the total AMUSING++ sample) are certainly SF--driven and 32 present nuclear ionization of AGN or a combination of AGN-SF. The fraction of bonafide AGN in the sample, corresponds to 8\,\%. This value is not fully in disagreement with other complete galaxy surveys, although the AMUSING++ sample seems to be biased towards the inclusion of these objects.

{ Most of the outflows found in the sample seem to be biased towards high inclined systems, where they are easily detected. However, at this stage it is not feasible to estimate the possible bias introduced by the inclination given that the original sample is not a complete and statistical representative sample. On the other hand we find a fraction of outflows in low inclined galaxies, that are generally excluded from these explorations by primary selections \citep[e.g.][]{clc2018}}.

Despite these biases, a comparison of the distribution of galaxies hosting outflows with the main population indicates that these events are found in almost all galaxy types. Outflows are found in a wide range of stellar masses, from $8.8<\log\, \mathrm{ M_{\star}/M_{\odot}}<11.7$, with a peak around  $~10.9$. Outflows are also present in all morphologies
as shown in Table \ref{table:properties}, although there seems to be a tendency towards disk-dominated spirals (Sb-Sc). There is also a possible bias towards SFGs, with a peak around $\log\, \mathrm{SFR/(M_{\odot}\,yr^{-1})} \sim 0.8$, irrespective of the mechanisms that drives the outflow. 
This result is in many ways similar to that recently reported by \citet{clc2018}, using a well defined and statistically representative sample of galaxies at similar redshifts. The main difference is that in previous studies \citep{Ho2014,clc2018} investigators have used an inclination criteria (high-inclination galaxies) to facilitate the detection of outflows. In our particular case, no inclination criteria was included. Thus, the previous results are here validated and do not seem to be affected by the selection process. 

It is worth noting that outflows do not seem to be  preferentially found in extreme SFGs (in terms of their location with respect to the SFMS), neither the AGN-driven nor SF-driven ones (as already noted by \citealt{clc2018}).
Indeed, outflow
galaxies present a deficit (enhancement) of up to a factor with 10 respect to the expected SFR given its stellar mass, that is relative to the SFMS.  
As noted in \cite{SaraE2018}, this relative excess (deficit) is related  with an overall increase (decrease) of the radial 
distribution of the star formation surface density ($\Sigma$SFR). Although it has been pointed that high values of $\Sigma$SFR are needed to drive outflows \cite[e.g.,][]{Heckman2001,Heckmanetal2002}, this condition seems to be not the main driver. Therefore, it is not conclusive that the amount of SF defines the presence or absence of outflows in galaxies. 
Thus, previous explorations of outflows reporting them preferentially in extreme starbursts \citep[e.g.,][]{Heckman1990,Lehnert1996b,Veilleux2003,Heckman2015,Rupke2018}, were clearly biased in their conclusion due to their target selection.

It seems that the requirements for the presence of outflow events have more to do with the ability of the considered driving mechanism to inject enough energy to the gas being expelled (or at least elevated) above the plane of the galaxy. Therefore, the ratio between the injected energy and the strength of the gravitational potential is probably more relevant than the absolute amount of energy injected. Now that we have a well defined sample of galaxies hosting outflows and a proper comparison sample, we can address this exploration, which will be presented in future studies.

In addition to the outflows, among the explored objects we report a group of galaxies hosting extended filaments and collimated structures of ionized gas with high values  of the \nii/\ha\ ratio. However, they do not fulfill all the requirements for being considered outflows. It is noticeable that all of these galaxies are massive ellipticals ($10.9<\mathrm{\log M_{\star}/M_{\sun}}<11.6$), hosting an AGN (weak or strong) in their nucleus, and lie well below the SFMS in the retired galaxies region. { As discussed in appendix \ref{appendix:individual_objects},  most of these objects are located in the center of galaxy clusters. The excess of \nii\ and \ha\ emission in the filaments resemble the cooling flows observed in the dominant cD elliptical galaxies  in galaxy clusters \citep[e.g.,][]{Heckman1989CoolingFlows,Fabian1994}. As observed in Fig. \ref{fig:appendix_eliptics}, the line ratios of these filaments lie in LINER-like region of the BPT diagrams. Indeed, shocks are the standard explanation to the ionization observed along these filaments \citep{Heckman1989CoolingFlows}.}
Very low \EWha\ are dominant in the filaments, which reflects the complex nature of shocks  being able to reproduce values similar to the ones produced by ionization by old stars. Although it was not the primarily goal of this study, we will continue our exploration of this sample of galaxies in comparison with those ellipticals not hosting such processes to understand their nature.

In summary, we present here a large compilation of 635 galaxies  in the nearby universe observed using the integral-field spectrograph MUSE, the instrument that offers (currently) the best spatial and spectral resolution and the largest FoV. Using this AMUSING++ compilation we developed a new procedure to select outflows without requiring the pre-selection of highly-inclined galaxies. Based on that technique we find a sample of 54 galaxies from which we are able to explore the nature of outflows and the required conditions to produce them: problems that we will further address in forthcoming studies.

\begin{longrotatetable}
\begin{deluxetable*}{l l c c c c c c c c c c}
\tablecaption{Main properties of the galaxies hosting outflows found in AMUSING++.\label{table:properties}
}
\tablewidth{700pt}
\tabletypesize{\scriptsize}
\tablehead{
\colhead{AMUSING++} & \colhead{Galaxy} & \colhead{z} & \colhead{Morph.} & \colhead{$i$} & \colhead{PA} & \colhead{R$_e$} &
\colhead{$\log$ M$_\star$}  & \colhead{$\log$ SFR} & \colhead{Nuclear} &\colhead{$W_{90}$} & References \\
 &  &  &  & \colhead{($^\circ$)} & \colhead{($^\circ$)} & \colhead{(arcsec)} &
\colhead{(M$_\odot$)}  & \colhead{(M$_{\odot}$\,yr$^{-1}$)} & \colhead{Ionization} &\colhead{(\kms)}
} 
\startdata
ASASSN14ko&ESO\,253-G003&0.0426&Sa&57.5&43.1&44.9&11.0&0.84&SF-AGN&57&1\\
ASASSN14li&PGC043234&0.0205&E?&33.5&152.1&11.1&9.7&-2.18&Retired&23&2\\
IC\_P1&IC5063&0.0115&S0-a&36.4&27.6&55.4&11.1&0.17&AGN&77&3\\
IRAS&ESO\,362-18&0.0122&S0-a&39.0&42.2&22.5&10.5&0.21&AGN&46&4\\
NGC\,0613&NGC\,0613&0.0049&Sbc&62.1&36.6&72.6&10.6&0.07&SF-AGN&24&5\\
NGC\,839&NGC\,839&0.0128&S0-a&65.1&174.6&57.6&10.7&1.04&SF&45&6\\
NGC\,838&NGC\,838&0.0128&S0-a&42.2&168.9&53.5&10.7&1.14&SF&59&6\\
NGC\,2992&NGC\,2992&0.0078&Sa&58.3&108.5&45.0&10.8&0.25&AGN&49&3\\
NGC\,4941&NGC\,4941&0.0037&SABa&55.2&106.1&60.5&10.1&-1.14&AGN&22&7\\
NGC\,5728&NGC\,5728&0.0093&Sa&57.0&117.2&65.8&11.1&0.19&AGN&42&8\\
NGC\,&NGC\,1365&0.0054&Sb&44.9&131.2&52.1&10.9&0.89&SF-AGN&46&3\\
SN2002jg\_2&NGC\,7253&0.0154&\nodata&59.8&33.9&49.9&10.1&-0.18&SF&37&new\\
SN2006et&NGC\,0232&0.0227&SBa&39.6&128.4&34.4&11.3&1.45&SF-AGN&32&9\\
SN2012hd&IC1657&0.0119&SBbc&75.0&79.9&68.0&10.6&0.22&SF-AGN&37&new\\
PS15mb&SDSSJ085940.13+151113.6&0.0291&S?&75.6&20.7&18.5&10.3&-0.70&SF&10&new\\
SN2008fp\_2&ESO\,428-14&0.0056&S0&44.9&52.4&68.7&10.3&-0.16&AGN&21&10\\
ESO\,157-49&ESO\,157-49&0.0056&Sc&68.6&120.4&47.7&9.6&-0.54&SF&17&new\\
NGC\,6810&NGC\,6810&0.0067&Sab&58.8&83.2&67.2&10.9&0.49&SF&29&11\\
JO204&ZwCl1012-0047&0.0425&\nodata&66.6&52.9&31.8&10.9&0.37&AGN&51&12\\
JO135&2MASXJ12570425-3022305&0.0545&S?&60.6&128.7&22.6&10.9&0.57&AGN&46&12\\
ASASSN14mw&PGC006240&0.0271&E-S0&46.9&26.8&23.2&10.9&0.51&AGN&39&new\\
IC1481&IC1481&0.0207&Sd&26.8&12.2&4.8&9.4&-0.40&Retired&37&13\\
SN2017ffm&NGC\,7592&0.0246&S0-a&33.6&105.7&30.7&11.0&1.15&SF&36&14\\
ESO\,\_P2&ESO\,343-13&0.0192&\nodata&53.6&45.6&33.0&10.9&0.51&SF&34&15\\
NGC\,1068&NGC\,1068&0.0038&Sb&38.5&118.2&86.5&11.4&1.35&AGN&41&3\\
NGC\,6240&NGC\,6240&0.0235&S0-a&54.1&124.5&70.4&11.7&0.34&AGN&98&16\\
NGC\,3256&NGC\,3256&0.0095&Sbc&43.3&1.9&57.6&11.2&1.69&SF&36&17\\
LSQ14aeg&2MASXJ10193682+1933131&0.0648&E&54.1&89.2&11.6&10.8&-0.04&AGN&45&new\\
AM0044-521&ESO\,194-39&0.0278&\nodata&47.6&160.9&31.4&10.9&0.78&SF&34&new\\
AM1331-231&ESO\,509-66&0.0344&\nodata&46.3&114.6&14.9&10.7&0.28&AGN&96&new\\
AM2113-341&ESO\,402-21&0.0300&SBa&74.5&65.3&51.4&11.0&-0.28&AGN&52&new\\
ESO\,148-IG002&ESO\,148-2&0.0450&Sm&62.2&80.8&37.6&11.0&1.26&SF&82&15\\
ESO\,338-IG04&ESO\,338-4&0.0097&S?&66.1&162.5&37.2&10.1&0.64&SF&20&18\\
HE0351+0240&2MASXJ03540948+0249307&0.0355&\nodata&53.0&115.3&12.4&10.8&0.79&AGN&32&19\\
HE1353-1917&ESO\,578-9&0.0349&Sbc&76.1&117.8&37.6&10.9&0.24&SF-AGN&33&20\\
NGC\,1705&NGC\,1705&0.0020&E-S0&36.1&137.2&37.3&8.9&-0.54&SF&21&21\\
NGC\,4945&NGC\,4945&0.0019&SBc&41.1&123.9&74.3&9.7&-0.42&SF&44&22\\
NGC\,7582&NGC\,7582&0.0052&SBab&72.3&69.1&94.1&10.5&0.42&SF-AGN&30&31\\
ASASSN14lp\_1\_new&NGC\,4666&0.0050&SABc&70.4&133.5&96.2&10.9&0.73&Retired&19&23\\
NGC\,5253&NGC\,5253&0.0013&SBm&63.9&129.2&55.7&8.8&-0.20&SF&27&24\\
NGC\,7130&NGC\,7130&0.0162&Sa&40.5&169.7&38.0&11.2&0.97&SF-AGN&17&25\\
ESO\,097-013&ESO\,097-013&0.0019&Sb&54.7&115.0&94.1&11.0&0.63&SF-AGN&\nodata&3\\
ESO\,\_339-G011&ESO\,339-11&0.0192&SBb&38.6&170.1&28.5&11.3&1.12&AGN&19&new\\
1414&ESO\,286-35&0.0180&Sc&72.5&117.6&44.6&10.5&0.80&SF&85&new\\
HE2302-0857&Mrk926&0.0472&Sbc&37.6&12.1&26.7&11.4&0.81&AGN&184&new\\
ESO\,\_353-G020&ESO\,353-20&0.0161&S0-a&67.8&162.1&38.0&10.9&0.82&SF-AGN&54&new\\
3C\,227&3C\,227&0.0866&\nodata&43.4&61.2&11.4&11.1&0.83&SF-AGN&48&32\\
3C\,277.3&3C\,277.3&0.0859&E&8.1&94.9&14.9&11.2&0.64&AGN&66&33\\
Centaurus&NGC\,5128&0.0017&S0&57.0&11.5&66.6&9.8&0.21&AGN&35&37\\
ESO\,\_440-IG058&MCG-05-29-017&0.0232&Sd&62.5&145.9&26.9&10.6&0.86&SF&48&34\\
HCG90bd&NGC\,7174&0.0090&Sb&81.1&171.6&20.6&10.9&-0.77&SF&29&new\\
NGC\,0034&NGC\,0034&0.0192&S0-a&35.9&144.4&34.5&10.8&1.25&SF-AGN&37&new\\
NGC\,5010&NGC\,5010&0.0099&S0-a&63.1&30.0&37.6&9.9&-0.67&SF&22&new\\
SCG0018\_fieldB1&NGC\,89&0.0110&S0-a&67.3&45.9&36.9&9.9&-0.02&SF-AGN&13&new\\
\hline
Abell&IC4374&0.0217&E-S0&38.2&17.9&37.7&10.9&-0.44&SF-AGN&68&26\\
R0338&2MASXJ03384056+0958119&0.0346&E&48.9&43.0&20.2&11.3&0.54&SF-AGN&45&27\\
PGC015524&PGC015524&0.0328&E&44.3&86.8&55.2&11.6&-0.44&SF-AGN&38&new\\
UGC\,09799&UGC\,09799&0.0344&E&44.1&127.4&42.2&11.3&-0.47&SF-AGN&32&28\\
NGC\,4936&NGC\,4936&0.0108&E&34.6&68.3&21.5&11.4&-0.69&SF-AGN&49&new\\
NGC\,4486&NGC\,4486&0.0042&E&17.4&68.3&37.5&11.1&-1.40&SF-AGN&70&29\\
LSQ13cmt&ESO\,541-13&0.0568&E&57.6&112.8&38.1&11.7&-1.26&Retired&27&new\\
A2626&WINGSJ233630.49+210847.3&0.0547&\nodata&56.3&125.8&39.4&11.5&-1.03&SF-AGN&20&new\\
MCUBE&M84&0.0034&E&31.0&36.4&44.8&10.7&-1.86&SF-AGN&27&30\\
3C\,318.1&NGC\,5920&0.0443&S0&50.5&14.0&34.5&11.3&-1.23&Retired&\nodata&35\\
N4325&NGC\,4325&0.0255&E&48.5&92.0&34.5&11.1&-0.81&SF-AGN&34&36\\
S555&ESO\,364-18&0.0442&E&42.5&52.4&26.4&11.4&-0.35&SF-AGN&59&new\\
\enddata
\tablecomments{AMUSING++ identification (col. 1), galaxy names (col. 2), redshift derived with the SSP analysis (col. 3), Hubble type from Hyperleda  (col. 4), inclination (col. 5),
position angle (col. 6), effective radius (col. 7), stellar mass derived with the SSP analysis (col. 8) source of the outflow based on the nuclear ionization (col. 9), $W_{90}$ of the absolute value of the median on the asymmetry map (col. 10) References at the observed outflow  (col. 10). References: (1) \citealt{Bonatto1997,Yuan_2010}; (2) \citealt{Prieto2016};  (3) \citealt{Mingozzi2019};  (4) \citealt{Mulchaey1996,Fraquelli2000,Humire2018}; (5) \citealt{TIMER}; (6) \citealt{Rich2010,Vogt2013}; (7) \citealt{Barbosa2009}; (8) \citealt{Durre2018a,Durre2018b}; (9) \citealt{LopezCoba2017}; (10) \citealt{Falcke1998}; (11) \citealt{Venturi2018}; (12) \citealt{Poggianti2019}; (13) \citealt{clc2018}; (14) \citealt{Rafanelli1992}; (15) \citealt{Rich2015}; (16) \citealt{Muller2018,Treister2018}; (17) \citealt{Rich2011}; (18) \citealt{Bik2015}; (19) \citealt{Powell2018}; (20) \citealt{Husemann2019}; (21)\citealt{Menacho2019}; (22) \citealt{Venturi2017}; (23) \citealt{Dahlem1997}; (24) \citealt{Heckman2015}; (25) \citealt{Davies2014}; (26) \citealt{Farage2012,Canning2013,Olivares2019}; (27) \citealt{Donahue2007};
(28) \citealt{Balmaverde2018}; (29) \citealt{Jarvis1990,Sparks1993,Gavazzi2000}; (30) \citealt{Bower1997}; (31) \citealt{Storchi-Bergmann1991}; (32) \citealt{Prieto1993}; (33) \citealt{Solorzano2003}; (34) \citealt{Monreal2010}; (35) \citealt{Edwards2009}; (36) \citealt{McDonald2011}; (37) \citealt{Santoro2015}.
}
\end{deluxetable*}
\end{longrotatetable}

\acknowledgments
We thank the anonymous referee for her/his useful comments.
CLC, SFS and JKBB are grateful for the support of a CONACYT grant CB-285080 and FC-2016-01-1916, and funding from the DGAPA-UNAM  IA101217 and PAPIIT: IN103318 projects. CLC acknowledges a CONACYT (Mexico) Ph.~D. scholarship. ICG acknowledges support from DGAPA-UNAM grant IN113417. LG was funded by the European Union's Horizon 2020 research and innovation programme under the Marie Sk\l{}odowska-Curie grant agreement No. 839090. TRL acknowledges financial support through the grants (AEI/FEDER, UE) AYA2017-89076-P, AYA2016-77237-C3-1-P and AYA2015-63810-P, as well as by the Ministerio de Ciencia, Innovaci\'on y Universidades (MCIU), through the State Budget and by the Consejer\'\i a de Econom\'\i a, Industria, Comercio y Conocimiento of the Canary Islands Autonomous Community, through the Regional Budget. TRL is supported by a MCIU Juan de la Cierva - Formaci\'on grant (FJCI-2016-30342). Support for JLP is provided in part by FONDECYT through the grant 1191038 and by the Ministry of Economy, Development, and Tourism’s Millennium Science Initiative through grant IC120009, awarded to The Millennium Institute of Astrophysics (MAS).

Based on observations collected at the European Southern Observatory under ESO programmes: 0100.A-0607(A), 0100.A-0779(A), 0100.B-0116(A), 0100.B-0573(A),
 0100.B-0769(A), 0100.D-0341(A), 0101.A-0168(A), 0101.A-0772(A),
 0101.B-0368(B), 0101.B-0603(A), 0101.B-0706(A), 0101.C-0329(C),
 0101.C-0329(D), 0101.D-0748(A), 0101.D-0748(B), 0102.B-0048(A),
 0103.A-0637(A), 0103.B-0834(B), 094.A-0205(B), 094.A-0859(A),
 094.B-0225(A), 094.B-0298(A), 094.B-0321(A), 094.B-0345(A),
 094.B-0592(A), 094.B-0592(C), 094.B-0612(A), 094.B-0711(A),
 094.B-0733(B), 094.B-0745(A), 094.B-0921(A), 095.A-0159(A),
 095.B-0015(A), 095.B-0023(A), 095.B-0042(A), 095.B-0049(A),
 095.B-0127(A), 095.B-0295(A), 095.B-0482(A), 095.B-0532(A),
 095.B-0624(A), 095.B-0686(A), 095.B-0934(A), 095.D-0091(B),
 096.A-0365(A), 096.B-0019(A), 096.B-0054(A), 096.B-0062(A),
 096.B-0223(A), 096.B-0230(A), 096.B-0309(A), 096.B-0325(A),
 096.B-0449(A), 096.B-0951(A), 096.D-0263(A), 096.D-0786(A),
 097.A-0366(A), 097.A-0366(B), 097.A-0987(A), 097.B-0041(A),
 097.B-0165(A), 097.B-0313(A), 097.B-0427(A), 097.B-0518(A),
 097.B-0640(A), 097.B-0761(A), 097.B-0766(A), 097.B-0776(A),
 097.D-0408(A), 097.D-1054(B), 098.A-0364(A), 098.B-0240(A),
 098.B-0619(A), 098.C-0484(A), 099.A-0023(A), 099.A-0862(A),
 099.A-0870(A), 099.B-0137(A), 099.B-0148(A), 099.B-0193(A),
 099.B-0242(A), 099.B-0294(A), 099.B-0384(A), 099.B-0411(A),
 099.D-0022(A), 196.B-0578(A), 196.B-0578(B), 196.B-0578(C),
 296.B-5054(A), 296.D-5003(A), 60.A-9100(H), 60.A-9194(A),
 60.A-9304(A), 60.A-9308(A), 60.A-9310(A), 60.A-9312(A),
 60.A-9313(A), 60.A-9314(A), 60.A-9317(A), 60.A-9328(A),
 60.A-9332(A), 60.A-9333(A), 60.A-9337(A), 60.A-9339(A) and
 60.A-9349(A).




\appendix

\section{Derivation of isophotal parameters}
\label{appendix:isophote_params}

\begin{figure*}
    \centering
        \includegraphics{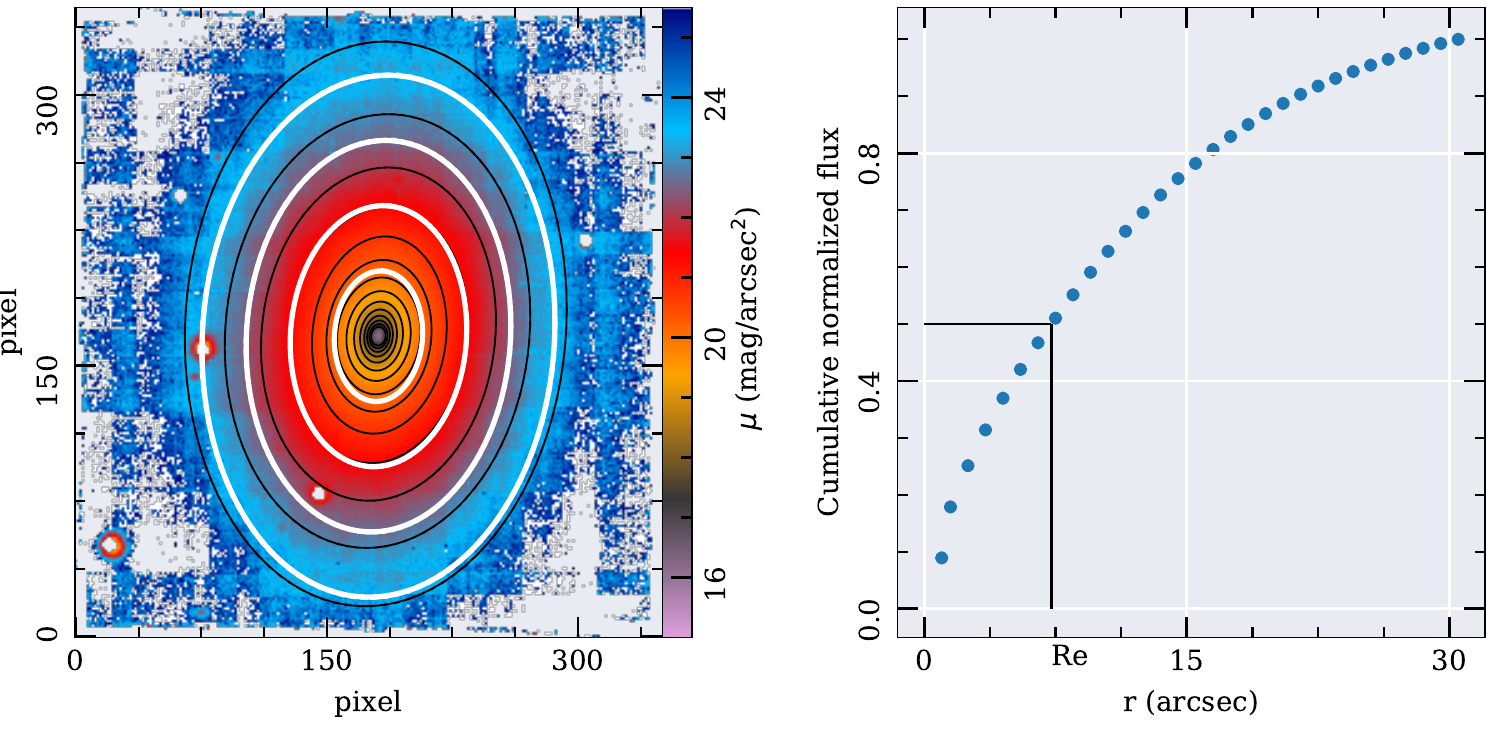}
    \caption{Isophotal analysis applied to the V--band images of all AMUSING++ galaxies to estimate the effective radius (Re), inclination ($i$) and position angle (PA). {\it Left panel:} surface brightness distribution of ESO\,197-18. Black elliptical annulii represent isophotes at different radii, with the outermost annulus corresponding to a surface brightness of 25 mag\,arcsec$^{-2}$ ($\mathrm{R_{25}}$). White annulus represent 1, 2, 3  and 4 effective radii (from innermost to outermost respectively).
    {\it Right panel:} Cumulative flux distribution along the galactocentric distances of the isophotes shown in the left panel (black solid-lines). The effective radius is marked with  black straight lines.}
    \label{fig:isophote_ana}
\end{figure*}

As part of the characterization of the AMUSING++ compilation we performed an isophotal analysis on all galaxies in the sample. For this purpose we use {\sc Photutils}, an {\sc Astropy} package for detection and photometry extraction of astronomical sources \citep{Bradley_2019_2533376}. This package mimics the algorithms included in {\sc SExtractor} \citep[e.g.,][]{sextractor}. We perform the isophotal analysis on the V--band images produced from the data cubes. Prior to this analysis, we mask the field stars. Then, multiple isophotes are estimated at differet galactocentric distances down to a surface brightness of 25 mag\,arcsec$^{-2}$ (i.e., at the isophotal radius, $\mathrm{R}_{25}$). The position angle and inclination of the last isophote are adopted for further inclination corrections. Inclination $i$ is defined in terms of the ellipticity, $\epsilon$, as cos\,$i = 1 -\epsilon$. The values of both parameters are listed in Table \ref{tab:all_amusing_table}. The effective radius is derived based on the cumulative distribution of fluxes within these isophotes, estimated as the radius at which half of the total light is contained. Figure~\ref{fig:isophote_ana} illustrates this procedure. 

As mentioned in section \ref{sec:amusing_presentation}, for $20 \%$ of the objects the isophotal radius is larger than the FoV of MUSE. For these cases, $\mathrm{R}_{25}$ was estimated by an extrapolation of the cumulative flux curve. The value of Re in these cases is therefore an approximation and should be used with care.

\section{Comments on individual outflow host galaxies}
\label{appendix:individual_objects}

In here it is addressed the individual outflow host galaxies found in AMUSING++. The corresponding figures of each galaxy can be found in the electronic version of this article. 
\\
\\
{ ESO\,253-G003:} 
This is a merging galaxy classified as a Seyfert 2 \citep[e.g.,][]{Bonatto1997,Yuan_2010}. It presents an extended emission in \oiii. The information provided by the emission-line image and the diagnostic diagrams show that the extended emission is located in the AGN region. Its nuclear ionization falls in the border of the K01 curves, suggesting an AGN as responsible for the high \oiii/\hb\ ratio observed across this structure. High velocity dispersion ($>70$ \kms) is observed throughout this structure which correlates with the \nii/\ha\ ratio. 

{ PGC\,043234:}
This is a post startbust galaxy ($z = 0.0205$) result of a late merger event \citep[e.g.,][]{Prieto2016}. PGC\,043234 shows clear fillamentary ionized structures visible mainly in \oiii\ emission. This galaxy has been recently studied (using the same data as here) by \citet{Prieto2016}. 
They found that the ionized filaments extend to 5 kpc from the nucleus and multiple small structures up to 10 kpc away. The ionization of these filaments falls above the K01 curves in the spatially resolved diagrams. These line ratios could be explained in principle by shocks or AGN photoinization. Except for \sii, which is not detected, the central ionization is consistent with an AGN.
\cite{Prieto2016} favour the AGN ionization, given the low velocity dispersion found along the \oiii\ fillaments ($\sigma<40$ \kms).
We also do not detect  significant asymmetries in the cross correlation function ($|\Delta\mathrm{v}| <15$ \kms) that could support the scenario of shock ionization. 

{ IC\,5063, NGC\,2992, NGC\,1365, NGC\,1068 \& ESO\,097-013:}
These galaxies have been recently analysed (with MUSE data) as part of the MAGNUM survey \citep{Mingozzi2019}. They all host well known outflows with conical structures, all of them driven by a central AGN. Our spatially resolved diagnostic diagrams reveal in detail the ionization across the outflowing gas.

{ ESO\,362-18 or IRAS0177:}
ESO\,362-18 is a Syfert 1.5 S0 galaxy at $z = 0.0125$ \citep[e.g.,][]{Bennert2006}. This galaxy presents an extended ionized gas emission  outflowing from its nucleus. This structure was first revealed in narrow band images in \oiii\ \citep[e.g.,][]{Mulchaey1996,Fraquelli2000}.
Recent studies using GMOS-IFS ($3.5\arcsec \times5\arcsec$) revealed the ionization cone of an outflow driven by the central AGN \citep[e.g.,][]{Humire2018}.

The nucleus of this galaxy lies above the K01 curves in the spatially resolved diagrams while the ionization at the cone falls in the AGN--driven wind region, confirming the origin of this outflow. We detect asymmetries in the cone, $|\Delta \mathrm{v}|$ $\sim 30 - 60$ \kms,  as well as in the nuclear region. The presence of multi component kinematics in the emission lines was also detected by \cite{Humire2018}.

{ NGC\,0613:}
This is a barred SBbc galaxy at 18.32 Mpc. Evidence of an outflow in this galaxy was first detected in radio waves. NGC\,0613 hosts a radio jet in the central region, with an optical counterpart seen in \ha\ and \oiii\ emission \citep[e.g.,][]{Hummel1987}. The MUSE data of this object was presented first by the TIMER project. NGC\,0613 shows a clear bi-conical outflow more intense in \nii\ (reddish color in the emission-line image). Its ionization is located in the upper region in the diagnostic diagrams without a clear trend of being AGN or SF driven outflow, possibly both. We detect some asymmetries, with $\mathrm{|\Delta v| <70}$ \kms\ around the nucleus and the cone structure.

The disk is traced by \ha\ emission coming from \hii\ regions. Curiously, a significant excess of \nii\ emission in the inter arms regions is observed. Part of these ``\nii\ arms''are dominated by ionization of old stars and is located in the LINER region in the diagnostic diagrams. This gas seems to lie over the disk, rather than having an extraplanar origin. More studies to confirm the origin of this excess in \nii\ are required.

{ NGC\,838 \& NGC\,839:}
These galaxies are part of the Hickson Compact Group 16 (HCG 16c,d) \citep{Hickson1989}. They both present SF driven outflows with clear bi-conical structure of ionized gas \citep[e.g.,][]{Vogt2013}.
NGC\,839 was studied before by \cite{Rich2010} using the Wide Field Spectrograph (WiFeS). \ha\ and \nii\ filaments are observed  ouflowing from the north and south of the nucleus. The inner areas are located in the SF region in the diagnostic diagrams. In contrast, the gas in the fillaments is compatible with being ionized by shocks produced by a SF-wind. Asymmetries in the emission lines are found in both ionized cones with values of $\mathrm{|\Delta v| <110}$, consistent with the presence of multiple kinematic components.

{ NGC\,5728:}
It is a Seyfert II galaxy that presents two ionization cones produced by a central AGN \cite[e.g.,][]{Wilson1993}. MUSE observations of this object have been presented in \cite{Durre2018a,Durre2018b}. Our emission-line images reveal two ionized cones extending toward the SE and NW from the optical nucleus. These cones are regions of high excitation reaching values of \oiii/\hb\ $\sim 10$ in the inner areas. The cone nebulae is consistent of being ionized by an AGN wind in the spatially resolved maps. We detect strong asymmetries at the edges of the two cones, $|\Delta\mathrm{v}| <100$ \kms.

{ NGC\,6810:}
An analysis of the MUSE data of this galaxy has been partially presented in \cite{Venturi2018}. NGC\,6810 presents a clear conical outflow, observed
mainly in \ha. Given the line ratios observed along the ionized gas cone, the origin of this outflow is most probably related to SF processes. The asymmetry map of this galaxy reveals high values at the location of the ionized cone, confirming the precense of multiple kinematic components.

{ NGC\,7253:}
NGC\,7253 (Northern galaxy from the corresponding image in the electronic version) is an interacting galaxy with UGC\,11985. The center of NGC\,7253 shows a collimated \nii\  structure outflowing from its nucleus. 
The spatially resolved diagrams place this structure in the shock region, compatible with a SF--driven outflow according to the \cite{Sharp2010} demarcations lines. 

{ NGC\,0232:}
The collimated \oiii\ structure observed in this galaxy has been recently reported in \cite{LopezCoba2017}. The collimated structure falls in the AGN region in the diagnostic diagrams, with clear asymmetries along this structure. The \oiii\ jet--like structure is produced most probably by the central AGN.

{ SDSSJ085940.13+151113.6:}
This is a highly inclined disk galaxy  ($i=75^{\circ}$), located at $z = 0.0292$ (the scale is 0.6 kpc/arcsec at the distance of this object). The large inclination of this galaxy facilitates the separation between the gas in the disk (mostly \ha\ emission) with respect to the extraplanar one. 
A \nii\ enhancement is observed in the nucleus, revealing the presence of a small bi-conic structure. The extension of this structure on both sides of the disk is the order of $6.5\arcsec$ ($\sim 4$ kpc at the distance of this object), and is clearly resolved. Even though just tens spaxels fall in the bi-cones, these  are well identified in the spatially resolved diagrams above the K01 curves. The velocity dispersion in the cones reaches 150 \kms, much larger than the values found in the disk. Asymmetries between 20-50 \kms\ are detected inside the cones, a signature of the presence of multiple components associated to shocks.

{ ESO\,428-14:}
It is a Seyfert 2 galaxy at $z = 0.0056$. A collimated ionized gas structure aligned to an inner radio jet was already reported in this galaxy \citep[e.g.,][]{Falcke1998}. The MUSE data reveals a very bright \oiii\ nucleus, clearly ionized by the central AGN.

{ UGC\,11723:}
This is an edge on galaxy located at 70 Mpc \citep[e.g.,][]{Mendel2011}. Their disk is dominated by \ha\ emission, with some filamentary structures in the SW side. A slab of \nii\ emission is distributed at both sides of the disk plane. The \nii/\ha\ line ratio increases at larger extra-planar distances, indicating the existence of an extra source of ionizing photons different to that provided by \hii\ regions in the disk plane. Although HOLMES can reproduce the observed line ratios, the large \EWha\ values do not favour this hypothesis. Shocks may be responsible for the enhancement of the \nii/\ha\ ratio. The spatially resolved diagrams of the \nii\ slab are consistent with this kind of ionization, as well as the observed increase in the velocity dispersion. 

{ JO135 \& JO204:}
Also named 2MASXJ\,12570425-3022305 and 2dFGRS\,TGN288Z210 respectively. These are two Jellyfish galaxies located in 
the A3530 and A957 clusters, respectively. These galaxies has been analysed with MUSE data as part of the GASP survey \citep[e.g.,][]{Poggianti2019}.
Both galaxies are classified as Seyfert 2 and host  ionization cones produced by central AGN. The ionization cones are observed in the emission-line image as an extended blue nebulae (\oiii\ emission). Multiple asymmetries are detected through the cones
as well as in regions around its respective nuclei.

{ PGC\,006240:}
ESA/Hubble images of PGC\,006240 show concentric shells which are most probably the product of a merger event in the past. An arc structure on the NW side of the MUSE continuum image is observed. The ionized gas distribution of this galaxy is also complex. Two filaments - one at the NW and other at SE from the nucleus - are observed, together with an arc shape \ha\ structure (green in the emission-line image) extending to the SW and coming from the nucleus. This arc structure presents low values of the \nii/\ha\ ratio consistent with ionization by \hii\ regions in the BPT diagram. The nucleus as well the filaments present ratios consistent with LINER ionization, nevertheless, this latter component is dominated by \EWha\ $>3$\,\AA, and is therefore most compatible with shock ionization.

{ IC\,1481:}
This galaxy was classified as an outflow candidate in \cite{clc2018}. Unfortunately, a bad sky subtraction in this galaxy prevented us from recovering clean maps of the emission lines. A new analysis of this object is required.

{ NGC\,6240:}
This galaxy hosts a well known outflow that is result of a merger \citep[e.g.,][]{Muller2018}. NGC\,6240 exhibits multiple \ha\ and \oiii\ filaments. The advanced merger stage of these galaxies did not allow us to properly estimate the line velocity at each spaxel. A bad estimation of the velocity gives as result an incorrect identification of the emission lines. \cite{Treister2018} has obtained emission line MUSE maps much better than those reported in this work.

{ NGC\,7592 \&  ESO\,343-13:}
These systems exhibit galactic outflows result of ongoing interactions. Both systems present biconical gas structures (seen in purple colors in the emission-line images) with line ratios consistent with SF-wind outflows.

{ NGC\,3256, NGC\,4666 \& NGC\,7130:}
These galaxies present regions where is observed as an excess in \nii\ decoupled from \ha\ gas distribution. The line ratios associated with this \nii\ emission locates them in regions above the K01 curves and they are characterized with \EWha $>3$\AA, and relatively high dispersion $>50$ \kms. All these suggest the idea of SF--driven outflows in these galaxies.

{ 2MASXJ10193682+1933131:}
This is an elliptical galaxy at $z = 0.0648$. It present an excess of \oiii\ emission at PA = 180. The \nii/\ha\ map together with the velocity dispersion shows enhancements towards the Northern and Southern parts of the nucleus. The central ionization, as well as the strong \oiii\ emission, supports the idea of an AGN-wind, at being far above from the K01 curves.

{ ESO\,194-39:}
This galaxy is in interaction with 6dFGS\,gJ004705.6-520301. Two \oiii\ ionized cones seems to be outflowing from the nucleus of ESO\,194-39. As indicated by the blue color of these cones, their ionization is located in the upper region in the spatially resolved diagrams, consistent with being ionized by a central AGN. These ionized filaments are accompanied with an increase in the velocity dispersion ($> 75$ \kms) as well as the asymmetries ($>30$ \kms).

{ ESO\,509-66:}
This is an interacting system, with its closest companion (6dFGS\,gJ133440.8-232645) at 10.5 kpc \citep{Koss2012}. ESO\,509-66 exhibits an impressive \oiii\ ionized cone that resembles that of well known cone of Circinus. High \oiii/\hb\ ratios are predominant in the cone nebula, that locates it the  AGN-driven wind region in the spatially resolved diagnostic diagrams.

{ ESO\,402-21:}
The high inclination of this galaxy ($i = 75^{\circ}$) favours the detection of a biconical outflow evident in \oiii\ emission. Extended \oiii\ fillaments are observed at both sides of the disk plane of this galaxy. The observed line ratios
in the cone nebulae are compatible of with the ionization produced by an AGN. Their location in the BPT diagram falls in the AGN-wind locus. 

{ ESO\,338-4}
This is a relative low mass galaxy ($\log\,\mathrm{M}_{\star}/\mathrm{M}_{\odot} = 10.1$ ). This starbust galaxy shows ionized \ha\ cones result of a recent or ongoing SF. The emission-line image is dominated by \oiii\ emission resulting in an excess in the blue color. The ionized gas in this galaxy is entirely produced by young stellar cluster. This is observed in the upper part of the spatially resolved diagrams below the K01 curves, compatible with low gas metallicities.  This galaxy has been previously analysed with MUSE data in \cite{Bik2015}.

{ 2MASXJ03540948+0249307 \& ESO\,578-9: }
These two galaxies exhibit ionized \oiii\ filaments outflowfing from the nuclear regions. The emission line-image shows the decoupling of the \oiii\ gas from the overall distribution of ionized gas traced by the \ha\ and \nii\ emission. The \oiii\ fillaments also shows a spatial decoupling with the continuum emission. An extensive analysis with MUSE data of these galaxies has been performed in \citet{Powell2018} and \citep{Husemann2019} respectively.

{ NGC\,1705}
This galaxy has been analysed with MUSE data in \cite{Menacho2019}, the driving mechanism of the observed outflow is by stellar feedback.

{ NGC\,4945:}
The relative low redshift of this galaxy (z = 0.0019) makes that just a small fraction of their optical extent fits into the FoV of MUSE. The continuum image shows the central region of NGC\,4945. The emission-line image reveals an ionized \nii\ cone on top of the \ha\ distribution tracing the disk. Spaxels belonging to this ionized cone lie in the SF-wind region in the diagnostic diagrams. This galaxy has been partially presented with MUSE data in \cite{Venturi2017}.

{ ESO\,339-11}
This is a LIRG with a Seyfert 2 nuclei \citep{Yuan_2010}. It shows an ionized cone extended in the NW direction. The ionized cone, (seen in pinky colors in the emision-line image), is located at the right side from the AGN-SF wind  bisector in the diagnostic diagrams. A correlation between the \nii/\ha\ ratio and $\sigma$ is observed in the cone, indicative that the gas is shocked.

{ NGC\,7582:}
The \oiii\ ionized cone observed in this galaxy was revealed with narrow band images in \cite{Storchi-Bergmann1991}. The MUSE emission-line image reveals this ionized cone with bluish colors. Indeed, it is appreciated a second cone attenuated by the dust trough the disk. This second cone is also evident in the velocity dispersion map, with velocities larger than 100 \kms. The correlation between \nii/\ha\ ratio and $\sigma$ is observed only in the cone nebulae. Spaxels belonging the cone lie in the AGN-wind region traced by the \citet{Sharp2010} bisector.
The conical \oiii\ outflow is clearly present in the \ha\ velocity map, producing clear deviation from the regular rotation pattern.

{ ESO\,286-35:}
ESO\,286-35 is an ULIRG ($\log\,L_{IR}/L_\odot = 11.25$) \citep[e.g,][]{Tateuchi2015}
classified as Sb galaxy at $z = 0.07$.  It presents highly disturbed ionized gas with \ha\ and \nii\ filaments extending to the NW and SE direction. The \nii\ filaments (redish colors in the emission-line image) lie all above the K01 curves and in the SF-wind region according to the \citep{Sharp2010} demarcation.

{ HE\,2302-0857 or Mrk\,926:}
This galaxy presents an ionized cone visible in \oiii\ emission. Their bright nucleus is located in the AGN region in the diagnostic diagrams suggesting an AGN as responsible of the observed cone. The nucleus shows higher velocity dispersion ($\sigma_{nucleus}> 100$ \kms) than found in the cone nebula ($\sigma_\mathrm{[OIII]\,cone} <80$ \kms).

{ ESO\,353-20:}
This a LIRG ($\log\,L_{IR}/L_\odot = 11.06$) \citep[e.g,][]{Lu2017} at $z=0.0161$.
The emission-line image reveals two ionized cones in \nii\ emission at both sides of the galaxy disk. Spaxels belonging to these cones (pinky colors), lie completely above the K01 curves and at the SF-wind region in the spatially resolved diagrams.  Both the \nii/\ha\  ratio as well as the velocity dispersion shows a positive correlation in the cone nebulae. Asymmetries $>50$ \kms\ are also observed, indicative of the precense of multicomponents in the emission lines associated most probably to shocked gas.

{ 3C\,227:}
This is a  powerful radio galaxy \citep{Black1992} at $z = 0.085$ that exhibit extended \oiii\ emission at large distances, up to 20 kpc from the bright nucleus. This extended gas was reported in \citep{Prieto1993} through the use of narrow band images at \oiii\ and  \ha\ + \nii. The MUSE emission-line image reveals a detail picture of the \oiii\ filaments. Its high ionization (\oiii/\hb\ $\sim 10$) points the cental AGN as responsible of the ionization.  Nevertheless pure photoionization by the AGN could not explain the relative constant high values of the \oiii/\hb\ ratio at kiloparsec scales. The \ha\ kinematics shows that the gas is more  or less ordered, with velocities ranging from  $\pm  320$ \kms.

{ 3C\,277.3:}
Similar to 3C\,227, 3C\,277.3 shows \oiii\ blobs but connected with \ha\ filaments (see the emission-line image). Past integral field studies of this object has been made using the INTEGRAL spectrograph \citep[e.g.,][]{Solorzano2003}, but with a more limited FoV ($14.6\arcsec \times 11.3\arcsec$), even so a blob of \oiii\ and \ha\ are clearly observed to the south of the nucleus. The wide FoV of MUSE reveals in the emission-line image three major \oiii\ blobs, two at the SE and one to  the NW, the three almost aligned in the same PA. The \oiii\ blobs shows high values of the \oiii/\hb\ ratio being well above the K01 curves. The \ha\ kinematics is more or less ordered, with velocities ranging from  $\pm  330$ \kms. 

{ NGC\,5920:}
NGC\,5920 is the brightest galaxy from the MKW3s group. Past integral field studies of this object with GMOS (FoV $5\arcsec \times  7\arcsec$) has revealed an elongated \ha\ $+$ \nii\ structure close to the nucleus \citep[e.g.,][]{Edwards2009}. The MUSE emission-line image reveals two elongated structures, one directly  connected to the nucleus of NGC\,5920,  and other structure to the South that seems to 
spatially disconnected from the previous. The brightest \ha\ spot in the Southern structure is located $~6\arcsec$ far from the optical nucleus of the closest galaxy companion,  NFP\,J152152.3+074218  (15h21m52.27s, $+$07d42m17.7s). The ionized gas kinematics of both structures do not show a regular rotation pattern.

{ MCG-05-29-017:}
This is an interacting system with its closest companion at the Northern, ESO\,440-58, separated by $11.8\arcsec$ (scale 0.469 kpc/\arcsec). MCG-05-29-017 is a LIRG ($\log\,L_{IR}/L_\odot = 11.37$) \citet{Monreal2010}. VLT-VIMOS observations of this object is presented in \citet{Monreal2010}. The emission-line image shows \nii\ fillaments  emanating from the central region. These filaments present a higher ionization than the gas distributed across the galaxy disk. This is more clear when it is observed the spatially resolved diagrams, where the separation of colors (green to reddish) indicated the different ionization condition in the gas. Fillaments are located above the K01 curves. The increase in the velocity dispersion ($>60$ \kms) and the high asymmetry values ($>50$ \kms) in the filaments reveal the precense of shocked gas.  

{ NGC\,7174:}
It belongs to the HCG90 \citep{Hickson1989}. The continuum image reveals a tidal tail in NGC\,7174 probably due by the interaction with it companions (NGC\,7176, NGC\,7174).  The emission-line image shows  filaments of ionized gas,  \nii mostly, at both sides of the disk plane. This gas  is consistent ob being ionized by shocks given its location in the spatially resolved diagrams and its high velocity dispersion ($50-100$ \kms). It is not conclusive  that the observed filaments are related to a SF-outflow event or is product of the tidal forces.

{ NGC\,4325:}
It is an elliptical galaxy ($\log \mathrm{M/M\odot} = 11.1$) at $z = 0.0255$. \ha\ images reveals radial filaments \citet{McDonald2011}. Integral  field studies of the central region
is presented in \cite{Hammer2016}. The MUSE emission-line image reveals extended \nii\ filaments. The  filamentary structure rule out HOLMES as responsible of the observed ionization, even though it present \EWha~$<3$ \AA.  Shocks seems  to be the most likely explanation to the observed line ratios.

{ NGC\,0034:}
It is a LIRG result of a past merger event, the continuum image reveals a tidal tail. IR observation suggest it present a central starburst \citep[e.g.,][]{Esquej2012}. A strong neutral outflow has been detected in this object with outflow velocities  (blueshifted) $> 1000$ \kms\ \citep[e.g.,][]{Schweizer2007}. A nuclear  cone nebular and arc shape structure is observed in the MUSE emission-line image (most prominent in  \nii). This gas is  consistent of  being ionized  by shock. It can be the optical counterpart of the previously detected neutral outflow.

{  NGC\,5010:}
This is an edge one galaxy  that exhibits what is seems a biconical  outflow. The extraplanar ionized gas (observed in red colors in the corresponding figure) present \EWha~$>3$  \AA.  The low values of the \EWha\ rule out the possible interpretation of a extraplanar diffuse ionized gas. A correlation between $\sigma$ and the \nii/\ha\ ratio is observed in this region, supporting the idea of shock ionization. 

{ ESO364-18: E}
This elliptical galaxies shows extended \nii\ filaments with low values of the \EWha ($<3$ \AA). Although the filaments are located in the shock region in the diagnostic diagrams, ionization by HOLMES is not totally ruled out. The filaments present a disordered kinematic and high velocity dispersion ($>50$ \kms). These high vales in $\sigma$ could support the the scenario of shocked gas.

{ NGC\,89:}
This is a spiral galaxy  classified as S0-a at $z = 0.011$. It shows a bright \ha\ nucleus, with collimated ionized gas filaments outflowing from its central region (reddish colors in the corresponding figure). Its nucleus shows a composite ionization  SF-AGN.

{ IC\,4374:}
This is an elliptical galaxy  at $z = 0.0217$. It is the brightest galaxy from the Abell 3581 cluster, and is classified as Fanaroff-Riley I radio galaxy. Multi-wavelength studies of this object has been performed \citep[e.g.,][]{Johnstone2005,Farage2012,Canning2013,Olivares2019}. 
IC\,4374 shows two radio lobes (at 1.4 GHz) extending to the east and west from its nucleus with size $\sim 4.6$ kpc \citep{Farage2012}. These radio lobes coincide spatially with two cavities in X-ray images which support the idea that a radio jet displaces the hot gas as it expand over the intracluster medium \citep{Johnstone2005}. In the optical it  presents fillamentary ionized emission \citep{Farage2012}. These filaments, mainly observed in \ha\ and \nii, seem to emanate from the nucleus, and they are apparently coincident with the previously detected X-ray bubbles \citep[e.g.,][]{Canning2013}. 
The MUSE emission-line image reveals with unprecedented detail the fillamentary structure around IC\,4374. A clear arc shaped structure is observed extending to the NE from the optical nucleus, and a small fillament originated from the nucleus and toward the north is apprectiated. Filaments with low \SN\ are observed at the North and West sides.

Is not obvious what the ionization source across these filaments is. The spatial diagnostic diagrams point towards  the ionization being LINER like. This seems to be true for the spaxels close the nucleus where low \EWha\ are found. Nevertheless, along the filaments the ionization is not dominated by old stars. Shock ionization can reproduce the observed line ratios. The mechanising driving the gas can not be due to SF-driven winds, neither AGN-winds with the LINER assumption. The interaction of the radio jet with the ISM can produce shocks that explain the observed ratios. 
The \ha\ velocity map reveals a quite complex kinematics in the filaments, without a clear rotation pattern. The asymmetries map reveals  important changes in the $|\Delta\mathrm{v}|$ map ( $<50$ \kms), which could indicate the presence of multi components along the filaments.

{ NGC\,4486:}
The well known elliptical galaxy M87 shows ionized gas filaments, detected before in narrow band filters around the \ha\ line  \citep[e.g.,][]{Jarvis1990,Sparks1993,Gavazzi2000}. As the spatially resolved diagnostic diagrams show, 
the \ha\ + \nii\ filaments present line ratios well above the K01 curves. Interestingly, even though the \EWha in the filaments is lower than 3\AA, and therefore should classified as HOLMES ionization, given the large separation to the nucleus of M87 (up to 15 kpc, \cite{Gavazzi2000}), post AGB stars being responsible for the low equivalent widths is 
unlikely. Shock ionization is a plausible explanation to the observed ratios. This suggests that in absence of any other ionization source, shocks can present \EWha\ $< 3$\AA.
A complex kinematics is observed along the filaments, without a clear patter of regular rotation. No signs of double components are observed.

{ NGC\,4936:}
NGC\,4936 is an elliptical galaxy surrounded by fillamentary ionized gas, mainly detected in \nii. Its nucleus as well the gas in the filaments fall in the LINER region of the diagnostic diagrams, and is characterized for having low values of the \EWha. The \ha\ kinematics show a regular rotation pattern, with no clear deviation in the asymmetries.

{ NGC\,4941:}
Using the GMOS-IFU, \cite{Barbosa2009} detected a compact outflow associated to a radio jet. This outflow is detected in the emission-line image as pinky structure, following closelly a spiral arm. Velocity dispersion larger than $50$ \kms\ is predominant over this structure. 
The high values of the \oiii/\hb\ ratio throughout this structure position it in the AGN region in the diagnostic diagrams, which suggest this last as responsible of the observed  line ratios.

{ PGC\,013424:}
PGC\,013424 is the brightest galaxy from the  2A 0335+096 galaxy cluster. It is a massive elliptical galaxy M$_\star  = 10^{11.3}$ M$_\sun$.
Ionized gas filaments around PGC\,013424 were first revealed with narrow band \ha\ and \nii\  images \citep[e.g.,][]{Donahue2007}. Integral field studies of this object have been made to study the ionized gas component \citep[e.g.,][]{Farage2012}. The MUSE emission-line image reveals extended  filaments of \nii\ with null contribution
of  HOLMES in its ionization. Therefore a suitable explanation can be shocks.

{ PGC\,015524:}
This is a massive elliptical galaxy (M$_\star  = 10^{11.3}$ M$_\sun$) and is the brightest galaxy in the A946 cluster \citep[e.g.,][]{Lin2004}. It presents multiple filaments of \nii\ and \ha\ emission. The  line-of-sight velocity \ha\ velocity do not allow to reveal if these filaments are being ejected from the nucleus or are in-falling filaments of gas. Its ionization is LINER like, although shocks can also reproduce the observed ratios.

{ UGC\,09799:}
UGC\,09799 is a radio galaxy located at the center of the  Abell cluster A2052. This galaxy present arc shape filaments  dominated by \nii\ emission. This galaxy has been studied in detail in \cite{Balmaverde2018}. They found that the filaments are the result of the expansion of the radio lobes produced by the central AGN. The nucleus as well the filaments falls in the LINER like region in the diagnostic diagrams. The ionization in the nucleus can be explained with HOLMES meanwhile shocks can reproduce the observed ratios in the filaments.

\section{outflows in AMUSING++}\label{appendix:outflows_amusing}

In this section the spatially resolved diagrams and kinematic maps for the outflows detected in AMUSING++ are shown.
We present four examples of outflows driven by different mechanism: AGN--driven, SF--driven, stellar feedback and merger--driven outflows. The rest of maps for the outflow host galaxies are accessible via the electronic version of this paper.

\begin{figure*}
    \centering
    \includegraphics[]{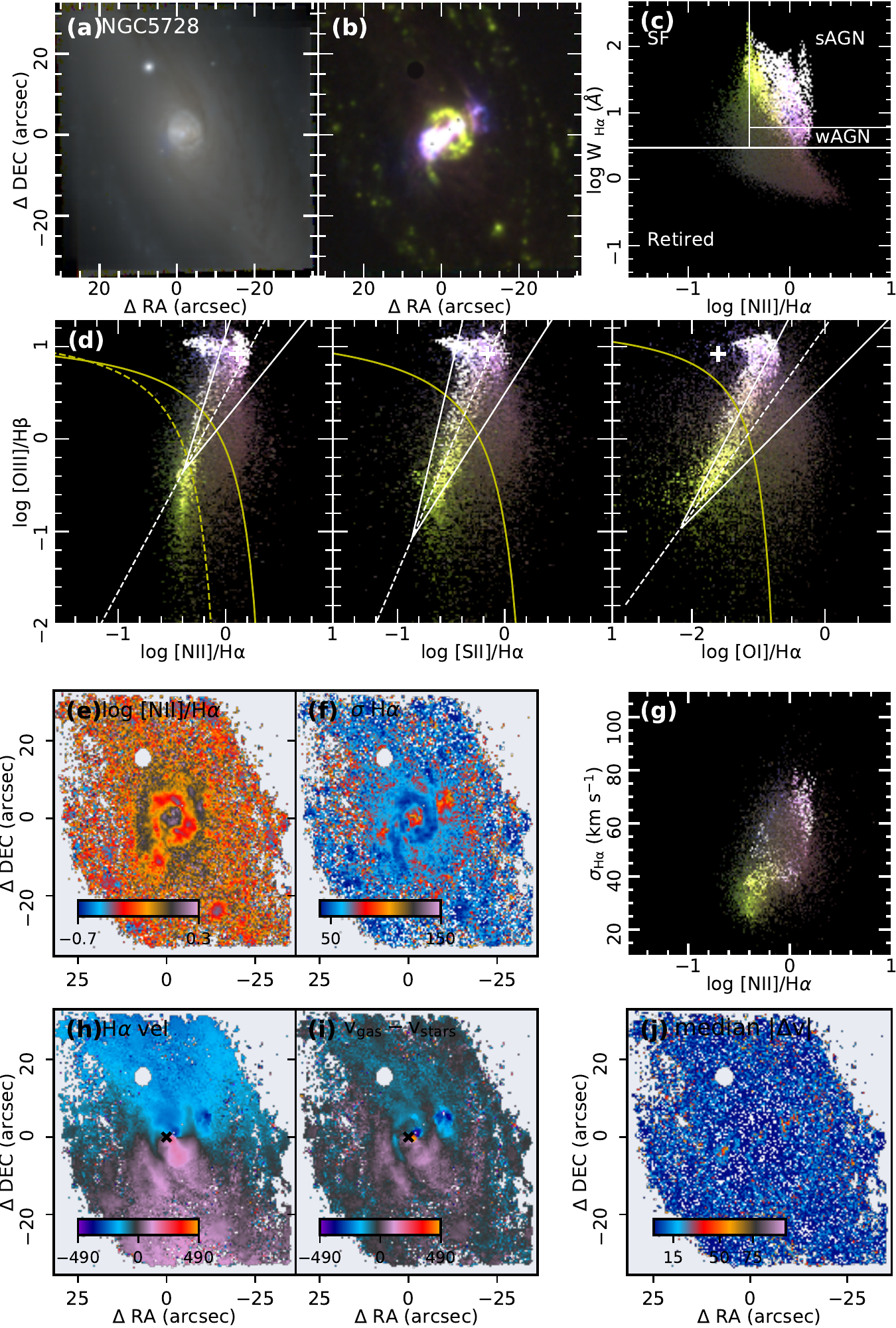}
        \caption{Galactic outflows found in the AMUSING++ compilation. (a) Reconstructed continuum image (red: $i$--band, green: $r$--band and blue: $g$--band); (b) RGB emission-line image (red: \nii, green: \ha\ and blue: \oiii); (c) WHAN diagram color coded with the emission-line image from panel (b);
    (d) Spatially resolved diagnostic diagrams color coded with the emission-line image from panel (b). Demarcation lines have the same meaning that those presented in Fig.~\ref{fig:eline_SN2012hd}; (e) \nii/\ha\ line ratio map with a cut in the signal to noise in both emission lines, \SN $>4$. (f) Velocity dispersion map estimated from the emission line fitting analysis; (g) Spatially resolved $\sigma$ vs.  $\log$ \nii/\ha\ diagram; (h) \ha\ velocity map; (i) Stellar velocity derived with the SSP analysis; (j) 2D map of the absolute value of the asymmetries ($|\mathrm{\Delta v}|$). The outflow in this galaxy is associated to an  AGN--driven wind}. 
    \label{fig:appendix_1}
\end{figure*}
\addtocounter{figure}{-1}
\begin{figure*}
    \centering
    \includegraphics[]{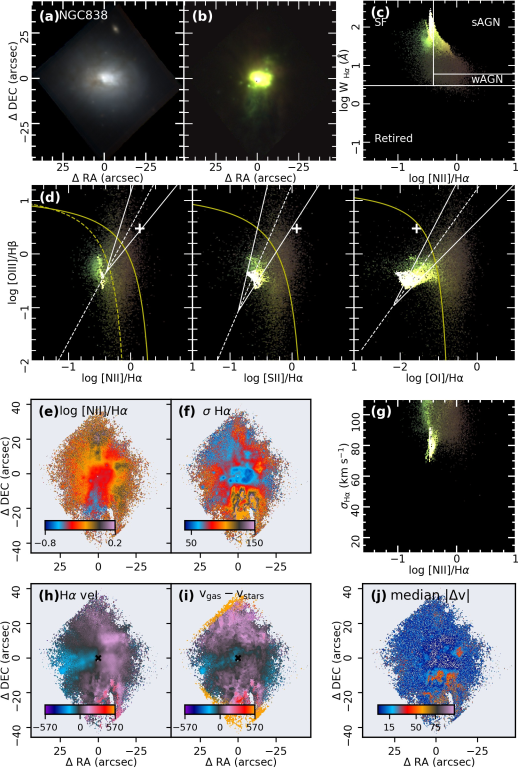}
    \caption{({\it continued}). The outflow in this galaxy is associated to a  SF--driven wind.}
\end{figure*}
\addtocounter{figure}{-1}
\begin{figure*}
    \centering
    \includegraphics[]{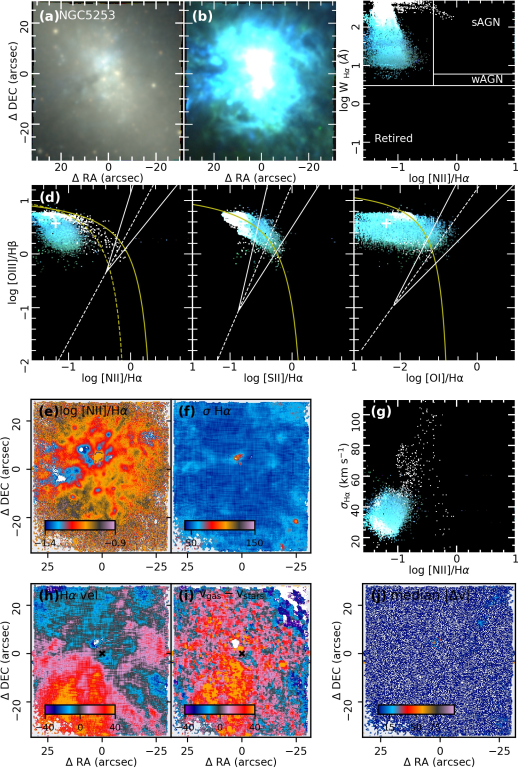}
    \caption{({\it continued}). The outflow in this galaxy is associated with stellar feedback.}
\end{figure*}
\addtocounter{figure}{-1}
\begin{figure*}
    \centering
    \includegraphics[]{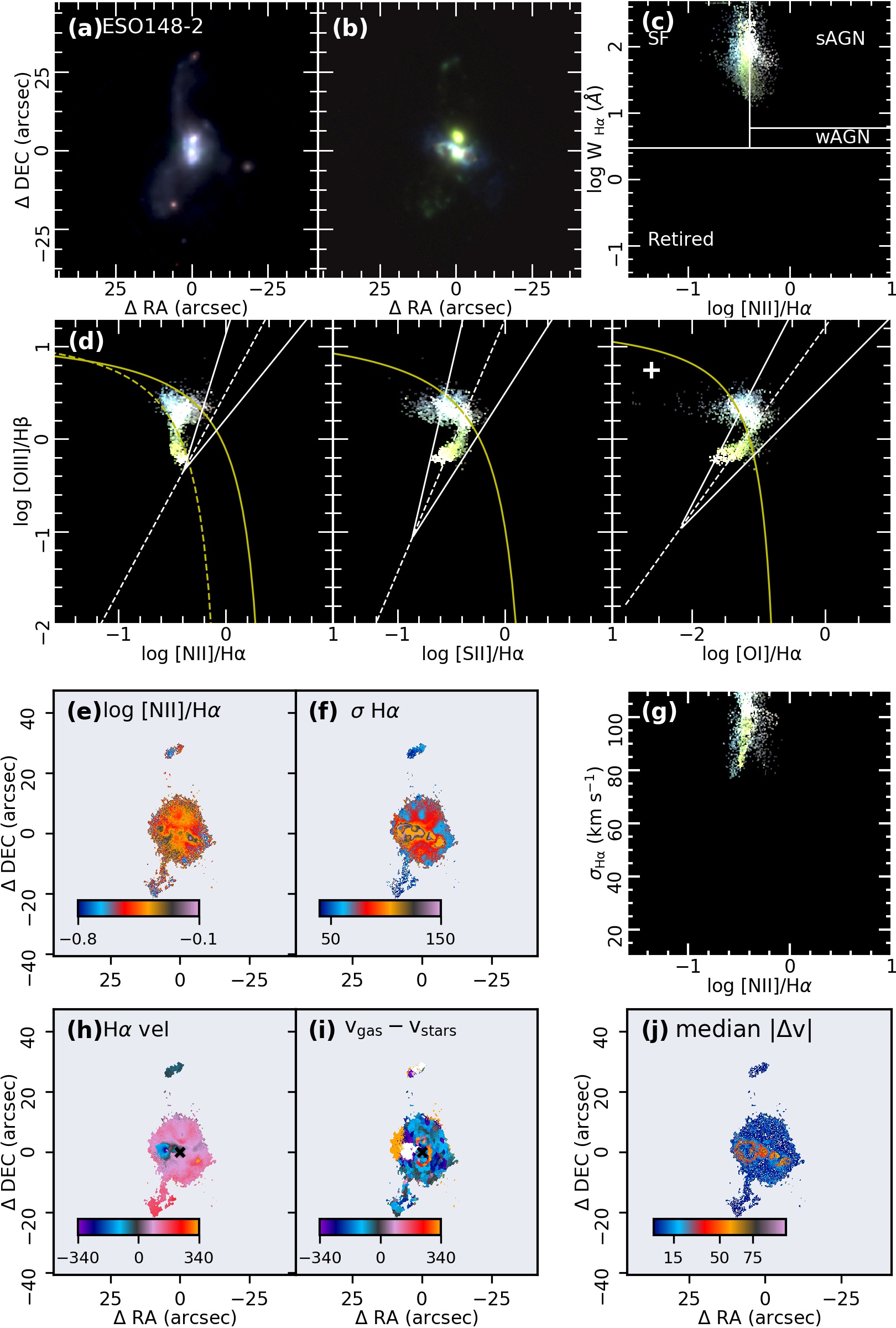}
    \caption{({\it continued}).The outflow in this galaxy is merger--driven. }
\end{figure*}
\begin{figure*}
    \centering
    \includegraphics[]{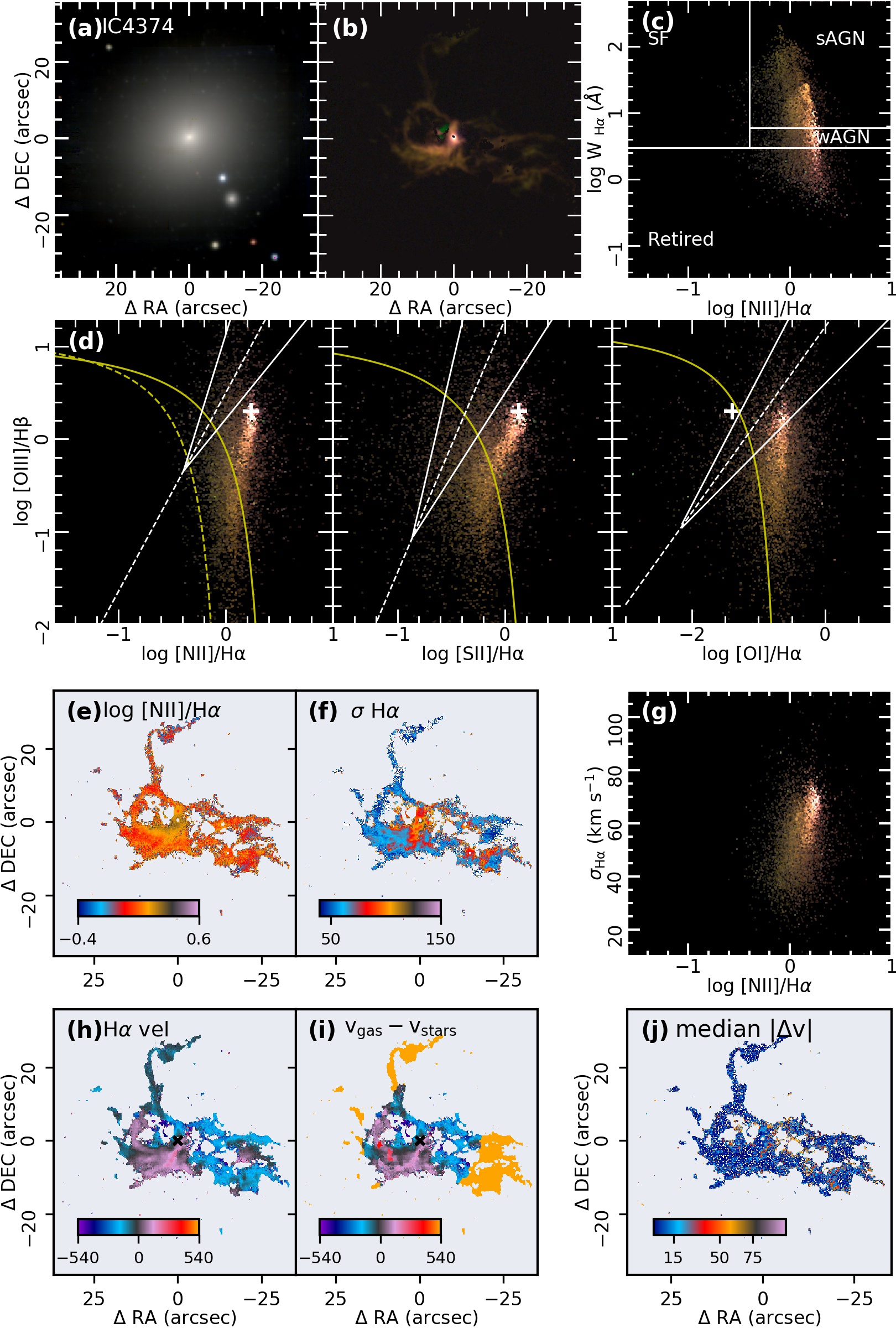}
    \caption{Elliptical galaxies with extended emission  that do not fulfil the outflow requirements. The remaining objects are found in the electronic version.}
    \label{fig:appendix_eliptics}
\end{figure*}

\begin{longrotatetable}
\begin{deluxetable*}{llcccccccccccrc}
\tablecaption{Isophotal parameters derived for the AMUSING++ galaxies. The remaining objects are found in the electronic version. \label{tab:all_amusing_table}}
\tablewidth{700pt}
\tabletypesize{\scriptsize}
\tablehead{
\colhead{index} & \colhead{AMUSING++} & \colhead{galaxy} & \colhead{RA} & \colhead{DEC} & \colhead{z} & \colhead{scale} & \colhead{Hubble} & \colhead{log SFR} & \colhead{log Mass} & \colhead{incl.} & \colhead{PA} & \colhead{R$_e$} & \colhead{$r$} & \colhead{$|v_{gas}-v_{\star}|$} \\
\colhead{} & \colhead{id} & \colhead{} & \colhead{h:m:s} & \colhead{d:m:s} & \colhead{} & \colhead{(kpc/$\arcsec$)} & \colhead{type} & \colhead{(M$_{\odot}$\,yr$^{-1}$)} & \colhead{(M$_{\odot}$)} & \colhead{(deg)} & \colhead{(deg)} & \colhead{($\arcsec$)} & \colhead{} & \colhead{(\kms)}
}
\startdata
0 & 2MIG & NGC3546 & 01:10:34.06 & -52:33:23.508 & 0.0248 & 0.5 & E-S0 & -1.2341 & 11.26 & 54.5 & 29.1 & 5.5 & \nodata,-0.14 & \nodata \\
1 & 3C029 & UGC595 & 00:57:34.88 & -1:23:27.564 & 0.0451 & 0.888 & E & -1.4966 & 11.45 & 11.3 & 114.7 & 6.3 & 0.05,0.14 & 94.0 \\
2 & Abell & IC4374 & 14:07:29.8 & -27:01:5.988 & 0.0217 & 0.44 & E-S0 & -0.4368 & 10.91 & 38.2 & 17.9 & 10.3 & 0.04,0.25 & 234.0 \\
3 & Antennae & ARP244 & 12:01:50.638 & -18:52:10.956 & 0.0056 & 0.114 & None & 0.316 & 10.25 & 70.1 & 91.5 & 7.1 & 0.31,0.04 & 87.0 \\
4 & ASASSN13an & PGC170294 & 13:45:36.527 & -7:19:32.196 & 0.0243 & 0.49 & Sa & 1.3905 & 10.73 & 27.2 & 138.8 & 3.9 & 0.31,0.3 & 20.0 \\
5 & ASASSN13bb\_1 & UGC1395 & 01:55:23.067 & 06:36:24.012 & 0.0172 & 0.35 & Sb & -2.009 & 5.13 & 41.6 & 60.5 & 14.6 & 0.18,0.3 & 57.0 \\
\enddata
\tablecomments{AMUSING++ identification (col. 2), galaxy names (col. 3), right ascension (col. 4), declination (col 5.), redshift derived with the SSP analysis (col. 6), angular scale (col. 7), Hubble type from Hyperleda  (col. 8), integrated SFR (col. 9), integrated stellar mass (col. 10), inclination (col. 11), position angle (col. 12), effective radius (col. 13),  { correlation coefficients between $\sigma$ and the $\log$~\nii/\ha\ ratio for spaxels lying bellow and above the \citet{kewley01} curve (col. 14), W90 of the difference between the gas and stellar kinematics for spaxels liying above the \citet{kewley01} curve (col. 15).}
For those galaxies where the optical diameter is not covered entirely by the FoV of MUSE, the values R$_{25}$ and R$_e$ are estimated by extrapolation up to R$_{25}$, therefore for those cases, R$_e$ should be considered just as an approximated value, as well as their integrated properties such as stellar mass, and SFR. Those galaxies are marked with an $(*)$.  The position angle is measured from the West.}
\end{deluxetable*}
\end{longrotatetable}

\section{Spatially resolved diagrams for AMUSING++ galaxies}
\begin{figure*}
    \centering
    \includegraphics[]{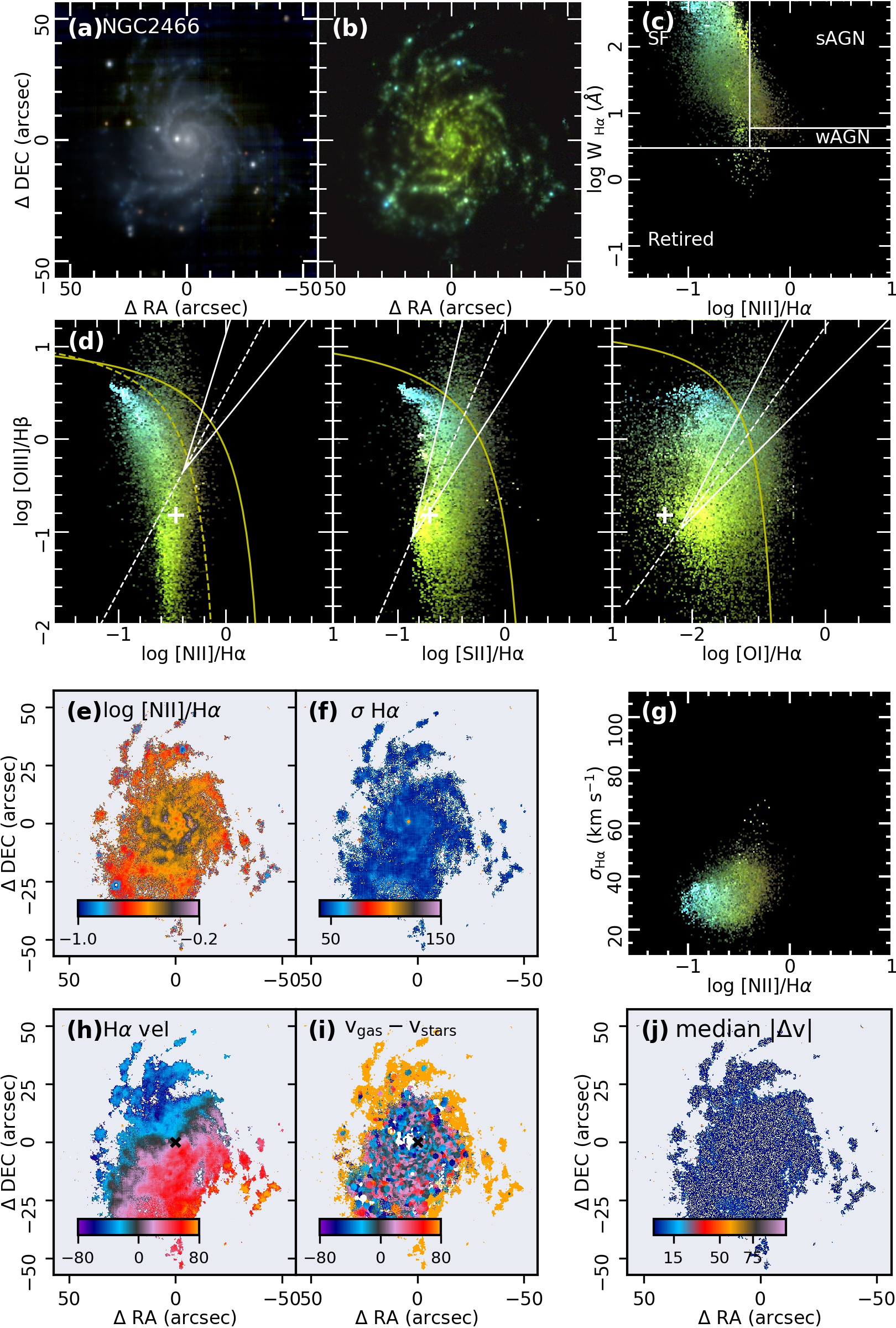}
    \caption{Spatially resolved diagrams for the AMUSING++ galaxies. (a) Reconstructed continuum image (red: $i$--band, green: $r$--band and blue: $g$--band); (b) RGB emission-line image (red: \nii, green: \ha\ and blue: \oiii); (c) WHAN diagram color coded with the emission-line image from panel (b);
    (d) Spatially resolved diagnostic diagrams color coded with the emission-line image from panel (b). Demarcation lines have the same meaning that those presented in Fig.~\ref{fig:eline_SN2012hd}; (e) \nii/\ha\ line ratio map with a cut in the signal to noise in both emission lines, \SN $>4$. (f) Velocity dispersion map estimated from the emission line fitting analysis; (g) Spatially resolved $\sigma$ vs.  $\log$ \nii/\ha\ diagram; (h) \ha\ velocity map; (i) Stellar velocity derived with the SSP analysis; (j) 2D map of the absolute value of the asymmetries ($|\mathrm{\Delta v}|$). The remaining galaxies of the AMUSING++ compilation can be found in the web page: \url{http://ifs.astroscu.unam.mx/AMUSING++/}.}
    \label{fig:appendix_all_AMUSING}
\end{figure*}

In this section the spatially resolved diagrams and kinematic maps for all the AMUSING++ galaxies are shown.
As example we show  NGC\,2466 in Fig.~\ref{fig:appendix_all_AMUSING}. The remaining maps of the AMUSING++ galaxies are accessible via the following web page: \url{http://ifs.astroscu.unam.mx/AMUSING++/}.
 \label{appendix:spatial_resolved_AMUSING}





\end{document}